\documentclass[11pt]{article}
\usepackage{latexsym}
\usepackage{amsmath}
\usepackage{amssymb}
\usepackage{amsfonts}
\usepackage{color}
\usepackage{showlabels}
\def\utw{\smash{\rlap{\lower5pt\hbox{$\sim$}}}}
\def\udtw{\smash{\rlap{\lower6pt\hbox{$\approx$}}}}

\def\bbbone{{\mathchoice {\rm 1\mskip-4mu l} {\rm 1\mskip-4mu l}
{\rm 1\mskip-4.5mu l} {\rm 1\mskip-5mu l}}}
\def\bbbc{{\mathchoice {\setbox0=\hbox{$\displaystyle\rm C$}\hbox{\hbox
to0pt{\kern0.4\wd0\vrule height0.9\ht0\hss}\box0}}
{\setbox0=\hbox{$\textstyle\rm C$}\hbox{\hbox
to0pt{\kern0.4\wd0\vrule height0.9\ht0\hss}\box0}}
{\setbox0=\hbox{$\scriptstyle\rm C$}\hbox{\hbox
to0pt{\kern0.4\wd0\vrule height0.9\ht0\hss}\box0}}
{\setbox0=\hbox{$\scriptscriptstyle\rm C$}\hbox{\hbox
to0pt{\kern0.4\wd0\vrule height0.9\ht0\hss}\box0}}}}
\def\bbbe{{\mathchoice {\setbox0=\hbox{\smalletextfont e}\hbox{\raise
0.1\ht0\hbox to0pt{\kern0.4\wd0\vrule width0.3pt
height0.7\ht0\hss}\box0}}
{\setbox0=\hbox{\smalletextfont e}\hbox{\raise
0.1\ht0\hbox to0pt{\kern0.4\wd0\vrule width0.3pt
height0.7\ht0\hss}\box0}}
{\setbox0=\hbox{\smallescriptfont e}\hbox{\raise
0.1\ht0\hbox to0pt{\kern0.5\wd0\vrule width0.2pt
height0.7\ht0\hss}\box0}}
{\setbox0=\hbox{\smallescriptscriptfont e}\hbox{\raise
0.1\ht0\hbox to0pt{\kern0.4\wd0\vrule width0.2pt
height0.7\ht0\hss}\box0}}}}
\def\bbbq{{\mathchoice {\setbox0=\hbox{$\displaystyle\rm Q$}\hbox{\raise
0.15\ht0\hbox to0pt{\kern0.4\wd0\vrule height0.8\ht0\hss}\box0}}
{\setbox0=\hbox{$\textstyle\rm Q$}\hbox{\raise
0.15\ht0\hbox to0pt{\kern0.4\wd0\vrule height0.8\ht0\hss}\box0}}
{\setbox0=\hbox{$\scriptstyle\rm Q$}\hbox{\raise
0.15\ht0\hbox to0pt{\kern0.4\wd0\vrule height0.7\ht0\hss}\box0}}
{\setbox0=\hbox{$\scriptscriptstyle\rm Q$}\hbox{\raise
0.15\ht0\hbox to0pt{\kern0.4\wd0\vrule height0.7\ht0\hss}\box0}}}}
\def\bbbt{{\mathchoice {\setbox0=\hbox{$\displaystyle\rm
T$}\hbox{\hbox to0pt{\kern0.3\wd0\vrule height0.9\ht0\hss}\box0}}
{\setbox0=\hbox{$\textstyle\rm T$}\hbox{\hbox
to0pt{\kern0.3\wd0\vrule height0.9\ht0\hss}\box0}}
{\setbox0=\hbox{$\scriptstyle\rm T$}\hbox{\hbox
to0pt{\kern0.3\wd0\vrule height0.9\ht0\hss}\box0}}
{\setbox0=\hbox{$\scriptscriptstyle\rm T$}\hbox{\hbox
to0pt{\kern0.3\wd0\vrule height0.9\ht0\hss}\box0}}}}
\def\bbbs{{\mathchoice
{\setbox0=\hbox{$\displaystyle     \rm S$}\hbox{\raise0.5\ht0\hbox
to0pt{\kern0.35\wd0\vrule height0.45\ht0\hss}\hbox
to0pt{\kern0.55\wd0\vrule height0.5\ht0\hss}\box0}}
{\setbox0=\hbox{$\textstyle        \rm S$}\hbox{\raise0.5\ht0\hbox
to0pt{\kern0.35\wd0\vrule height0.45\ht0\hss}\hbox
to0pt{\kern0.55\wd0\vrule height0.5\ht0\hss}\box0}}
{\setbox0=\hbox{$\scriptstyle      \rm S$}\hbox{\raise0.5\ht0\hbox
to0pt{\kern0.35\wd0\vrule height0.45\ht0\hss}\raise0.05\ht0\hbox
to0pt{\kern0.5\wd0\vrule height0.45\ht0\hss}\box0}}
{\setbox0=\hbox{$\scriptscriptstyle\rm S$}\hbox{\raise0.5\ht0\hbox
to0pt{\kern0.4\wd0\vrule height0.45\ht0\hss}\raise0.05\ht0\hbox
to0pt{\kern0.55\wd0\vrule height0.45\ht0\hss}\box0}}}}

\def\bbbz{{\mathchoice {\hbox{$\sf\textstyle Z\kern-0.4em Z$}}
{\hbox{$\sf\textstyle Z\kern-0.4em Z$}}
{\hbox{$\sf\scriptstyle Z\kern-0.3em Z$}}
{\hbox{$\sf\scriptscriptstyle Z\kern-0.2em Z$}}}}

\def\diameter{{\ifmmode\oslash\else$\oslash$\fi}}
\def\init{\setcounter{equation}{0}}

\newtheorem{theoreme}{Theorem }[section]
\newtheorem{proposition}[theoreme]{Proposition}
\newtheorem{lemma}[theoreme]{Lemma}
\newtheorem{definition}[theoreme]{Definition}

\newtheorem{remark}[theoreme]{Remark}

\def\rr{\mathbb{R}}
\def\cc{\mathbb{C}}
\def\nn{\mathbb{N}}
\def\zz{\mathbb{Z}}
\def\one{\bbbone}

\def\fin{{\rm fin}}

\def\Wick{{\rm Wick}}

\def\ie{{\it i.e., }}
\def\fld{\rightarrow}

\def\e{{\rm e}}

\def\i{{\rm i}}
\def\d{{\rm d}}
\def\12{\frac{1}{2}}
\def\cinf{C^{\infty}}
\def\ie{{\it i.e.,  }}
\def\proof{\noindent{\bf  Proof. }}
\def\slim{\hbox{\rm s-}\lim}

\def\coinf{C_{0}^{\infty}}

\def\qed{$\Box$}

\def\supp{{\rm supp\,}}
\def\cH{{\cal H}}
\def\cS{{\cal S}}
\def\G{\Gamma}

\def\cD{{\cal D}}

\def\ch{{\mathfrak h}}

\def\p{\partial}
\def\s{{\rm s}}

\def\x{\langle x\rangle}

\def\hc{{\rm h.c.}}

\def\pfi2{P(\varphi)_{2}}

\def\Op{{\rm Op}}
\def\Gh{\Gamma(\ch)}
\def\ad{{\rm ad}}
\def\tD{\langle D\rangle}
\newcommand{\beq}{\begin{equation}}
\newcommand{\eeq}{\end{equation}}
\newcommand{\bet}{\begin{theoreme}}
\newcommand{\eet}{\end{theoreme}}
\newcommand{\bel}{\begin{lemma}}
\newcommand{\eel}{\end{lemma}}
\newcommand{\bep}{\begin{proposition}}
\newcommand{\eep}{\end{proposition}}

\newcommand{\bear}[1]{\begin{array}{#1}}
\newcommand{\ear}{\end{array}}

\begin{document}
\title{Spectral and scattering theory for \\ space-cutoff
$P(\varphi)_{2}$ models with variable metric}\author{C. G\'erard, A. Panati, \\ Laboratoire de math\'ematiques, Universit\'e de Paris XI,\\
91\,405 Orsay Cedex France}\date{June 2008}
\maketitle
\begin{abstract}
We consider space-cutoff $P(\varphi)_{2}$ models with a variable metric
of the form
\[
H= \d\G(\omega)+ \int_{\rr}g(x):\!P(x, \varphi(x))\!:\d x,
\]
on the bosonic Fock space $L^{2}(\rr)$, where the kinetic energy
$\omega= h^{\12}$ is the square root of a real second order
differential operator
\[
h= Da(x)D+ c(x),
\]
where the coefficients $a(x), c(x)$ tend respectively to $1$ and
$m_{\infty}^{2}$ at $\infty$ for some  $m_{\infty}>0$.

The interaction term $\int_{\rr}g(x):\!P(x, \varphi(x))\!:\d x$ is
defined using a bounded below polynomial in $\lambda$ with
variable coefficients $P(x, \lambda)$ and a positive function $g$
decaying fast enough at infinity.

We extend in this paper the results of   \cite{DG} where $h$
had constant coefficients and $P(x, \lambda)$ was independent of $x$.

We describe the essential spectrum of $H$, prove a Mourre estimate
outside a set of thresholds and prove the existence of asymptotic
fields. Our main result is the {\em asymptotic completeness} of the
scattering theory, which means that the CCR representation given by
the asymptotic fields is of Fock type, with the asymptotic vacua equal
to bound states of $H$. As a consequence $H$ is unitarily equivalent
to a collection of second quantized Hamiltonians.

An important role in
the proofs is
played by the {\em higher order estimates}, which allow to control
powers of the number operator by powers of the resolvent. To obtain
these estimates some conditions on the eigenfunctions and generalized
eigenfunctions of $h$ are necessary. We also discuss similar models in
higher space dimensions where the interaction has an ultraviolet
cutoff.
\end{abstract}
\section{Introduction}\init\label{sec0}
\subsection{Space-cutoff
$P(\varphi)_{2}$ models with variable metric}\label{sec0.1}
The $P(\varphi)_{2}$ {\em model} describes a self-interacting   field of scalar
 bosons in 2 space-time dimensions with the
interaction given by a  bounded below polynomial $P(\varphi)$ of
degree at least 4.  Its  construction in the seventies by Glimm
and Jaffe (see e.g. \cite{GJ}) was one of the early successes of
constructive field theory.  The first step of the construction
relied on the consideration  of a {\em spatially cutoff}
$P(\varphi)_{2}$ interaction, where the cutoff is defined with a
positive coupling function $g(x)$ of compact support.  The formal
expression
\[
H =\d\G(\omega)+ \int_{\rr}g(x):\!P(\varphi(x))\!:\d x,
\]
where $\omega= (D^{2}+ m^{2})^{\12}$ for $m>0$ and $:\ \ :$
denotes the {\em Wick ordering},   can be given a rigorous
meaning as a bounded below selfadjoint Hamiltonian on the Fock space
$\G(L^{2}(\rr))$.

 The spectral and scattering theory of $H$ was studied in \cite{DG}
by adapting methods originally developped for $N-$particle Schr\"{o}dinger operators.

Concerning spectral theory, an HVZ theorem describing the
essential spectrum of $H$ and a Mourre positive commutator estimate
were proved in \cite{DG}. As consequences of the Mourre estimate, one
obtains as usual the local finiteness of point spectrum outside of the
threshold set and, under additional assumptions, the limiting absorption
principle.

The scattering theory of $H$ was treated in \cite{DG} by the standard
approach consisting in constructing first the {\em asymptotic fields},
which roughly speaking are the limits
\[
\lim_{t\to \pm\infty}\e^{\i tH}\phi(\e^{-\i t \omega}h)\e^{-\i tH}=:
\phi^{\pm}(h), \ \ h\in L^{2}(\rr),
\]
where $\phi(h)$ for $h\in L^{2}(\rr)$ are the Segal field operators. Since the model is massive, it is rather easy to see that the two
CCR representations
\[
h\mapsto \phi^{\pm}(h)
\]
are unitarily equivalent to a direct sum of Fock representations.
The central problem of scattering theory becomes then the description
of the space of {\em vacua} for these asymptotic representations. The
main result of \cite{DG} is the {\em asymptotic completeness}, which
says that the asymptotic vacua coincide with the bound states of $H$.
It implies that under time evolution any initial state eventually
decays into a superposition of bound states of $H$ and  a finite
number of asymptotically free bosons.

Although the Hamiltonians $H$ do not describe any real physical
system,  they played an important role in the development of
constructive field theory. Moreover they have the important property
that the interaction is {\em local}. As far as we know, the
$P(\varphi)_{2}$ models and the (non-relativistic) {\em Nelson model}
are the only models with local interactions which can be constructed
on Fock space by relatively easy arguments.

\medskip

Our goal in this paper is to extend the results of \cite{DG} to the
case where both the one particle kinetic energy $\omega$ and the
polynomial $P$ have {\em variable coefficients}. More precisely we consider
 Hamiltonians
\[
H= \d\G(\omega)+ \int_{\rr}g(x):\!P(x, \varphi(x))\!:\d x,
\]
on the bosonic Fock space $L^{2}(\rr)$, where the kinetic energy
$\omega= h^{\12}$ is the square root of a real second order
differential operator
\[
h= Da(x)D+ c(x),
\]
$P(x, \lambda)$ is  a variable coefficients
polynomial
\[
P(x,\lambda)=\sum_{p=0}^{2n}a_{p}(x)\lambda^{p}, \ \ a_{2n}(x)\equiv
a_{2n}>0,
\]
and $g\geq 0$ is a function decaying fast enough at infinity. We assume that $a(x), c(x)>0$ and $a(x)\to 1$ and $c(x)\to m_{\infty}^{2}$ when $x\to
\infty$. The constant $m_{\infty}$ has the meaning of the {\em mass at
infinity}. Most of the time we will assume that $m_{\infty}>0$.

As is well known, the Hamiltonian $H$ appears  when  one tries to quantize the
following non linear Klein-Gordon equation with variable coefficients:
\[
\p^{2}_{t}\varphi(t, x)+ (Da(x)D+ c(x))\varphi(t, x)+
g(x)\frac{\p P}{\p \lambda}(x, \varphi(x, t))=0.
\]
Note that in \cite{Di}, Dimock has considered perturbations of the 
full (translation-invariant) $\varphi^{4}_{2}$ model by lower order
perturbations $\rho(t,x):\! \varphi(t,x)\!:$ where $\rho(t,x)$ has
compact support in space-time.

We now describe in more details the content of the paper.
\subsection{Content of the paper}\label{sec0.2}
The first difference between the  $P(\varphi)_{2}$ models with a variable
metric considered in this
paper and the constant coefficients ones considered in \cite{DG} is
 that  the polynomial $P(\lambda)$ is replaced by a variable
coefficients polynomial $P(x, \lambda)$ in the
interaction. The second  is that  the constant coefficients one
particle energy
$(D^{2}+ m^{2})^{\12}$
is replaced by a variable coefficients energy
$(Da(x)D+c(x))^{\12}$.

Replacing $P(\lambda)$ by $P(x, \lambda)$ is rather easy.
Actually,  conditions on the function $g$ and coefficients
$a_{p}$ needed to make sense of the Hamiltonian can be found in
\cite{Si1}.

On the contrary replacing  $(D^{2}+ m^{2})^{\12}$ by
$(Da(x)D+c(x))^{\12}$ leads to new difficulties. The
construction of the Hamiltonian $H$ is still rather easy, using
hypercontractivity arguments.

However an essential tool to study the spectral and scattering theory
of $H$ is the so called {\em higher order estimates}, originally
proved by Rosen \cite{Ro}, an example being the bound
\[
N^{2p}\leq C_{p}(H+b)^{2p}, \ \ p\in \nn.
\]
These bounds are very important to control  various error terms and are a
subsitute for the lack of knowledge of the domain of $H$.

An substantial part of this paper is devoted to the proof of the
higher order estimates in the variable metric case.

\medskip

Let us now describe  in more details the content of the paper.

In Sect. \ref{sec1} we recall various well-known results, like
standard Fock space notations, the notion of Wick polynomials and
results on contractive and hypercontractive semigroups. We also recall
some classical results on pseudodifferential calculus.

The space-cutoff $P(\varphi)_{2}$ model with a variable metric is
described in Sect. \ref{sec2} and its existence and basic properties
are proved in Thms. \ref{basic} and \ref{basic-massless}.

In the massive case $m_{\infty}>0$ we show using standard arguments on
perturbations of hypercontractive semigroups that $H$ is essentially
selfadjoint and bounded below.  The necessary properties of the
interaction
\[
V=\int_{\rr}g(x):\!P(x, \varphi(x))\!:\d x
\]
as a multiplication operator are proved  in Subsect. \ref{sec2.2} using
pseudodifferential calculus and the analogous results known in the
constant coefficients case.

The massless case $m_{\infty =0}$ leads to serious difficulties, even
to obtain the existence of the model. In fact the free semigroup
$\e^{-t\d\G(\omega)}$ is no more hypercontractive if $m_{\infty}=0$
but only $L^{p}-$contractive. Using a result from Klein and Landau
\cite{KL2} we can show that $H$ is essentially selfadjoint
if for example $g$ is compactly supported.  Again the necessary
properties of the interaction are proved in Subsect.
\ref{sec2.massless}.
 The property that $H$ is bounded below remains an open question and
massless models will not be further considered in this paper.

Sect. \ref{spect-scatt} is devoted to the spectral and scattering
theory of $P(\varphi)_{2}$ Hamiltonians with variable metric.
It turns out  that many arguments of \cite{DG} do not rely on the
detailed properties of $P(\varphi)_{2}$ models but can be extended to
an abstract framework.

In \cite{GP} we
consider abstract bosonic QFT Hamiltonians of the form
\[
H = \d\G(\omega)+ \Wick(w),
\]
acting on a bosonic Fock space $\G(\ch)$, where the one particle
energy $\omega$ is a selfadjoint operator on the one particle Hilbert space $\ch$
and the interaction term $\Wick(w)$ is a {\em Wick polynomial} associated to
some kernel $w$.  The spectral and scattering theory of such Hamiltonians
is studied in \cite{GP} under a rather general set of conditions.

The first type of conditions requires that $H$ is essentially
selfadjoint and bounded below and satisfies {\em higher order
estimates}, allowing to bound $\d\G(\omega)$ and powers of the number
operator $N$ by sufficiently high powers of $H$.

The second type of conditions concern the one-particle energy
$\omega$. Essentially one requires that $\omega$ is {\em massive }
i.e. $\omega\geq m>0$ and has a nice spectral and scattering
theory.

The last type of conditions concern the kernel $w$ of the interaction
$\Wick(w)$ and requires some decay properties of $w$ at infinity.

The core of the present paper consists in proving that our
$P(\varphi)_{2}$ Hamiltonians satisfy the hypotheses of \cite{GP}, so
that the results here follow from the abstract theorems in \cite{GP}.

The essential spectrum of $H$ is described in Thm. \ref{mainmain}.
As a consequence one obtains that $H$ has a ground state. The
Mourre estimate  is shown in Thm. \ref{mainim}. We do not prove
the limiting absorption principle, but note that for example 
the absence of singular
continuous spectrum will follow from  unitarity of the wave
operators and asymptotic completeness.

The scattering theory and asymptotic completeness of wave operators,
formulated as explained in Subsect. \ref{sec0.1} using asymptotic
fields, is proved in Thm. \ref{ima}.

Note that even in the constant coefficients case, we
improve the results of \cite{DG}. No smoothness of the
coupling function $g$ is required and  we can  remove an unpleasant
technical assumption on the coupling function $g$ (condition {\it
(Bm)} in \cite[Subsect. 6.2]{DG}) which  excluded for example
 compactly supported $g$.

Analogous results for higher dimensional models where the interaction
has also an ultraviolet cutoff are described in Sect. \ref{higher}.

The properties of the interaction $\int_{\rr}g(x):\!P(x,
\varphi(x))\!:\d x$  needed in Sect. \ref{sec2}
are proved in Sect. \ref{kernel}. In this section the interaction is
considered as a Wick polynomial.

In Sect. \ref{lowersec} we prove some lower bounds on perturbations of
$P(\varphi)_{2}$ Hamiltonians which will be needed in Sect.
\ref{sec4}.

Sect. \ref{sec4} is devoted to the proof of the higher order
estimates. It turns out that the method of Rosen \cite{Ro} uses in
an essential way the fact that $D^{2}+ m^{2}$ has the  family
$\{\e^{\i kx}\}_{k\in\rr}$ as a basis of generalized eigenfunctions and
that the functions $\e^{\i kx}$ are {\em uniformly bounded} both in
$x$ and $k$. In our case we have to use instead of $\{\e^{\i
kx}\}_{k\in\rr}$ a family of eigenfunctions and generalized
eigenfunctions for $Da(x)D+ c(x)$. It is necessary to impose
some bounds on these functions to substitute for the uniform
boundedness property in the constant coefficients case. These bounds
are stated  in Sect. \ref{sec4}  as  conditions {\it (BM1)}, {\it
(BM2)} and deal respectively with the eigenfunctions and generalized
eigenfunctions of $Da(x)D+ c(x)$.
Corresponding assumptions on the coupling function $g$
and the polynomial $P(x, \lambda)$ are described in condition {\it
(BM3)}.

Fortunately as we show in Appendices \ref{sec3}  and \ref{urk}, these conditions hold
for a large class of second order differential operators.

Appendices \ref{sec3} and \ref{urk} are devoted to
conditions {\it (BM1)}, {\it (BM2)}. In Appendix \ref{sec3} we discuss
condition {\it (BM1)} and show
that we can always reduce ourselves to the case where $h$ is a
Schr\"{o}dinger operator $D^{2}+ V(x)$, where $V(x)\to m_{\infty}^{2}$ at $\pm\infty$.
We also prove that it is possible to
find generalized eigenfunctions such that the associated unitary
operator diagonalizing $h$ on the continuous spectral subspace is {\em
real}. This property is important in connection with Sect. \ref{sec4}.

Appendix \ref{urk} is devoted to condition {\it (BM2)}. It turns out
that {\it (BM2)} is actually a condition on the behavior of
generalized eigenfunctions $\psi(x, k)$ for $k$  near $0$. If  $h=
D^{2}+ V(x)$ and $V(x)\in O(\langle x\rangle^{-\mu})$ for some
$\mu>0$,  it is well
known that the two cases $\mu >2$ and $\mu \leq 2$ lead to different
behaviors of generalized eigenfunctions near $k=0$.

We discuss condition {\it (BM2)} if $\mu>2$ using standard arguments
based on {\em Jost solutions} which we recall for the reader's convenience.
The case $0<\mu \leq 2$ is discussed using  {\em quasiclassical
solutions} by adapting results of Yafaev \cite{Ya2}.

Finally Appendix \ref{appc} contains some technical estimates.

\medskip

{\bf Acknowledgements} We thank Fritz Gestesy, Erik Skibsted, Martin
Klaus and especially Dimitri Yafaev for very helpful correspondence
on generalized eigenfunctions for one dimensional Schr\"{o}dinger
operators.

\section{Preparations}\label{sec1}\init
In this section we collect various well-known results which will be
used in the sequel.
\subsection{Functional calculus}\label{sec1.4}
If $\chi\in \coinf(\rr)$, we denote by $\tilde{\chi}\in \coinf(\cc)$ an
almost analytic extension of $\chi$, satisfying
\[
\begin{array}{l}
\tilde{\chi}_{\mid \rr}=\chi,\\[2mm]
|\p_{\,\overline z}\tilde{\chi}(z) |\leq C_{n}|Imz|^{n},\ \  \: n\in \nn.
\end{array}
\]
We use the following functional calculus
formula for $\chi\in \coinf(\rr)$ and $A$ selfadjoint:
\beq
\chi(A)=\frac{\i}{2\pi}\int_{\cc}\partial_{\,\overline z}\tilde{\chi}(z)
(z-A)^{-1}\d z\wedge \d\,\overline z.
\label{HS}
\eeq
\subsection{Fock spaces}\label{sec1.1}
In this subsection we recall  various definitions on bosonic
Fock spaces.

\medskip

{\bf Bosonic Fock spaces.}

\medskip

If $\ch$ is  a Hilbert space then
\[
\Gamma(\ch):=\bigoplus_{n=0}^\infty\otimes_{\rm s}^{n}\ch,
\]
is the {\em bosonic Fock space} over $\ch$. $\Omega\in \G(\ch)$ will denote the
{\em vacuum vector}.

In all this paper the one-particle space $\ch$ will be equal to
$L^{2}(\rr, \d x)$. We denote by ${\cal F}:L^{2}(\rr, \d x)\to
L^{2}(\rr, \d k)$ the unitary Fourier
transform
\[
{\cal F}u(k)= (2\pi)^{-\12}\int\e^{-\i x.k}u(x)\d x.
\]
The {\em number operator} $N$ is defined as
\[
N\Big|_{\bigotimes_{\rm s}^{n}\ch}=n\one.
\]
We define the space of {\em finite particle vectors}:
\[
\Gamma_{\rm fin}(\ch)=\cH_{\rm comp}(N):=\{u\in \Gamma(\ch) \: |
\hbox{ for some }\ n\in\nn,\ \ \one_{[0,n]}(N)u=u\},
\]
The {\em creation-annihilation} operators on $\G(\ch)$ are denoted by
$a^{*}(h)$ and $a(h)$. The {\em field operators} are
\[
\phi(h):=\frac{1}{\sqrt{2}}(a^{*}(h)+ a(h)),
\]
which are essentially selfadjoint on $\G_{\rm fin}(\ch)$,
and the {\em Weyl operators} are
\[
W(h):=\e^{\i \phi(h)}.
\]
\medskip

{\bf $\d\G$ operators.}

\medskip

If $r:\ch_{1}\to \ch_{2}$ is an operator one sets:
\[
\begin{array}{rl}
\d \Gamma(r)&:\Gamma(\ch_{1})\to\Gamma(\ch_{2}),\\[2mm]
\d\Gamma(r)\Big|_{\bigotimes_\s^n\ch}&
:=\sum\limits_{j=1}^n\one^{\otimes(j-1)}
\otimes r\otimes \one^{\otimes(n-j)},
\end{array}
\]
with domain $\G_{\rm fin}(\cD(r))$. If $r$ is closeable, so is
$\d\G(r)$.
\medskip

{\bf $\G$ operators.}
\medskip

If  $q:\ch_{1}\mapsto \ch_{2}$
is  bounded  one sets:
\[
\begin{array}{l}\G(q):
\G(\ch_{1})\mapsto \G(\ch_{2})\\[2mm]
\G(q)\Big|_{\bigotimes_{\rm s}^{n}\ch_{1}}= q\otimes\cdots \otimes q.
\end{array}
\]
$\G(q)$ is bounded iff $\|q\|\leq 1$ and then $\|\G(q)\|=1$.

\subsection{Wick polynomials}
We now recall the definition of Wick polynomials
We set
\[
B_{\rm fin}(\Gamma(\ch)):=\{B\in B(\Gh)\: | \hbox{ for some }n\in \nn\
\ \one_{[0, n]}(N)B\one_{[0, n]}(N)=B\}.
\]
Let $w\in B(\otimes_\s^p\ch,\otimes_\s^q\ch)$. The {\em Wick monomial}
associated to the symbol $w$ is:
\[
\Wick(w):\Gamma_\fin(\ch)\to \Gamma_\fin(\ch)
\]
defined as
\beq
\Wick(w)\Big|_{\bigotimes_\s^n\ch}:
=\frac{\sqrt{n!(n+q-p)!}}{(n-p)!}w\otimes_\s\one^{\otimes(n-p)}.
\label{sec.wick.e1}\eeq
This definition extends to $w\in B_\fin(\Gamma(\ch))$ by linearity.
The operator $\Wick (w)$ is called a {\em Wick polynomial} and
the operator $w$ is called the {\em symbol} of
$\Wick (w)$.

 For example if $h_{1}, \dots, h_{p},
g_{1},\dots, g_{q}\in \ch$ then:
\[
\Wick\left(
|\ g_1\otimes_\s\cdots\otimes_\s g_q)
(h_p\otimes_\s\cdots\otimes_\s h_1)|\right)
=a^*(q_1)\cdots a^*(g_q)
a( h_p)\cdots a( h_1).
\]
If $\ch= L^{2}(\rr, \d k)$ then any $w\in B(\otimes_{\s}^{p}\ch,
\otimes_{\s}^{q}\ch)$ is a bounded operator from $\cS(\rr^{p})$ to
$\cS'(\rr^{q})$, where $\cS(\rr^{n})$, $\cS'(\rr^{n})$ denote the
Schwartz spaces of functions and temperate distributions. It has hence
a distribution kernel
\[
w(k_1,\dots,k_q,k'_p,\dots,k'_1)\in S'(\rr^{p+q}),
\]
which is separately symmetric in the variables $k$ and $k'$. It is
then customary to denote the Wick monomial $\Wick(w)$ by:
\[
\int w(k_1,\dots,k_q,k'_p,\dots,k'_1)
a^*(k_1)\cdots a^*(k_q)a(k'_p)\cdots a(k'_1)\d k_1\cdots \d k_q
\d k'_p\cdots \d k'_1.
\]
If $\ch= L^{2}(\rr, \ d x)$, we will use the same notation, tacitly
identifying  $L^{2}(\rr, \d x)$ and $L^{2}(\rr, \d k)$ by Fourier
transform. 
\subsection{$Q-$space representation of Fock space}
Let $\ch$ be a Hilbert space and $c:\ch\to \ch$ a {\em conjugation} on
$\ch$, \ie an anti-unitary involution.  If $\ch= L^{2}(\rr, \d x)$, we
will take  the standard conjugation $c: u\to \overline{u}$.

We denote by $\ch_{c}\subset \ch$ the real subspace of real
vectors for $c$ and  ${\mathfrak M}_c\subset B(\Gamma(\ch))$ be
the abelian Von Neumann algebra generated by the Weyl operators
$W(h)$ for $h\in \ch_c$. The following  result  follows from the
fact that $\Omega$ is a cyclic vector for ${\mathfrak M}_c$ (see
e.g. \cite{SHK}). \bet There exists a compact Hausdorff space $Q$,
a probability measure $\mu$ on $Q$ and a unitary map
 $U$ such that
\[
\begin{array}{l}
U: \G(\ch)\fld L^{2}(Q, \d\mu),\\[2mm]
U\Omega= 1, \\[2mm]
U{\mathfrak M}_c U^{*}= L^{\infty}(Q, \d\mu).
\end{array}
\]
where $1\in L^{2}(Q, \d\mu)$ is the constant function equal to $1$ on
$Q$.
Moreover:
\[
U\G(c)u= \overline{Uu}, \: u\in \G(\ch).
\]
\label{p.3}
\eet
The space $L^{2}(Q, \d\mu)$ is called the $Q-${\em space representation}
of the Fock space $\G(\ch)$ associated to the conjugation $c$.

\subsection{Contractive and hypercontractive semigroups}
\label{subsecp1}
We collect now some standard results  on contractive and
hypercontractive semigroups.

We fix  a  probability space $(Q, \mu)$.
\begin{definition}
Let $H_{0}\geq 0$ be a selfadjoint operator on $\cH =L^{2}(Q, \d\mu)$.

The semigroup $\e^{-tH_{0}}$ is $L^{p}-${\em contractive}  if
$\e^{-tH_{0}}$ extends as a contraction in $L^{p}(Q, \d\mu)$ for all
$1\leq p\leq \infty$ and $t\geq 0$.

The semigroup $\e^{-tH_{0}}$ is {\em hypercontractive} if

{\it i)} $\e^{-tH_{0}}$ is a contraction on $L^{1}(Q, \d\mu)$ for all
$t>0$,

{\it ii)} $\exists \: T, C$ such that
\[
\|\e^{-TH_{0}}\psi\|_{L^{4}(Q, \d\mu)}\leq C\|\psi\|_{L^{2}(Q, \d\mu)}.
\]
\label{p.1}
\end{definition}
If $\e^{-tH_{0}}$ is {\em positivity preserving} (i.e. $f\geq 0$
a.e. implies $\e^{-tH_{0}}f\geq 0$ a.e.) and $\e^{-tH_{0}}1\leq 1$
then $\e^{-tH_{0}}$ is $L^{p}-$contractive (see e.g. \cite[Prop.
1.2]{KL1})
\subsection{Perturbations of hypercontractive semigroups}
The abstract result used to construct the $P(\varphi)_{2}$
Hamiltonian is the following theorem, due to Segal (\cite{Se}).
\bet
Let $\e^{-tH_{0}}$ be a hypercontractive semigroup. Let $V$ be a real
 measurable
function on $Q$ such that $V\in L^{p}(Q, \d\mu)$ for some $p>2$ and
$\e^{-tV}\in L^{1}(Q, \d\mu)$ for all $t>0$. Let $V_{n}= \one_{\{|V|\leq
n\}}V$ and $H_{n}= H_{0}+V_{n}$. Then the semigroups
$\e^{-tH_{n}}$ converge strongly on $\cH$ when $n\fld \infty$
to a strongly continuous
semigroup on $\cH$ denoted by $\e^{-tH}$.
Its infinitesimal generator $H$
 has the
following properties:

{\it i)}    $H$ is the closure of $H_{0}+V$ defined on $\cD(H_{0})\cap
\cD(V)$,

{\it ii)}  $H$ is bounded below:
\[
H\geq -c -\ln \|\e^{-\delta V}\|_{L^{1}(Q, \d\mu)},
\]
where $c$ and $\delta$ depend only on the constants $C$ and $T$ in Def.
\ref{p.1}.
\label{p.2}
\eet
We will also need the following result \cite[Thm. 2.21]{SHK}.
\begin{proposition}
\label{shk}
Let $\e^{-tH_{0}}$ be a hypercontractive semigroup. Let $V, V_{n}$ be
real measurable functions on $Q$ such that $V_{n}\to V$ in $L^{p}(Q,
\d\mu)$ for some $p>2$, $\e^{-tV}, \e^{-tV_{n}}\in L^{1}(Q, \d \mu)$ for each
$t>0$ and $\|\e^{-tV_{n}}\|_{L^{1}}$ is uniformly bounded in $n$ for each
$t>0$. Then for $b$ large enough
\[
(H_{0}+ V_{n}+b)^{-1}\to (H_{0}+ V+ b)^{-1}\hbox{ in norm.}
\]
\end{proposition}
The following lemma (see \cite[Lemma V.5]{Si1} for a proof)
will be used later to show that a given function $V$ on $Q$ verifies
$\e^{-tV}\in L^{1}(Q, \d\mu)$.
\begin{lemma}
Let for $\kappa\geq 1$, $V_{\kappa}, V$ be functions on $Q$ such that
for some $n\in \nn$:
\beq
\begin{array}{l}
\|V-V_{\kappa}\|_{L^{p}(Q, \d\mu)}\leq
C_{1}(p-1)^{n}\kappa^{-\epsilon}, \:\forall \: p\geq 2,
\\[2mm]
V_{\kappa}\geq -C_{2}-C_{3}(\ln \kappa)^{n}.
\end{array}
\label{ep.4}
\eeq
Then there exists constants $\kappa_{0}$, $C_{4}$ and $\alpha>0$ such that
\[
\mu\{q\in Q|V(q)\leq -C_{4}(\ln \kappa)^{n}\}\leq
\e^{-\kappa^{\alpha}}, \ \ \forall \kappa\geq \kappa_{0}.
\]
 Consequently $\e^{-tV}\in L^{1}(Q, \d\mu),\:
\forall t>0$ with a norm depending only on $t$ and the constants $C_{i}$ in
(\ref{ep.4}).
\label{p.6}
\end{lemma}

The following theorem of Nelson (see \cite[Thm. 1.17]{Si1})
establishes a connection between contractions on
$\ch$ and hypercontractive semigroups on  the $Q-$space representation
$L^{2}(Q, \d\mu)$ associated to a conjugation $c$.
\bet
Let $r\in B(\ch)$ be a selfadjoint contraction commuting with $c$.
Then

{\it i)} $U\G(r)U^{*}$ is a positivity preserving
contraction on $L^{p}(Q, \d\mu)$, $1\leq p\leq \infty$.

{\it ii)} if $\|r\|\leq (p-1)^{\12}(q-1)^{-\12}$ for $1<p\leq q<\infty$
then $U\G(r)U^{*}$ is a contraction from $L^{p}(Q, \d\mu)$ to $L^{q}(Q,
d\mu)$.
\label{p.4}
\eet
Combining Thm. \ref{p.4} with Thm. \ref{p.2}, we obtain the following
result.
\bet
Let $\ch$ be a Hilbert space with a conjugation $c$. Let $a$ be a
selfadjoint operator on $\ch$ with
\beq
[a, c]=0, \: a\geq m>0.
\label{ep.3}
\eeq
Let $L^{2}(Q, \d\mu)$ be the $Q-$space representation of $\G(\ch)$ and
let $V$ be a real function on $Q$ with
$V\in L^{p}(Q, \d\mu)$ for some $p>2$ and $\e^{-tV}\in L^{1}(Q, \d\mu)$
for all $t>0$. Then:

{\it i)} the operator sum $H=d\G(a)+ V$ is essentially selfadjoint on
$\cD(d\G(a))\cap \cD(V)$.

{\it ii)} $H\geq -C$, where $C$ depends only on $m$ and
$\|\e^{-V}\|_{L^{p}(Q, \d\mu)}$, for some $p$ depending only on $m$.
\label{XII}
\eet
Note that by applying Thm. \ref{p.4} to $a=(q-1)^{-\12}\one_{\ch}$
for $q>2$, we obtain the following lemma about the $L^{p}$ properties
of finite vectors in $\G(\ch)$ (see \cite[Thm. 1.22]{Si1}).
\begin{lemma}
Let $\psi\in \otimes_{\rm s}^{n}\ch$ and $q\geq 2$.
Then
\[
\|U\psi\|_{L^{q}(Q, \d\mu)}\leq (q-1)^{n/2}\|\psi\|.
\]
\label{p.5}
\end{lemma}
\subsection{Perturbations of $L^{p}-$contractive semigroups}
The following theorem is shown in \cite[Sect. II.2]{KL2}.
\begin{theoreme}\label{contract}
Let $\e^{-tH_{0}}$ be an $L^{p}-$contractive semigroup and $V$ a real
measurable function on $Q$ such that $V\in L^{p_{0}}(Q, \d \mu)$ for some
$p_{0}>2$ and $e^{-\delta V}\in L^{1}(Q, \d \mu)$ for some $\delta>0$. Then
$H_{0}+V$ is essentially selfadjoint on ${\cal A}(H_{0})\cap L^{q}(Q,
\d \mu)$ for any $(\12-\frac{1}{p_{0}})^{-1}\leq q<\infty$ where
${\cal A}(H_{0})$ is the space of analytic vectors for $H_{0}$.
\end{theoreme}

\subsection{Pseudodifferential calculus on $L^{2}(\rr^{d})$}\label{sec1.10}
We denote by $\cS(\rr^{d})$ the Schwartz class of functions on $\rr^{d}$ and by
$\cS'(\rr^{d})$ the Schwartz class of tempered distributions on $\rr^{d}$. We
denote by $H^{s}(\rr^{d})$ for $s\in \rr$ the Sobolev spaces on $\rr^{d}$.

We set as usual $D= \i^{-1}\p_{x}$ and $\langle s\rangle=(s^{2}+1)^{\12}$.

For $p,m\in \rr$ and $0\leq \epsilon<\12$,we denote by $S_{\epsilon}^{p, m}$ the class of symbols $a\in
\cinf(\rr^{2d})$ such that
\[
|\p^{\alpha}_{x}\p^{\beta}_{k}a(x,k)|\leq C_{\alpha,
\beta}\langle k\rangle^{p-|\beta|}\langle
x\rangle^{m-(1-\epsilon)|\alpha|+ \epsilon||\beta|}, \ \ \alpha,
\beta\in \nn^{d}.
\]
The symbol class $S^{p, m}_{0}$ will be simply denoted by $S^{p,m}$.
The symbol classes above are equipped with the toplogy given by the
seminorms equal to the best constants in the estimates above.

For $a\in S^{p,m}_{\epsilon}$, we denote by $\Op^{1, 0}(a)$ (resp.
$\Op^{0,1}(a)$) the {\em Kohn-Nirenberg} (resp {\em anti
Kohn-Nirenberg}) quantization of $a$ defined by:
\[
\Op^{1, 0}(a)(x,D)u(x):=(2\pi)^{-d}\int\int\e^{\i (x-y)k}a(x, k)u(y)\d y\d
k,
\]
\[
\Op^{0,1}(a)(x,D)u(x):=(2\pi)^{-d}\int\int\e^{\i (x-y)k}a(y, k)u(y)\d y\d
k,
\]
which are well defined as continous maps from $S(\rr^{d})$ to
$S'(\rr^{d})$.
We denote by  $\Op^{\rm w}(a)$ the {\em Weyl} quantization of $a$
defined by:
\[
\Op^{\rm w}(a)(x,D)u(x):=(2\pi)^{-1}\int\int\e^{\i
(x-y)k}a(\frac{x+y}{2}, k)u(y)\d y\d k.
\]
We recall that as operators from $\cS(\rr^{d})$ to $\cS'(\rr^{d})$:
\[
 \Op^{0,1}(m)^{*}= \Op^{1, 0}(\overline{m}), \ \  \Op^{\rm w}(m)^{*}=
\Op^{\rm w}(\overline{m}).
\]
We will also need the following facts (see \cite[Thm. 18.5.4]{Ho}):
\beq\label{calc}
[\Op^{\rm w}(b_{1}), \i \Op^{\rm w}(b_{2})]= \Op^{\rm w}(\{b_{1}, b_{2}\})+ \Op^{\rm w}(S^{p_{1}+
p_{2}-3, m_{1}+ m_{2}-3(1-2\epsilon)}_{\epsilon}),
\eeq
\beq\label{calc-bis}
\Op^{\rm w}(b_{1})\Op^{\rm w}(b_{2})+ \Op^{\rm w}(b_{2})\Op^{\rm w}(b_{1})= 
2\Op^{\rm w}(b_{1}b_{2})+ \Op^{\rm w}(S^{p_{1}+
p_{2}-2, m_{1}+ m_{2}-2(1-2\epsilon)}_{\epsilon}),
\eeq
if  $b_{i}\in S^{p_{i}, m_{i}}_{\epsilon}$ and $\{ \: , \:\}$ denotes
the Poisson bracket.

The following two propositions will be proved in Appendix
\ref{appc}.
\begin{proposition}\label{1.1}
Let $b\in S^{2, 0}$ a real symbol such that for some $C_{1},
C_{2}>0$
\[
b(x, k)\geq C_{1}\langle k\rangle^{2}-C_{2}.
\]
Then:

{\it i)} $\Op^{\rm w}(b)(x,D)$ is  selfadjoint and bounded below on
$H^{2}(\rr^{d})$.

{\it ii)} Let  $C$ such that $\Op^{\rm w}(b)(x, D)+ C>0$ and $s\in \rr$.
Then there exist $m_{i}\in S^{2s, 0}$ for $i=1, 2,3$ such that
\[
(\Op^{\rm w}(b)(x, D)+ C)^{s}= \Op^{\rm w}(m_{1})(x,D)= \Op^{1,
0}(m_{2})(x,D)= \Op^{0,1}(m_{3})(x,D).
\]
\end{proposition}

\begin{proposition}\label{exemple1}
Let $a_{ij}, c$ are real such that:
\beq
\begin{array}{l}
[a_{ij}](x)\geq
c_{0}\one, \ c(x)\geq c_{0}\hbox{ for some }c_{0}>0, \\[2mm]
 [a_{ij}]-\one,
\ c(x)-m_{\infty}^{2}\in S^{0, -\mu} \hbox{ for some }	m_{\infty}, \mu>0.
\end{array}
\label{hippopo}
\eeq 
Set:
\[
b(x, k):= \sum_{1\leq i,j \leq d}k_{i}a_{ij}(x)k_{j}+ c(x),
\]
and
\[
h := \sum_{1\leq i,j \leq d}D_{i}a_{ij}(x)D_{j}+ c(x)=\Op^{\rm w}(b).
\]
Then:
\[
i) \ \  \omega:= h^{\12} = \Op^{\rm w}(b^{\12})+ \Op^{\rm w}(S^{0, -1-\mu}).
\] 
{\it ii)}  there exists $0<\epsilon<\12$ such that:
\[
[\omega, \i [\omega, \i \x]]= \Op^{\rm w}(\gamma)^{2}+ \Op^{\rm
w}(r_{-1-\epsilon}), \hbox{ for }\gamma\in S_{\epsilon}^{0,-\12}, \
r_{-1-\epsilon}\in S_{\epsilon}^{0, -1-\epsilon}.
\]
\end{proposition}

\section{The space-cutoff $P(\varphi)_{2}$ model with variable metric}
\init\label{sec2}
In this section we define the space-cutoff $P(\varphi)_{2}$
Hamiltonians with variable metric and we prove some of their basic
properties.
\subsection{The  $P(\varphi)_{2}$ model with variable metric}
\label{sec2.1}
For $\mu\in \rr$ we denote by $S^{\mu}$ the class of symbols  $a\in \cinf(\rr)$ such that
\[
|\p^{\alpha}_{x}a(x)|\leq C_{\alpha}\langle x\rangle^{\mu-\alpha}, \ \
\alpha\in \nn.
\]
Let  $a, c$ two real symbols such that for some $\mu>0$:
\beq\label{e1.1} a-1\in S^{-\mu}, \ \ a(x)>0, \ \
c-m^{2}_{\infty}\in S^{-\mu}, \ \ c(x)>0, \eeq where the constant
$m_{\infty}\geq  0$ has the meaning of the {\em mass at infinity}.
For most of the paper we will assume that the model is {\em
massive} i.e. $m_{\infty}>0$. The existence of the Hamiltonian in
the massless case $m_{\infty}=0$ will be proved in Thm.
\ref{basic-massless}.

We consider the second order differential operator
\[
h=Da(x)D+ c(x),
\]
 which is selfadjoint on $H^{2}(\rr)$.  Clearly $h\geq m$ for some
$m>0$ if $m_{\infty}>0$ and for $m=0$ if $m_{\infty}=0$.
 Note that $h$ is a {\em real }operator i.e. $[h,
c]=0$, if $c$ is the standard conjugation.

The {\em one particle space} is
\[
\ch= L^{2}(\rr, \d x),
\]
and the {\em one particle energy} is
\[
 \omega:= (Da(x)D+ c(x))^{\12}, \hbox{acting on }\ch.
\]
The kinetic energy of the field is
\[
H_{0}:= \d\Gamma(\omega), \hbox{ acting on }\G(\ch).
\]
To define the  interaction we
fix a real polynomial with $x-$dependent coefficients:
\beq\label{defdeP}
P(x,\lambda)=\sum_{p=0}^{2n}a_{p}(x)\lambda^{p}, \ \ a_{2n}(x)\equiv
a_{2n}>0,
\eeq
and  a measurable function $g$ with:
\[
g(x)\geq 0, \ \ \forall\:  x\in \rr.
\]
and set for $1\leq \kappa<\infty$ an UV-cutoff parameter:
\[
 V_{\kappa}:=\int g(x):\! P(x,\varphi_{\kappa}(x))\! :\d x,
\]
where $:\ \ :$ denotes the {\em Wick ordering}  and
$\varphi_{\kappa}(x)$  are the {\em UV-cutoff fields}.

In the massive case, they  are defined as:
\beq
\varphi_{\kappa}(x):=\phi(f_{\kappa, x}),
\label{e2.1}
\eeq
for
\beq\label{defde}
f_{\kappa,x}=\sqrt{2}\omega^{-\12}\chi(\frac{\omega_{\infty}}{\kappa})\delta_{x},
\ \ x\in\rr.
\eeq
Here $\chi\in \coinf(\rr)$ is a cutoff function equal to $1$ near $0$,
$\omega_{\infty}= (D^{2}+ m_{\infty}^{2})^{\12}$,
 and $\delta_{x}$ is the $\delta$ distribution centered at $x$.

In the massless case we take:
\[
f_{\kappa,x}=\sqrt{2}\omega^{-\12}\chi(\frac{\omega}{\kappa})\delta_{x},
\ \ x\in\rr.
\]
Note that one can also use the above definition in the massive case
(see Lemma \ref{troud}).

Note also that since $\omega$ is a real operator, $f_{\kappa, x}$ are real
vectors, which implies that $V_{\kappa}$ is affiliated to ${\mathfrak
M}_{c}$. Therefore in the $Q-$space representation associated to $c$,  $V_{\kappa}$
becomes a measurable function on $(Q, \mu)$.

We will see later that  under appropriate conditions on the functions
$ga_{p}$ (see Thms. \ref{basic} and \ref{basic-massless}) the
functions  $V_{\kappa}$  converge
in $L^{2}(Q, \d\mu)$  when $\kappa\to \infty$ to a function $V$ which will be denoted by
\[
 V:=\int_{\rr} g(x):\!P(x,\varphi(x))\! :\d x.
\]
\subsection{Existence and basic properties}
We consider first the massive case $m_{\infty}>0$.
\begin{theoreme}\label{basic}
Let $\omega= (D a(x)D + c(x))^{\12}$ where $a, c>0$ and $a-1$,
$c-m_{\infty}^{2}\in S^{-\mu}$ for some $\mu>0$. Assume that
\[
m_{\infty}>0.
\]
Let:
\[
P(x,\lambda)=\sum_{p=0}^{2n}a_{p}(x)\lambda^{p}, \ \ a_{2n}(x)\equiv
a_{2n}>0.
\]
Assume:
\beq\label{ut1}
\begin{array}{l}
ga_{p}\in L^{2}(\rr), \hbox{
for }0\leq p\leq 2n, \ \ g\in L^{1}(\rr), \ \ g\geq 0, \\[2mm]
   g(a_{p})^{2n/(2n-p)}\in L^{1}(\rr) \hbox{
for }0\leq p\leq 2n-1.
\end{array}
\eeq
Then
\[
H= \d\G(\omega)+ \int_{\rr}g(x):\!P(x, \varphi(x))\!:\d x= H_{0}+V
\]
is essentially selfadjoint and bounded below on $\cD(H_{0})\cap
\cD(V)$.
\end{theoreme}
\proof We apply Thm. \ref{XII} to $a= \omega$. We need to show
that $V\in L^{p}(Q)$ for some $p>2$ and $\e^{-tV}\in L^{1}(Q)$ for
all $t>0$.  The first fact follows from Lemma \ref{2.2} and Lemma
\ref{p.5}. To prove that $\e^{-tV}\in L^{1}(Q)$ we use Lemma
\ref{p.6}: we know from Lemma \ref{2.2} {\it i)} that
$\|V-V_{\kappa}\|_{L^{2}(Q)}\in O(\kappa^{-\epsilon})$ for some
$\epsilon>0$. Since $V\Omega$ and $V_{\kappa}\Omega$ are finite
particle vectors, we deduce from Lemma \ref{p.5} that for all
$p\geq 2$ one has
\[
\|V-V_{\kappa}\|_{L^{p}(Q)}\leq C(p-1)^{n}\kappa^{-\epsilon}.
\]
Hence the first estimate of (\ref{ep.4}) is satisfied. The second
follows from Lemma \ref{lower1}.  \qed

\medskip

We now consider the massless case $m_{\infty}=0$. For simplicity we
assume that $a(x)\equiv 1$.
\begin{theoreme}\label{basic-massless}
Let $\omega= (D^{2} + c(x))^{\12}$ where $ c>0$ and
$c\in S^{-\mu}$ for some $\mu>0$.
Let:
\[
P(x,\lambda)=\sum_{p=0}^{2n}a_{p}(x)\lambda^{p}, \ \ a_{2n}(x)\equiv
a_{2n}>0.
\]
Assume:
\beq\label{ut2}
g\hbox{ is compactly supported},
\eeq
\beq\label{ut3}
\begin{array}{l}
ga_{p}\in L^{2}(\rr), \hbox{
for }0\leq p\leq 2n,  \ \ g\geq 0, \\[2mm]
   g(a_{p})^{2n/(2n-p)}\in L^{1}(\rr) \hbox{
for }0\leq p\leq 2n-1.
\end{array}
\eeq
Then
\[
H= \d\G(\omega)+ \int_{\rr}g(x):\!P(x, \varphi(x))\!:\d x= H_{0}+V
\]
is essentially selfadjoint  on ${\cal A}(H_{0})\cap
L^{q}(Q, \d \mu)$ for $q$ large enough,  where ${\cal A}(H_{0})$ is the space of analytic vectors for
$H_{0}$.
\end{theoreme}
\begin{remark}
It is not necessary to assume that $g$ is compactly supported.  In
fact if we replace  the cutoff function $\chi$ in the proof of Lemma \ref{mass1} by
the function $\x^{-\mu/2}$ we see that
Lemma \ref{mass1} still holds if:
\beq\label{mini}
c(x)\geq C\x^{-\mu}, \hbox{ for some }C>0.
\eeq
Similarly Lemma \ref{mass2} {\it ii)} still holds if we replace the conditions
\[
ga_{p}\in L^{2}(\rr), \ \ g\hbox{ compactly supported},
\]
by
\[
ga_{p}\x^{p\mu/2}\in L^{2}(\rr).
\]
The estimate {\it iii)} in Lemma \ref{mass2} is replaced by:
\[
\x^{-\mu/2} \omega^{-\12}F(\frac{h}{k^{2}})\delta_{x}\in O(({\rm
ln}\kappa)^{\12}), \hbox{ uniformly in }x\in \rr.
\]
Following the proof of Lemma \ref{lower1}, we see   that Thm.
\ref{basic-massless} still holds  if we assume (\ref{mini}), $g\in
L^{1}(\rr)$ and
if conditions (\ref{ut3}) hold with $a_{p}$ replaced by
$a_{p}\x^{p\mu/2}$.
\end{remark}
\begin{remark}
We believe that $H$ is still  bounded below in the massless case. For
example using arguments similar to those in Lemma
\ref{mass1}, one can check that the second order term in formal perturbation theory of the
ground state energy $E(\lambda)$ of $H_{0}+ \lambda V$ is finite.
\end{remark}
\proof Since $\omega\geq 0$ is a real operator, we see from Thm.
\ref{p.4} that $\e^{-tH_{0}}$ is an $L^{p}-$contractive semigroup.
Applying Thm. \ref{contract}, it suffices to  show that
$V\in L^{p}(Q)$ for some $p>2$ and $\e^{-\delta V}\in L^{1}(Q)$ for
some $\delta>0$.
 The first fact follows from Lemma \ref{mass2} and Lemma
\ref{p.5}. To prove that $\e^{-tV}\in L^{1}(Q)$ we use again Lemma
\ref{p.6}: the fact that for all $p\geq 2$
\[
\|V-V_{\kappa}\|_{L^{p}(Q)}\leq C(p-1)^{n}\kappa^{-\epsilon},
\]
follows as before from Lemma \ref{mass2}.
 The second condition in (\ref{ep.4}) follows from Lemma \ref{mass2}
{\it iii)}, arguing as in the proof of
Lemma \ref{lower1}.  \qed

\section{Spectral and scattering theory of $P(\varphi)_{2}$
Hamiltonians}\label{spect-scatt}\init
In this section, we state the main results of this paper.
We consider a $P(\varphi)_{2}$ Hamiltonian as in Thm. \ref{basic}.
We need first to state some conditions on the eigenfunctions and
generalized eigenfunctions of $h= \omega^{2}$. These conditions will
be needed to obtain {\em higher order estimates} in Sect. \ref{sec4},
an important ingredient in the proof of Thms. \ref{mainmain},
\ref{mainim} and \ref{ima}.

We will say that the families
$\{\psi_{l}(x)\}_{l\in I}$ and $\{\psi(x, k)\}_{k\in \rr}$ form a
basis of (generalized) eigenfunctions of $h$ if:
\[
\begin{array}{l}
\psi_{l}(\cdot)\in L^{2}(\rr), \ \ \psi(\cdot, k)\in \cS'(\rr), \\[2mm]
h \psi_{l}= \epsilon_{l}\psi_{l}, \ \ \epsilon_{l}\leq m^{2}_{\infty},
 \ \ l\in I,
\\[2mm]
h \psi(\cdot, k)= (k^{2}+ m_{\infty}^{2})\psi(\cdot, k), \ \ k\in \rr,
\\[2mm]
\sum_{l\in I}|\psi_{l})(\psi_{l}|+
\frac{1}{2\pi}\int_{\rr}|\psi(\cdot, k))(\psi(\cdot, k)|\d k=\one.
\end{array}
\]
Here $I$  is equal either to $\nn$ or to a finite subset of $\nn$. The
existence of such  bases follows easily from the spectral theory and
scattering theory of
the second order differential operator $h$, using hypotheses
(\ref{e1.1}).

Let $M:\rr\to [1 +\infty[$ a locally bounded Borel function.
We introduce the following assumption on such a basis:
\[
\begin{array}{l}
(BM1)\ \ \sum_{l\in I}\|M^{-1}(\cdot)\psi_{l}(\cdot)\|_{\infty}^{2}<\infty, \\[2mm]
(BM2)\ \ \|M^{-1}(\cdot)\psi(\cdot, k)\|_{\infty}\leq C, \ \ \forall
\: k\in \rr.
\end{array}
\]
For a given weight function $M$, we introduce the following hypotheses
on the coefficients of $P(x, \lambda)$:
\[
(BM3)\ \ ga_{p}M^{s}\in L^{2}(\rr), \ \ g(a_{p}M^{s})^{\frac{2n}{2n-p+s}}\in
L^{1}(\rr),
\ \ \forall \: 0\leq s\leq p\leq 2n-1.
\]

\begin{remark}\label{infin}
Hypotheses {\it (BMi)} for $1\leq i\leq 3$ have still  a meaning
if $M$ takes values in $[1, +\infty]$, if we use the convention that
$(+\infty)^{-1}=0$.  Of course in order for {\it (BM3)} to hold $M$
must take finite values on $\supp g$.
\end{remark}
\begin{remark}\label{remi}
The results below still hold   if  we replace {\it (BM2)} by
\[
(BM2')\ \ \|M^{-1}(\cdot)\psi(\cdot, k)\|_{\infty}\leq C\sup(1,
|k|^{-\alpha}), \ \ k\in \rr.
\]
for some $0\leq \alpha<\12$.
\end{remark}

The results of the paper are summarized in the following three theorems.
\begin{theoreme}\label{mainmain}[HVZ Theorem]

Let  $H$ be as in Thm. \ref{basic} and assume that there exists
a basis of eigenfunctions $\{\psi_{l}(x)\}_{l\in I}$ and generalized
eigenfunctions $\{\psi(x, k)\}_{k\in \rr}$ of $h$ such that conditions
{\it (BM1)}, {\it (BM2)}, {\it (BM3)} hold.
Then the essential spectrum of $H$ equals $[\inf \sigma(H)+
m_{\infty}, +\infty[$. Consequently $H$ has a ground state.
\end{theoreme}

\begin{theoreme}\label{mainim}[Mourre estimate]

Let  $H$ be as in Thm. \ref{basic} and assume in addition to the hypotheses of Thm. \ref{mainmain} that
\[
\x^{s}ga_{p}\in L^{2}(\rr), \ \ 0\leq p\leq 2n, \ \ s>1.
\]
let $a= \12(\langle D\rangle^{-1}D. x+ \hc)$ and $A= \d\G(a)$.
Let \[
\tau=\sigma_{\rm pp}(H)+ m_{\infty}\nn^{*}
\]
be the set of {\em
thresholds} of $H$. Then:

{\it i)} the quadratic form $[H, \i A]$ defined on $\cD(H)\cap \cD(A)$
uniquely extend to a bounded quadratic form $[H, \i A]_{0}$ on
$\cD(H^{m})$ for some $m$ large enough.

{\it ii)} if  $\lambda\in \rr\backslash\tau$ there exists
$\epsilon>0$, $c_{0}>0$ and a compact operator $K$ such that
\[
\one_{[\lambda-\epsilon, \lambda+ \epsilon]}(H)[H, \i
A]_{0}\one_{[\lambda-\epsilon, \lambda+ \epsilon]}(H)\geq
c_{0}\one_{[\lambda-\epsilon, \lambda+ \epsilon]}(H)+ K.
\]
{\it iii)} for all $\lambda_{1}\leq \lambda_{2}$ such that
$[\lambda_{1}, \lambda_{2}]\cap \tau=\emptyset$ one has:
\[
{\rm dim}\one_{[\lambda_{1}, \lambda_{2}]}(H)<\infty.
\]
Consequently $\sigma_{\rm pp}(H)$ can accumulate only at $\tau$, which
is a closed countable set.

{\it iv)} if $\lambda\in \rr\backslash(\tau\cup\sigma_{\rm pp}(H))$ there exists
$\epsilon>0$ and  $c_{0}>0$  such that
\[
\one_{[\lambda-\epsilon, \lambda+ \epsilon]}(H)[H, \i
A]_{0}\one_{[\lambda-\epsilon, \lambda+ \epsilon]}(H)\geq
c_{0}\one_{[\lambda-\epsilon, \lambda+ \epsilon]}(H).
\]
\end{theoreme}
\begin{theoreme}\label{ima}[Scattering theory]

Let  $H$ be as in Thm. \ref{basic} and assume that the hypotheses of
Thm. \ref{mainim} hold. Let us denote by $\ch_{\rm c}(\omega)$ the
continuous spectral subspace of $\ch$ for $\omega$. Then:
\begin{enumerate}
\item The {\em asymptotic Weyl operators}:
\[
W^{\pm}(h):=\slim_{t\pm\infty}\e^{\i tH }W(\e^{-\i t \omega}h)\e^{-\i
tH}\hbox{ exist for all }h\in \ch_{\rm c}(\omega),
\]
and define a regular CCR representation over $\ch_{\rm c}(\omega)$.

\item There exist unitary operators $\Omega^{\pm}$, called the {\em
wave operators}:
\[
\Omega^{\pm}: \cH_{\rm pp}(H)\otimes \G(\ch_{\rm c}(\omega))\to \G(\ch)
\]
such that
\[
\begin{array}{l}
W^{\pm}(h)=\Omega^{\pm} \one\otimes W(h)\Omega^{\pm*}, \ \ h\in
\ch_{\rm c}(\omega),\\[2mm]
H= \Omega^{\pm}(H_{|\cH_{\rm pp}(H)}\otimes \one + \one\otimes
\d\G(\omega))\Omega^{\pm*}.
\end{array}
\]
\end{enumerate}
\end{theoreme}
\begin{remark}
Appendices \ref{sec3} and \ref{urk} are devoted to
conditions {\it (BM1)}, {\it (BM2)}. For example condition {\it (BM1)}
is always satisfied for $M(x)=\langle x\rangle^{\alpha}$ if
$\alpha>\12$ and is satisfied for $M(x)=1$
if $h$ has a finite number of eigenvalues (see Prop. \ref{suff}).

Concerning condition {\it (BM2)}, we show in Lemma \ref{1.2} that it
suffices to consider the case where $a(x)\equiv 1$.  For example if
$c(x)-m_{\infty}^{2}\in
O(\langle x\rangle^{-\mu})$ for $\mu>2$ and $h$ has no zero energy
resonances, then {\it (BM2)} is satisfied for $M(x)=1$ (see Prop.
\ref{3.3bis}).

If $c(x)-m_{\infty}^{2}\in O(\langle x\rangle^{-\mu})$ for $0<\mu <2$,
 is {\em negative near infinity} and has no zero energy
resonances, then {\it (BM2)} is satisfied for
$M(x)= \langle x\rangle^{\mu/4}$ (see Prop. \ref{suffsuff}).

If
$c(x)-m_{\infty}^{2}$ is {\em positive near infinity}, holomorphic in
a conic neighborhood of $\rr$ and has no zero energy
resonances, then {\it
(BM2)} is satisfied for $M(x)=1$ in $\{|x|\leq R\}$ and $M(x)=+\infty$
in $\{|x|>R\}$ (see Prop. \ref{suffsuffsuff}).
\end{remark}

\begin{remark}
A typical situation in which all the assumptions are satisfied is when
$a(x)-1$, $c(x)-m_{\infty}^{2}$ and $g$, $a_{p}$ are all in the
Schwartz class $\cS(\rr)$.
\end{remark}

\medskip

{\it Proofs of Thms. \ref{mainmain}, \ref{mainim} and \ref{ima}.}

It suffices to check that $H$ belongs to the class  of abstract
QFT Hamiltonians considered in \cite{GP}. We check that $H$
satisfies all the conditions in  \cite[Thm. 4.1]{GP}, introduced in
\cite[Sect. 3]{GP}.

Since $\omega\geq m>0$, condition {\it (H1)} in \cite[Subsect.
3.1]{GP} is satisfied.  The interaction term $V$ is clearly a Wick
polynomial.
 By  Thm. \ref{basic}, $H$ is essentially selfadjoint and bounded
below on $\cD(H_{0})\cap \cD(V)$, i.e. condition {\it (H2)} in
\cite[Subsect. 3.1]{GP} holds. Next by Thm. \ref{4.1} the
higher order estimates hold for $H$, i.e. condition {\it (H3)} in
\cite[Subsect. 3.1]{GP} is satisfied.

The  second set of conditions  concern the one-particle energy
$\omega$. Conditions {\it (G1)} in \cite[Subsect. 3.2]{GP} are
satisfied for $\cS= \cS(\rr)$ and $\x= (x^{2}+1)^{\12}$. This follows
immediately from the fact that $\omega\in \Op(S^{1, 0})$ shown in
Prop. \ref{1.1} and pseudodifferential calculus. Condition {\it (G2)} 
in \cite[Subsect. 3.2]{GP} has been checked in Prop. \ref{exemple1}.

Let us now consider the conjugate operator $a$. To define $a$ without
ambiguity, we set $\e^{-\i t a}:= {\cal F}^{-1}u_{t}{\cal F}$, where
$u_{t}$ is the unitary group on $L^{2}(\rr, \d k)$ generated by
the vector field $-\frac{k}{\langle k\rangle}\cdot \p_{k}$. We see that  $u_{t}$
preserves the spaces $\cS(\rr)$ and ${\cal F}\cD(\omega)= \cD(\langle k\rangle)$. This
implies first that  $a$ is essentially selfadjoint
on $\cS(\rr)$, by Nelson's invariant subspace theorem.  Moreover
$\e^{\i ta}$ preserves $D(\omega)$ and $[\omega, a]$ is bounded on
$L^{2}(\rr)$. By
\cite[Prop. 5.1.2]{ABG}, $\omega\in C^{1}(a)$ and condition{\it (M1
i)}  in \cite[Subsect. 3.2]{GP} holds.

We see also that $a\in \Op(S^{0, 1})$, so conditions {\it (G3)} and
{\it (G4)} in \cite[Subsect. 3.2]{GP} hold. For
$\omega_{\infty}=(D^{2}+ m_{\infty}^{2})^{\12}$, we deduce as
above from pseudodifferential calculus that
\[
[\omega, \i a]_{0}=\omega_{\infty}^{-1}\langle D\rangle^{-1}D^{2}+ \Op(S^{0, -\mu}).
\]
Since $\chi(\omega)-\chi(\omega_{\infty})$ is compact, we obtain that
\[
\chi(\omega)[\omega, \i a]_{0}\chi(\omega)= 
\chi^{2}(\omega_{\infty})\omega_{\infty}^{-1}\langle D\rangle^{-1}D^{2} +K,\hbox{ where }K\hbox{ is compact}.
\]
This implies that $\rho^{a}_{\omega}\geq 0$ and
$\tau_{a}(\omega)=\{m_{\infty}\}$,
hence {\it (M1 ii)} in \cite[Subsect. 3.2]{GP} holds.

Property {\it (C)} in \cite[Subsect. 3.2]{GP} follows from the fact that
$\omega-\omega_{\infty}\in \Op(S^{1, -\mu})$ and
pseudodifferential calculus. Finally property {\it (S)} in
\cite[Subsect. 3.2]{GP} can be proved
as explained in \cite[Subsect. 3.2]{GP}.

The last set of conditions   concern the decay properties of the Wick kernel
of $V$. We see that  condition {\it (D)} in \cite[Subsect. 3.2]{GP} is satisfied, using
Lemma \ref{2.2} and the fact that $\x^{s}ga_{p}\in L^{2}(\rr)$ for all
$0\leq p\leq 2n$.

 Applying then \cite[Thm.
4.1]{GP} we obtain Thms. \ref{mainmain}, \ref{mainim} and
\ref{ima}. \qed

\section{Higher dimensional models}\label{higher}\init
In this section we briefly  discuss similar models in higher space
dimension, when the interaction term has an ultraviolet cutoff.

We work now on $L^{2}(\rr^{d}, \d x)$ for $d\geq 2$ and consider
\[
h=\sum_{1\leq i,j\leq d}D_{i}a_{ij}(x)D_{j}+ c(x), \ \omega= h^{\12}.
\]
where $a_{ij}, c$ satisfy (\ref{hippopo}). The free Hamiltonian is
as above
\[
H_{0}= \d\G(\omega),
\]
acting on the Fock space $\G(L^{2}(\rr^{d}))$.

Since  $d\geq 2$ it is necessary to add an
ultraviolet cutoff to make sense out of the formal expression
\[
\int_{\rr^{d}}g(x)P(x, \varphi(x))\d x.
\]
We set
\[
\varphi_{\kappa}(x):=
\phi(\omega^{-\12}\chi(\frac{\omega}{\kappa})\delta_{x}),
\]
where $\chi\in \coinf([-1, 1])$ is a cutoff function equal to $1$
on $[-\12, \12]$ and $\kappa\gg 1$ is an ultraviolet cutoff
parameter.  Since
$\omega^{-\12}\chi(\frac{\omega}{\kappa})\delta_{x}\in
L^{2}(\rr^{d})$, $\varphi_{\kappa}(x)$ is a well defined
selfadjoint operator on $\G(L^{2}(\rr^{d}))$.

If $P(x, \lambda)$
is as in (\ref{defdeP}) and $g\in L^{1}(\rr^{d})$, then
\[
V:=\int_{\rr^{d}}g(x)P(x, \varphi_{\kappa}(x))\d x,
\]
is a well defined selfadjoint operator on $\G(L^{2}(\rr^{d}))$.

\begin{lemma}\label{exemple5}
Assume that $g\geq 0$,  $g\in L^{1}(\rr^{d})\cap L^{2}(\rr^{d})$ and
$ga_{p}\in L^{2}(\rr^{d})$,
$ga_{p}^{2n(2n-p)}\in L^{1}(\rr^{d})$ for $0\leq p\leq 2n-1$. Then
\[
V\in \bigcap_{1\leq p<\infty}L^{p}(Q, \d \mu), \ \ V\hbox{ is bounded
below}.
\]
\end{lemma}
\proof It is easy to see that $\Omega\in \cD(V)$ hence $V\in L^{2}(Q,
\d\mu)$. Using that $V\Omega$ is a finite particle vector we
obtain by Lemma \ref{p.5} that  $V\in \bigcap_{1\leq p<\infty}L^{p}(Q,
\d \mu)$.

 To prove that $V$ is bounded below,
we use the inequality:
\[
a^{p}b^{n-p}\leq \epsilon b^{n}+ C_{\epsilon}a^{n}, \ \ a, b\geq 0,
\]
and obtain as an inequality between functions on $Q$:
\[
|a_{p}(x)\varphi_{\kappa}(x)^{p}|\leq \epsilon
\varphi_{\kappa}(x)^{2n}+ C_{\epsilon}|a_{p}(x)|^{2n/(2n-p)}.
\]
Integrating this bound for $\epsilon$ small enough we obtain that $V$
is bounded below. \qed

Applying then Thm. \ref{XII}, we obtain that:
\[
 H= \d\G(\omega)+ \int_{\rr^{d}}g(x)P(x, \varphi_{\kappa}(x))\d x
\]
is essentially selfadjoint and bounded below.

We have then the following theorem. As before we consider a
generalized basis $\{\psi_{l}(x)\}_{l\in I}$ and $\{\psi(x, k)\}_{k\in
\rr^{d}}$ of eigenfunctions of $h$.
\begin{theoreme}\label{exemple3}
Assume that:
\[
ga_{p}\in L^{2}(\rr^{d}), \: 0\leq p\leq 2n, \ \ g\in
L^{1}(\rr^{d}), \ \ g\geq 0, \ \   g(a_{p})^{2n/(2n-p)}\in
L^{1}(\rr^{d}), \: 0\leq p\leq 2n-1,
\]
\[
\x^{s}ga_{p}\in L^{2}(\rr^{d}) \ \ \forall \: 0\leq p\leq 2n,
\hbox{ for some }s>1.
\]
Assume moreover that for   a measurable function
$M:\rr^{d}\to \rr^{+}$ with  $M(x)\geq 1$
there exists a generalized basis of eigenfunctions of $h$
such that:
\[
\left\{
\begin{array}{l}
\sum_{l\in I}\|M^{-1}(\cdot)\psi_{l}(\cdot)\|_{\infty}^{2}<\infty, \\[3mm]
\|M^{-1}(\cdot)\psi(\cdot, k)\|_{\infty}\leq C, \ \ k\in \rr,
\end{array}
\right.
\]
\[
ga_{p}M^{s}\in L^{2}(\rr^{d}), \ \  g (a_{p}M^{s})^{2n/(2n-p+s)}\in
L^{1}(\rr^{d}), \ \ \forall \: 0\leq s\leq p\leq 2n-1.
\]
Then  the analogs of Thms. \ref{mainmain}, \ref{mainim} and
\ref{ima} hold for the Hamiltonian:
\[
H= \d\G(\omega)+ \int_{\rr^{d}}g(x)P(x, \varphi_{\kappa}(x))\d x.
\]
\end{theoreme}
\begin{remark}
As in the one-dimensional case, the hypotheses concerning generalized
eigenfunctions can be checked in some cases. An example is   if $d=3$, 
$[a_{ij}](x)= \one$  $c(x)-m_\infty^2 \in O(\x^{-3-
\epsilon}))$ and $h- m_{\infty}^{2}$ has no zero resonance or
eigenvalue, where we can take $M(x)\equiv 1$.  ( See eg \cite[Prop. 2.5 \textit{iv)}]{Teu}).
\end{remark}

We will sketch the proof of Thm. \ref{exemple3}, which  again consists in
showing that the conditions of \cite[Thm. 4.1]{GP} are satisfied.

The condition on the  
one-particle  operators can be checked exactly as in the
one-dimensional case, as can the decay  of the interaction kernel.
 To prove
 the higher order estimates, , we can argue as in Sect. \ref{sec4} working now
with the  family $\{\psi(x)_{l}\}_{l\in I}\cup\{\psi(x, k)\}_{k\in
\rr^{d}}$. The
various integrals in $k$ occurring in the proof of the higher order
estimates are convergent because the domain of integration is included
in $\{|k|^{2}\leq \kappa-m_{\infty}^{2}\}$
due to the energy cutoff $\chi(\kappa^{-1}\omega)$ in the definition
of $\varphi_{\kappa}(x)$.

\section{Properties of the interaction kernel}\label{kernel}\init
In this section prove some properties of the interaction
$V=\int_{\rr}g(x):\! P(x, \varphi(x))\!:\d x$, considering $V$ as a Wick polynomial.
\subsection{Massive case}\label{sec2.2}
In this subsection we consider the massive case $m_{\infty}>0$.
\begin{lemma}\label{2.1}
Let $g\in S(\rr)$. Then for $\kappa<\infty$:
\[
\begin{array}{rl}
&\int g(x):\!\varphi_{\kappa}(x)^{p}\! :\d x\\[2mm]
=&\sum\limits_{r=0}^{p}\left(\begin{array}{c}p\\r\end{array}\right) \int
 w_{p,\kappa}(k_{1},\ldots, k_{r}, k_{r+1}, \ldots, k_{p})
 a^{*}(k_{1})\cdots a^{*}(k_{r})
a(-k_{r+1})\cdots a(-k_{p})\d k_{1}\cdots \d k_{p},
\end{array}
\]
where: \beq\label{e2.2} w_{p,\kappa}(k_{1},\ldots,
k_{p})=(2\pi)^{-p/2}\int g(x)\prod_{j=1}^{p}\e^{-\i k_{j} \cdot
x}m_{\kappa}(x, k_{j})\d x \eeq and $m_{\kappa}(x,k)$  is the {\em anti
Kohn-Nirenberg } symbol of
$\omega^{-\12}\chi(\frac{\omega_{\infty}}{\kappa})$.
\end{lemma}
\proof If $m_{\kappa}(x,k)$ is the anti Kohn-Nirenberg symbol of
$\omega^{-\12}\chi(\frac{\omega_{\infty}}{\kappa})$ we have:
\[
{\cal
F}(\omega^{-\12}\chi(\frac{\omega_{\infty}}{\kappa})\delta_{x})(k)=
(2\pi)^{-\12}\e^{-\i x \cdot k}m_{\kappa}(x,k).
\]
Note that it follows from Prop. \ref{1.1} that $m_{\kappa}\in
S^{-r, 0}$ for each $r\in \nn$.
We observe moreover that
$\omega^{-\12}\chi(\frac{\omega_{\infty}}{\kappa})$ is a real operator
which implies that $m_{\kappa}(x,k)= \overline{m}_{\kappa}(x, -k)$ and
hence
\[
\varphi_{\kappa}(x)= (2\pi)^{-\12}\int \e^{-\i k \cdot x
}m_{\kappa}(x, k)(a^{*}(k)+ a(-k))\d k,
\]
from which the lemma follows. \qed

\medskip

We extend the above notation to $\kappa=\infty$ by denoting  by
$m_{\infty}(x,k)$  the anti Kohn-Nirenberg  symbol of $\omega^{-\12}$ and by
$w_{p,\infty}$ the function in (\ref{e2.2}) with $m_{\kappa}$ replaced
by $m_{\infty}$. Note that by Prop.  \ref{1.1} $m_{\infty}\in S^{-\12,
0}$ so $w_{p,\infty}$ is a well defined function on $\rr^{d}$ if $g\in
S(\rr)$.

To study the properties of $w_{\kappa, p}$
 it is convenient to introduce the following maps:
\[
\begin{array}{rl}
T_{\kappa}: &\cS(\rr)\to \cS'(\rr^{p}), \ \ 1\leq
\kappa\leq\infty,\\[2mm]
& g\mapsto w_{\kappa ,p}.
\end{array}
\]
\begin{lemma}\label{2.2}

{\it i)} $T_{\kappa}$ is bounded from $L^{2}(\rr)$ to
$L^{2}(\rr^{p})$
for each $1\leq \kappa\leq \infty$ and
there exists $\epsilon>0$ such that
\[
\|T_{\kappa}- T_{\infty}\|_{B(L^{2}(\rr), L^{2}(\rr^{p}))}\in
O(\kappa^{-\epsilon}).
\]
{\it ii)} the map $\langle D_{x_{i}}\rangle^{s}T_{\infty}\x^{-s}$  is
bounded from $L^{2}(\rr)$ to $L^{2}(\rr^{p})$
 for each $s\geq 0$  and $1\leq i\leq
p$.

{\it iii)} one has \[
\|f_{\kappa, x}\|\in 0((\ln \kappa)^{\12}), \hbox{ uniformly for }x\in \rr.
\]
\end{lemma}
\proof  The operator $T_{\kappa}$ has the
distribution kernel
\[
(2\pi)^{-p/2}\prod_{j=1}^{p}\e^{-\i k_{j}.x}m_{\kappa}(x, k_{j}),
\]
hence for $f\in S(\rr^{p})$ we have:
\[
T_{\kappa}^{*}f(x)= (2\pi)^{-p/2}\int\prod_{j=1}^{p}\e^{\i
k_{j}.x}\overline{m}_{\kappa}(x, k_{j})f(k_{1},\ldots, k_{p})\d
k_{1}\cdots \d k_{p}.
\]
If $R:\cinf(\rr^{p})\to \cinf(\rr)$ is the operator of restriction to
the diagonal
\[
Rf(x)=f(x,\ldots, x),
\]
we see that
\[
T_{\kappa}^{*}f= RM_{\kappa}{\cal F}_{p}^{-1}f,
\]
where
\[
M_{\kappa}= \prod_{j=1}^{p}\Op^{1, 0}(\overline{m}_{\kappa})(x_{j},
D_{x_{j}}),
\]
and we have denoted by ${\cal F}_{p}$ the unitary Fourier
transform on $L^{2}(\rr^{p})$. Since 
 ${\cal F} is $ the unitary Fourier transform on
$L^{2}(\rr)$, so we have with obvious identificaction $\Gamma({\cal F})={\cal F}_{p}$.  Since
$\Op^{1,0}(\overline{m})= \Op^{0,1}(m)^{*}$, we see that
\[
M_{\kappa}=
\Gamma(\chi(\frac{\omega_{\infty}}{\kappa})\omega^{-\12})_{\mid
\otimes^{p}_{\s}L^{2}(\rr)},
\]
where we have used the Fock space  notation.
This yields
\[
T_{\kappa}^{*}=
R\Gamma(\chi(\frac{\omega_{\infty}}{\kappa})\omega_{\infty}^{-\12}{\cal
F}^{-1})\Gamma({\cal F}\omega_{\infty}^{\12}\omega^{-\12}{\cal
F}^{-1})=: T^{0*}_{\kappa}\Gamma({\cal F}\omega_{\infty}^{\12}\omega^{-\12}{\cal
F}^{-1}),
\]
where $T^{0}_{\kappa}$ is the analog of $T_{\kappa}$ with $\omega$
replaced by $\omega_{\infty}$. This yields: \beq T_{\kappa}=
\Gamma({\cal F}\omega^{-\12}\omega_{\infty}^{\12}{\cal
F}^{-1})T^{0}_{\kappa}. \label{e2.3} \eeq By pseudodifferential
calculus, we know that $\omega^{-\12}\omega_{\infty}^{\12}\in
\Op(S^{0, 0})$ and hence is bounded on $\cD(\x^{s})$ for all $s$.
Therefore it suffices to prove {\it i)} and {\it ii)} for
$T^{0}_{\kappa}$, $T^{0}_{\infty}$. {\it i)} for $T^{0}_{\kappa}$ is shown in
\cite[Lemma 6.1]{DG}. To check {\it ii)} for $T^{0}_{\kappa}$ for
integer $s$ we use that
\[
T^{0}_{\infty}(g)(k_{1}, \dots, k_{p})= \hat{g}(k_{1}\cdots
+k_{p})\prod_{i=1}^{p}\omega_{\infty}^{-\12}(k_{i}).
\]
Then $\p^{s}_{k_{1}}T^{0}_{\infty}(k_{1}, \dots, k_{p})$ is a sum of
terms
\[
\p^{s_{1}}_{k_{1}}\hat{g}(k_{1}\cdots
+k_{p})\p_{k_{1}}^{s_{2}}\omega_{\infty}^{-\12}(k_{1})\prod_{i=2}^{p}\omega_{\infty}^{-\12}(k_{i})
\]
for $s_{1}+ s_{2}=s$. We note that $\p^{s}_{k}\omega_{\infty}^{-\12}\in
O(\langle k\rangle^{-\12-s})$ for all $s\in \nn$. This implies  that
if $\p_{k}^{s}\hat{g}\in L^{2}(\rr)$ then
$\p^{s}_{k_{1}}T^{0}_{\infty}(g)\in L^{2}(\rr^{p})$. This proves {\it
ii)} for integer $s$. We extend it to all $s\geq 0$ by interpolation.

 Finally a direct computation shows that
$\|\omega_{\infty}^{-\12}\chi(\frac{\omega_{\infty}}{\kappa})\delta_{x}\|=O((\ln
\kappa)^{\12})$, which implies {\it iii)} since
$\omega^{-\12}\omega_{\infty}^{\12}$ is bounded on $L^{2}(\rr)$. \qed

\medskip

The following proposition follows easily from Lemmas \ref{2.1} and
\ref{2.2}.
\begin{proposition}
{\it i)} Assume that $ga_{p}\in L^{2}(\rr)$ for $0\leq p\leq 2n$. Then
\[
\lim_{\kappa\to \infty}V_{\kappa}=: V\hbox{  exists in }\bigcap_{1\leq
p<\infty}L^{p}(Q, \d\mu).
\]
{\it ii)} $V$ is a Wick polynomial with a Hilbert-Schmidt symbol.
\end{proposition}
\proof From Lemma \ref{2.2} {\it i)} it follows that
$V_{\kappa}\Omega\to V\Omega$ in $L^{2}(Q, \d\mu)$. The
convergence in all $L^{p}$ spaces for $p<\infty$ follows from the
fact that $V, V_{\kappa}$ are finite particle vectors, using Lemma
\ref{p.5}. Part {\it ii)} follows also from Lemmas \ref{2.1} and
\ref{2.2}. \qed

\medskip

It will be useful later to define the interaction term using an alternative definition of the UV-cutoff
fields, namely:
\beq\label{defdede}
\varphi^{\rm mod}_{\kappa}(x):=\phi(f_{\kappa, x}^{\rm mod}),
\hbox{ for }f_{\kappa, x}^{\rm
mod}=\sqrt{2}\omega^{-\12}\chi(\frac{\omega}{\kappa})\delta_{x},
\eeq
leading to the UV-cutoff interaction
\[
V^{\rm mod}_{\kappa}= \int_{\rr}g(x):\! P(x,\varphi^{\rm
mod}_{\kappa}(x))\! :\d x.
\]
Clearly $V^{\rm mod}_{\kappa}$ is also affiliated to ${\cal
M}_{\rm c}$.
We will use later the following lemma.
\begin{lemma}\label{troud}
{\it i)} $V^{\rm mod}_{\kappa}$ converges to $V$ in $L^{2}(Q, \d \mu)$ when $\kappa\to \infty$.

{\it ii)}
\[
\|f_{\kappa, x}^{\rm mod}\|=O((\ln \kappa)^{\12}),
\]
uniformly for $x\in \rr$.
\end{lemma}
\proof Let us denote by $T^{\rm mod}_{\kappa}$ the
analog of $T_{\kappa}$  for the alternative definition of UV-cutoff fields.
We claim that
\beq
\slim_{\kappa\to \infty}T^{\rm mod}_{\kappa}= T_{\infty}.
\label{e2.4}
\eeq
which implies {\it i)}. In fact arguing as in the proof of Lemma \ref{2.2} we have
\[
T^{\rm mod}_{k}=\Gamma({\cal
F}\chi(\frac{\omega}{\kappa})\omega^{-\12}\omega_{\infty}^{\12}{\cal
F}^{-1})T^{0}_{\infty},
\]
which implies (\ref{e2.4}) since
$\chi(\frac{\omega}{\kappa})\omega^{-\12}\omega_{\infty}^{\12}$ is
uniformly bounded and converges strongly to
$\omega^{-\12}\omega_{\infty}^{\12}$ when $\kappa\to \infty$.
To prove {\it ii)} we write with obvious notation:
\[
\begin{array}{rl}
&\omega^{-\12} \chi(\frac{\omega}{\kappa})\delta_{x} \\[2mm]
=& \omega^{-\12} \chi(\frac{\omega}{\kappa})F(\omega_{\infty}\leq
C\kappa)\delta_{x}+ \omega^{-\12}
\chi(\frac{\omega}{\kappa})F(\omega_{\infty}\geq
C\kappa)\delta_{x}\\[2mm]
=&\chi(\frac{\omega}{\kappa})\omega^{-\12}\omega_{\infty}^{\12}\omega_{\infty}^{-\12}F(\omega_{\infty}\leq
C\kappa)\delta_{x}+ \omega^{-\12}
\chi(\frac{\omega}{\kappa})F(\omega_{\infty}\geq
C\kappa)\omega_{\infty}\omega_{\infty}^{-1}\delta_{x}.
\end{array}
\]
The first term in the last line is $O(({\rm ln}\kappa)^{\12})$
uniformly in $x$, the
second is $O(1)$ if  $C$ is large enough, using Lemma \ref{trouduc} and
the fact that $\omega_{\infty}^{-1}\delta_{x}$ is in $L^{2}(\rr)$
uniformly in $x$. \qed
\subsection{Massless case}\label{sec2.massless}
We consider now the massless case $m_{\infty}=0$. For simplicity
we will assume that $a(x)\equiv 1$, i.e.
\[
\omega=(D^{2}+ c(x))^{\12}, \: c(x)>0, \: c\in S^{-\mu}.
\]
We set as above
\[
h= D^{2}+ c(x), \ \ \omega_{1}= (h+1)^{\12}.
\]
\begin{lemma}\label{mass1}
Let $\chi\in \coinf(\rr)$. Then:
\[
i)\ \ \omega_{1}^{\12}\chi(x)\omega^{-\12}, \ \
\omega_{1}\chi(x)\omega^{-1}\hbox{ are bounded.}
\]
If $F\in \coinf(\rr)$ then
\[
ii) \ \ \omega_{1}^{\delta}[\chi(x), F(\frac{h}{\kappa^{2}})]
\omega^{-\12}\in O(\kappa^{\delta-3/2}) \ \ \forall\: 0\leq
\delta<3/2.
\]
\end{lemma}
\proof Set $\chi= \chi(x)$. Then $\chi D^{2}\chi= D
\chi^{2}D-\chi"\chi$ and hence $\chi D^{2}\chi\leq C D^{2}+
C\chi_{1}$, for $\chi_{1}\in \coinf(\rr)$. This implies that
\[
\chi (h+1)\chi\leq  C(D^{2}+ \chi_{1})\leq C h,
\]
since $c(x)>0$. Therefore $\omega_{1}\chi\omega^{-1}$ is bounded,
which proves the second statement of {\it i)}.  Since
$\omega_{1}\chi^{2}\omega_{1}\leq C (h+1)$, we also have
\[
\chi \omega_{1}\chi^{2}\omega_{1}\chi\leq C \omega^{2},
\]
which by Heinz theorem implies that $\chi \omega_{1}\chi\leq C\omega$
and  proves the first statement of {\it i)}.

To prove {\it ii)} we  write
using (\ref{HS}):
\[
\begin{array}{rl}
&\omega_{1}^{\delta}[\chi, F(\frac{h}{\kappa^{2}})]\omega^{-\12}\\[2mm]
=&\frac{\i}{2\pi\kappa^{2}}\int_{\cc}\partial_{\,\overline z}\tilde{F}(z)
(z-\frac{h}{\kappa^{2}})^{-1}\omega_{1}^{\delta}[\chi,
h]\omega^{-\12}(z-\frac{h}{\kappa^{2}})^{-1}\d z\wedge \d\,\overline z.
\end{array}
\]
Since $[\chi, h]= 2D \chi'-\chi"$ we see using {\it i)} that
$\omega_{1}^{\delta}[\chi, h]\omega^{-\12}= \omega_{1}^{\delta+
\12}B$, where $B$ is bounded. Using the bound $\langle
h\rangle^{\alpha}(z-\frac{h}{\kappa^{2}})^{-1}\in
O(\kappa^{-2\alpha})|{\rm Im}z|^{-1}$ for $z\in \supp\tilde{F}$, we
obtain {\it ii)}. \qed

\medskip

To define the interaction in the massless case, we set:
\[
\varphi_{\kappa}(x):=
\sqrt{2}\phi(\omega^{-\12}F(\frac{h}{\kappa^{2}})\delta_{x}) \ \ x\in \rr,
\]
where $F\in \coinf(\rr)$ equals $1$ near $0$, $\kappa\gg 1$ is
again an UV cutoff parameter, and:
\[
V_{\kappa}:=\int_{\rr}g(x):\!P(x, \varphi_{\kappa}(x))\!:\d x.
\]
\begin{lemma}\label{mass2}
Assume that $g$ is compactly supported and $ga_{p}\in L^{2}(\rr)$ for
$0\leq p\leq 2n$. Then:

{\it i)} $\omega^{-\12}F(\frac{h}{\kappa^{2}})\delta_{x}\in
L^{2}(\rr)$  for $x\in \supp g$ so the UV cutoff fields $\varphi_{\kappa}(x)$ are well
defined.

{\it ii)} $V_{\kappa}$ converges in $\bigcap_{1\leq p<\infty}L^{p}(Q,
\d \mu)$ to a real function $V$ and there exists $\epsilon>0$ such
that:
\[
 \|V-V_{\kappa}\|_{L^{p}(Q, \d \mu)}\leq C(p-1)^{n}\kappa^{-\epsilon},
\:\forall \: p\geq 2.
\]
 {\it iii)} one has
\[
\|\omega^{-\12}F(\frac{h}{\kappa^{2}})\delta_{x}\|\in O(({\rm
ln}\kappa)^{\12}), \hbox{ uniformly for }x\in \supp g.
\]
\end{lemma}

The function $V$ in Lemma \ref{mass2}  will be denoted by:
\[
V=:\int_{\rr}g(x):\!P(x, \varphi(x))\!:\d x.
\]
\proof
To simplify notation we set $F_{\kappa}= F(\frac{h}{\kappa^{2}})$.
We take $\chi\in \coinf(\rr)$ equal to $1$ on $\supp g$. Then for
$x\in \supp g$, we have
\[
\omega^{-\12}F_{\kappa}\delta_{x}=
\omega^{-\12}F_{\kappa}\chi \delta_{x}=
\omega^{\12}F_{\kappa} \omega^{-1}\chi \omega_{1}
\omega_{1}^{-1}\delta_{x} \in L^{2}(\rr),
\]
since $\omega_{1}^{-1}\delta_{x}\in L^{2}$ and $\omega^{-1}\chi
\omega_{1}$ is bounded by Lemma \ref{mass1} {\it i)}.

To prove {\it ii)}  we may assume that $P(x, \lambda)= \lambda^{p}$.
We express the kernel $w_{p, \kappa}(k_{1}, \dots, k_{p})$ as in Lemma
\ref{2.1} and set $w_{p, \kappa}=: T_{\kappa}g$.  Since $g= \chi^{p} g$,
we have $w_{p, \kappa}= T_{\kappa}\chi^{p} g$, and hence $w_{p, \kappa}=
\tilde{T}_{\kappa}g$, where:
\[
\tilde{T}_{\kappa}^{*}= R\Gamma(\chi
\omega^{-\12}F_{\kappa}{\cal F}^{-1})=
R\Gamma(\omega_{1}^{-\12})\G(a(\kappa){\cal F}^{-1}),
\]
for $a(\kappa)=
\omega_{1}^{\12}\chi F_{\kappa}\omega^{-\12}$.
We set also
\[
 \tilde{T}_{\infty}=R\G(\chi\omega^{-\12}{\cal F}^{-1}),
\]
and we claim that
\beq
\|\tilde{T}_{\kappa}^{*}-\tilde{T}_{\infty}^{*}\|\in
O(\kappa^{-\epsilon})\hbox{ for some }\epsilon>0,
\label{e.mass3}
\eeq
which clearly implies {\it ii)}.

If we set
\[
 a_{0}(\kappa)= F_{\kappa}\omega_{1}^{\12}\chi
\omega^{-\12},
\]
then using Lemma
\ref{mass1} {\it ii)}, we obtain:
\beq\label{e.mass2}
a(\kappa)= a_{0}(\kappa) + a_{1}(\kappa), \hbox{ and }
a_{0}(\kappa)\in O(1), \ \ a_{1}(\kappa)\in
O(\kappa^{-\delta}), \hbox{ for some } \delta>0.
\eeq
Clearly on $\otimes^{p}\ch$, one has:
\beq\label{e.mass1}
\G(a_{0}+ a_{1})= \sum_{I\subset \{1, \dots, p\}}
a_{I(1)}\otimes\cdots \otimes a_{I(p)}=: \G(a_{0})+ S(\kappa),
\eeq
for $I(j)= \one_{I}(j)$.
By (\ref{e.mass2}) the terms in (\ref{e.mass1}) for $I\neq\emptyset$ are $O(\kappa^{-\delta})$ hence $S(\kappa)$
is $O(\kappa^{-\delta})$. Since by Lemma \ref{2.2}
$R\G(\omega_{1}^{-\12})$ is bounded, it follows that
 $R\G(\omega_{1}^{-\12})S(\kappa)$ is  $O(\kappa^{-\delta})$.
Therefore we only have to estimate
\[
\begin{array}{rl}
&R\G(\chi\omega^{-\12})- R\G(\omega_{1}^{-\12})\G(a_{0}(\kappa))\\[2mm]
=& \left(R\G(\omega_{1}^{-\12})-
R\G(\omega_{1}^{-\12}F_{\kappa})\right)\G(\omega_{1}^{\12}\chi
\omega^{-\12}).
\end{array}
\]
By Lemma \ref{mass1} {\it i)}, $\G(\omega_{1}^{\12}\chi
\omega^{-\12})$ is bounded, and by Lemma \ref{2.2}
\[
R\G(\omega_{1}^{-\12})-
R\G(\omega_{1}^{-\12}F_{\kappa})\in O(\kappa^{-\epsilon}).
\]
This completes the proof of {\it ii)}.

It remains to prove {\it iii)}.  We write for $x\in \supp g$:
\beq\label{e.mass5}
\begin{array}{rl}
&\omega^{-\12}F_{\kappa}\delta_{x}= \omega^{-\12}F_{k}\chi
\delta_{x}\\[2mm]
=& \omega^{-\12}\chi F_{\kappa}\delta_{x}+ \omega^{-\12}[ F_{\kappa},
\chi]\omega_{1}\omega_{1}^{-1}\delta_{x}\\[2mm]
=& \omega^{-\12}\chi
\omega_{1}^{\12}\omega_{1}^{-\12}F_{\kappa}\delta_{x}+ \omega^{-\12}[ F_{\kappa},
\chi]\omega_{1}\omega_{1}^{-1}\delta_{x}.
\end{array}
\eeq By Lemma \ref{troud} {\it ii)},
$\omega_{1}^{-\12}F_{\kappa}\delta_{x}\in O(({\rm
ln}\kappa)^{\12})$, uniformly for $x\in \supp g$. Moreover by
Lemma \ref{mass1}, $\omega^{-\12}\chi \omega_{1}^{\12}$ is
bounded, hence the first term in the r.h.s. of (\ref{e.mass5}) is
$O({\rm ln}\kappa)^{\12}$. Next $\omega_{1}^{-1}\delta_{x}$ is in
$L^{2}(\rr)$ uniformly in $x$, so the second term is
$O(\kappa^{-\delta})$ for some $\delta>0$ by Lemma \ref{mass1}
{\it ii)}. This completes the proof of {\it iii)}. \qed

\section{Lower bounds}\label{lowersec}\init
In this section we prove some lower bounds on the UV cutoff
interaction $V_{\kappa}$. As explained in Sect. \ref{sec2},
$V_{\kappa}$ is now considered as a function on $Q$. In all this
section we assume that $m_{\infty}>0$.

As consequence we prove Prop. \ref{lower-pert}, which
will be needed in Sect. \ref{sec4}.

We recall from (\ref{defdeP}) that:
\[
P(x, \lambda)= \sum_{p=0}^{2n}a_{p}(x) \lambda^{p},
\]
for $a_{2n}(x)\equiv a_{2n}>0$.
\begin{lemma}\label{lower1}
Let $f_{\kappa, x}$ and $f_{\kappa, x}^{\rm mod}$ be defined in
(\ref{defde}), (\ref{defdede}).
Assume that
\[
g\geq 0, \ \ g\in L^{1}(\rr), \ \  ga_{p}^{\frac{2n}{2n-p}}\in
L^{1}(\rr), \ \ 0\leq p\leq 2n-1.
\]
Then there exists $C>0$ such that if
\[
D_{2}:= C(1+\sup_{0\leq p\leq 2n-1}\int
g(x)|a_{p}(x)|^{\frac{2n}{2n-p}}\d x), \ \ D_{3}=C(1+\int g(x)\d
x),
\]
one has
\[
\int g(x):\!P(x, \phi(f_{\kappa, x}))\!:\d x\geq -D_{2}-D_{3} (\ln
\kappa)^{n}, \ \ \forall \kappa\geq 2,
\]
and the analogous result for $f_{\kappa,x}$ replaced by $f^{\rm
mod}_{\kappa, x}$.
\end{lemma}
\proof We prove the lemma for $f_{\kappa, x}$, the proof for $f^{\rm
mod}_{\kappa, x}$ being the same, using Lemma \ref{troud} {\it ii)}
instead of Lemma \ref{2.2} {\it iii)}.
Note first from by Lemma \ref{2.2} {\it iii)}
$\|f_{\kappa, x}\|\in O((\ln \kappa)^{\12})$ uniformly in $x$. We will use the inequality
\beq
a^{p}b^{n-p}\leq \epsilon b^{n}+ C_{\epsilon}a^{n} \ \ \forall
\epsilon>0, \ \ a, b\geq 0,
\label{el1.1}
\eeq
valid for $n,p\in \nn$ with $p\leq n$. In fact (\ref{el1.1}) follows
from
\[
\lambda^{p}\leq \epsilon \lambda^{n} + C_{\epsilon}, \ \ \forall
\epsilon>0, \lambda\geq 0,
\]
by setting $\lambda= ba^{-1}$.

We recall the well-known {\em Wick identities}:
\beq
\label{wick}
:\!\phi(f)^{n}\!:=
\sum_{m=0}^{[n/2]}\frac{n!}{m!(n-2m!)}\phi(f)^{n-2m}\Bigl(-\12
\|f\|^{2} \Bigr)^{m}.
\eeq
We apply (\ref{wick}) to $f= f_{\kappa, x}$. Picking first $\epsilon$
small enough in  (\ref{el1.1}) we get:
\[
:\!\phi(f_{\kappa, x})^{2n}\!:\geq \12 (\phi(f_{\kappa,
x})^{2n}- C(\ln \kappa)^{n}).
\]
Using again (\ref{el1.1}) for $\epsilon=1$, we get also:
\[
|:\!\phi(f_{\kappa, x})^{p}\!:|\leq
C_{2}(|\phi(f_{\kappa, x})|^{p}+ (\ln \kappa)^{p/2}) \ \ 0\leq p<2n.
\]
which yields:
\[
\begin{array}{rl}
:\!P(x, \phi(f_{\kappa, x}))\!:\geq &\12( \phi(f_{\kappa,
x})^{2n}- C\sum_{p=0}^{2n-1}a_{p}(x)|\phi(f_{\kappa, x})|^{p})\\[2mm]
&-C((\ln \kappa)^{n}+ \sum_{p=0}^{2n-1} a_{p}(x)(\ln \kappa)^{p/2}).
\end{array}
\]
Using again (\ref{el1.1}), we get:
\[
a_{p}(x)|\phi(f_{\kappa, x})|^{p}=
a_{p}(x)^{\frac{2n-p}{2n-p}}|\phi(f_{\kappa, x})|^{p}\leq \epsilon
\phi(f_{\kappa, x})^{2n}+ C_{\epsilon}a_{p}(x)^{\frac{2n}{2n-p}},
\]
\[
a_{p}(x)(\ln \kappa)^{p/2}= a_{p}(x)^{\frac{2n-p}{2n-p}}(\ln
\kappa)^{p/2}\leq C((\ln \kappa)^{n}+ a_{p}(x)^{\frac{2n}{2n-p}}),
\]
which yields for $\epsilon$ small enough:
\[
:\!P(x, \phi(f_{\kappa, x}))\!:\geq-C\sum_{p=0}^{2n-1}
a_{p}(x)^{\frac{2n}{2n-p}} -C (\ln \kappa)^{n}.
\]
Integrating this estimate we obtain the lemma. \qed

\medskip

As a consequence of Lemma \ref{lower1}, we have the following
proposition, which allows to control a lower order polynomial by the
$P(\varphi)_{2}$ Hamiltonian $H$.
\begin{proposition}\label{lower-pert}
Let $P(x, \lambda)$ be as in (\ref{defdeP}).
Let
\[
H= \d\G(\omega)+ \int_{\rr}g(x)\:P(x, \varphi(x))\!:\d x
\]
and
\[
Q(x, \lambda)=\sum_{r=0}^{2n-1}b_{r}(x)\lambda^{r}
\]
where $gb_{r}\in L^{2}(\rr)$, $g b_{r}^{\frac{2n}{2n-r}}\in
L^{1}(\rr)$. Let $D>0$ such that
\[
\sup_{0\leq p\leq 2n}\|ga_{p}\|_{2}+\sup_{0\leq r\leq
2n-1}\|gb_{r}\|_{2}+ \|g\|_{1}+\sup_{0\leq p\leq 2n-1}\|g
a_{p}^{\frac{2n}{2n-p}}\|_{1}+ \sup_{0\leq
r\leq 2n-1}\|g b_{r}^{\frac{2n}{2n-r}}\|_{1}\leq D.
\]
Then
\[
\pm \int_{\rr} g(x):\! Q(x, \varphi(x))\!:\d x\leq H+C(D).
\]
\end{proposition}
\proof Set $R(x, \lambda)= P(x, \lambda)\pm
Q(x, \lambda)$ and
\[
W=\int_{\rr} g(x):\! R(x, \varphi(x))\!:\d x, \ \ W_{\kappa}=\int_{\rr} g(x):\!
R(x, \varphi_{\kappa}(x))\!:\d x.
\]
It follows from Lemma \ref{2.2}, Lemma \ref{p.5}, and Lemma
\ref{lower1} that $W, W_{\kappa}$ satisfy the conditions in Lemma
\ref{p.6} with constants $C_{i}$ depending only on $D$.
It follows then from Thm. \ref{p.2} that
\[
H\pm\int_{\rr} g(x):\! Q(x, \varphi(x))\!:\d x= H_{0}+ W\geq -C(D),
\]
for some constant $C(D)$ depending only on $D$. \qed

\section{Higher order  estimates}\init\label{sec4}
This section is devoted to the proof of {\em higher order estimates}
for variable coefficients $P(\varphi)_{2}$ Hamiltonians. Higher order
estimates are important for the spectral and scattering theory of $H$,
because they substitute for the lack of knowledge of the domain of
$H$.

The higher order estimates were originally proved by Rosen \cite{Ro}
in the constant coefficients case $\omega= (D^{2}+ m^{2})^{\12}$ for
$g\in \coinf(\rr)$
and $P(x, \lambda)$
independent on $x$. The proof
was later extended in \cite{DG} to the natural class $g\in L^{1}(\rr)\cap
L^{2}(\rr)$. The extension of these results to $x-$dependent polynomials
is straightforward.

 Analysing closely the proof of Rosen, one notes that a
crucial role is played by the fact that the generalized
eigenfunctions of the one-particle energy $(D^{2}+ m^{2})^{\12}$,
namely the exponentials $\e^{\i k \cdot x}$ are {\em
uniformly bounded } both in $x$ and $k$.

To extend Rosen's proof to the variable coefficients case, it is
convenient to diagonalize the one-particle energy $\omega$ in terms of
eigenfunctions and generalized eigenfunctions of $\omega^{2}=
Da(x)D +
c(x)$. However some bounds on eigenfunctions and generalized
eigenfunctions are needed to replace the uniform boundedness of the
exponentials in the constant coefficients case. These bounds are
 given  by conditions {\it (BM1)}, {\it (BM2)}.

In this section, we will prove the following theorem.
\begin{theoreme}\label{4.1}
Let $H$ be a variable coefficients $P(\varphi)_{2}$ Hamiltonian as in
Thm. \ref{basic}. Assume that hypotheses {\it (BM1)}, {\it (BM2)},
{\it (BM3)} hold.  Then there exists
$b>0$ such that for all $\alpha\in \nn$, the following
{\em higher order estimates} hold:
\beq
\begin{array}{l}
\|N^{\alpha}(H+b)^{-\alpha}\|<\infty,\\[2mm]
\|H_{0}N^{\alpha}(H+b)^{-n-\alpha}\|<\infty,\\[2mm]
\|N^{\alpha}(H+b)^{-1}(N+1)^{1-\alpha}\|<\infty.
\end{array}
\label{e4.0}
\eeq
\end{theoreme}
The rest of the section is devoted to the proof of Thm. \ref{4.1}.

\subsection{Diagonalization of $\omega$}\label{sec4.1}
Let $h$, $\omega$ as in Thm. \ref{basic}. By Subsect.
\ref{sec1.3}, $h$ is unitarily equivalent (modulo a constant term)
to a Schr\"{o}dinger operator $D^{2}+ V(x)$ for $V\in
S^{-\mu}$.

Applying then  standard results on the spectral theory of one
dimensional  Schr\"{o}dinger
operators, we know that there exists $\{\psi_{l}\}_{l\in I}$ and
$\{\psi(\cdot, k)\}_{k\in \rr}$ such that
\[
\begin{array}{l}
\psi_{l}(\cdot)\in L^{2}(\rr), \ \ \psi(\cdot, k)\in \cS'(\rr), \\[2mm]
h\psi_{l}= (\lambda_{l}+ m_{\infty}^{2})\psi_{l}, \ \ \lambda_{l}<0, \ \
\psi_{l}\in L^{2}(\rr),
\\[2mm]
h\psi(\cdot, k)= (k^{2}+ m_{\infty}^{2})\psi(\cdot, k), \ \ k\in
\rr^{*}, \\[2mm]
\sum_{l\in I}|\psi_{l})(\psi_{l}|+\frac{1}{2\pi}\int_{\rr}
|\psi(\cdot, k))(\psi(\cdot, k)|\d k=\one.
\end{array}
\]
Moreover using the results of Subsect. \ref{sec3.1} and the fact that
$h$ is a real operator we can assume that
\beq
\overline{\psi}_{l}= \psi_{l}, \ \ \overline{\psi}(x, k)=
\psi(x, -k).
\label{e4.1}
\eeq
The index set $I$ equals either  $\nn$ or a finite subset of $\nn$
depending on the number of negative eigenvalues of $D^{2}+ V$.

Let
\[
\tilde{\ch}:= l^{2}(I)\oplus L^{2}(\rr, \d k),
\]
and
\beq\label{e4.001}
\begin{array}{l}
W: L^{2}(\rr, \d x)\to \tilde{\ch}, \\[2mm]
Wu:=(( \psi_{l}|u))_{l\in I}\oplus
\frac{1}{\sqrt{2\pi}}\int_{\rr}\overline{\psi}(y, k) u(y)\d y.
\end{array}
\eeq
Clearly $W$ is unitary and
\[
W\omega W^{*}=: (\tilde{\omega}_{\rm d}\oplus
\tilde{\omega}_{\rm c}),
\]
for
\[
\tilde{\omega}_{\rm d}= \oplus_{l\in I}(\lambda_{l}+ m_{\infty}^{2})^{\12}, \ \ \tilde{\omega}_{\rm c}=
(k^{2}+ m_{\infty}^{2})^{\12}.
\]
If we set $\tilde{c}= W cW^{*}$, then it follows from (\ref{e4.1})
that
\[
\tilde{c}((u_{l})_{l\in
I}\oplus u(k))= (\overline{u}_{l})_{l\in I}\oplus \overline{u}(-k),
\]
i.e. $\tilde{c}$ is the direct sum of the canonical conjugation on
$l^{2}(I)$ and the standard conjugation on $L^{2}(\rr, \d k)$ used
for the constant coefficients $P(\varphi)_{2}$ model.
\subsection{Reduction of $H$}\label{sec4.2}
We will consider in the rest of this section the transformed
Hamiltonian:
\[
\tilde{H}:= \G(W)H\G(W)^{*}.
\]
In this subsection we determine the explicit form of $\tilde{H}$.

Let $(\tilde{Q}, \tilde{\mu})$ be the $Q-$space associated to the
couple $(\tilde{h}, \tilde{c})$. We can extend $\Gamma(W):
\Gamma(\ch)\to \Gamma(\tilde{\ch})$ to a unitary map
$T: L^{2}(Q, \d \mu)\to L^{2}(\tilde{Q}, \d \tilde{\mu})$.
\begin{lemma}\label{4.1b}
$T$ is an isometry from $L^{p}(Q, \ \mu)$ to $L^{p}(\tilde{Q}, \d
\tilde{\mu})$ for all $1\leq p\leq \infty$ and $T1=1$.
\end{lemma}
\proof 
If $F$ is a real measurable function on $Q$, and $m(F)$ the operator
of multiplication by $F$ on $\Gamma(\ch)$, then $m(TF)=
\Gamma(W)m(F)\Gamma(W)^{*}$, which shows that $T$ is positivity
preserving. Since $T1= T^{*}1= 1$, $T$ is doubly Markovian, hence a
contraction on all $L^{p}$ spaces (see \cite{Si1}). We use the same
argument for $T^{-1}$.  \qed

\medskip

Coming back to $\tilde{H}$ we have:
\[
\tilde{H}=\overline{\tilde{H}_{0}+\tilde{V}},
\]
for
\[
\tilde{H}_{0}:=\Gamma(W)H_{0}\Gamma(W)^{*}= \d\Gamma(\tilde{\omega}_{{\rm d}}\oplus
\tilde{\omega}_{\rm c}), \ \ \tilde{V}:= \Gamma(W)V\Gamma(W)^{*}.
\]
We know from Lemma \ref{troud} that  $V$ is the limit in $\bigcap_{1\leq p<\infty}L^{p}(Q, \d
\mu)$ of $V^{\rm mod}_{\kappa}$, where $V_{\kappa}$ is a sum of terms of the
form
\[
\int_{\rr} ga_{p}(x)\prod_{1}^{r}a^{*}(f^{\rm mod}_{\kappa,
x})\prod_{r+1}^{p}a(f^{\rm mod}_{\kappa,x})\d x,
\]
where $f^{\rm mod }_{\kappa, x}=
\omega^{-\12}\chi(\frac{\omega}{\kappa})\delta_{x}$. This implies
using Lemma \ref{4.1b} that
\[
\tilde{V}=\lim_{\kappa\to \infty}\tilde{V}_{\kappa},\hbox{ in }\bigcap_{1\leq p<\infty}L^{p}(Q, \d
\mu)
\]
where $\tilde{V}_{\kappa}$ is a sum of terms of the form
\[
\int_{\rr} ga_{p}(x)\prod_{1}^{r}a^{*}(Wf^{\rm mod}_{\kappa,
x})\prod_{r+1}^{p}a(Wf^{\rm mod}_{\kappa,x})\d x.
\]
Another useful expression of $\tilde{V}$ is
\beq
\tilde{V}=\int_{\rr} g(x) :\!P(x, \tilde{\varphi}(x))\!:\d x,
\label{e4.3}
\eeq
for
\[
\tilde{\varphi}(x)= \varphi(W\delta_{x}).
\]
Therefore we see that $\tilde{H}$ is very similar to a
$P(\varphi)_{2}$ Hamiltonian with constant coefficients, the only differences being that in
addition to the usual one-particle energy $(k^{2}+
m_{\infty}^{2})^{\12}$ we have the diagonal operator
$\tilde{\omega}_{\rm d}$, and in the interaction the delta function $\delta_{x}$ is
replaced by $W \delta_{x}$.

From now on we will work with $\tilde{H}$ and to simplify notation we
will omit the tildes on the objects $\tilde{Q}$, $\tilde{\mu}$,
$\tilde{H}$, $\tilde{H}_{0}$, $\tilde{V}$, $\tilde{\ch}$,
$\tilde{\omega}_{\rm d}$, $\tilde{\omega}_{\rm c}$. The
one-particle energy $\omega_{\rm d}\oplus \omega_{\rm c}$ will be
denoted simply by $\omega$.
\subsection{Cutoff Hamiltonians}\label{sec4.3}
We first recall some facts from \cite{DG}.

Let $\ch$ be a Hilbert space equipped with a
conjugation $c$. Let $\pi_{1}:\ch\to \ch_{1}$ be an orthogonal projection
on a closed subspace $\ch_{1}$ of $\ch$
with $[\pi_{1}, c]=0$. Let $\ch_{1}^{\perp}$ be the orthogonal
complement of $\ch_{1}$. In all formulas below we will consider
$\pi_{1}$ as an element of $B(\ch, \ch_{1})$. With this convention
the orthogonal projection on $\ch_{1}$,
considered as an element of $B(\ch, \ch)$, is
equal to $\pi_{1}^{*}\pi_{1}$.

Let $U:\G(\ch_{1})\otimes
\G(\ch_{1}^{\perp})\to \G(\ch)$  the canonical unitary map.
 We denote by $L^{2}(Q_{1}, \d\mu_{1})$,
$L^{2}(Q_{1}^{\perp}, \d \mu_{1}^{\perp})$ the $Q-$space
representations of $\G(h_{1})$, $\G(\ch_{1}^{\perp})$.
Recall that  by \cite[Prop.  5.3]{DG}, we may take as $Q-$space
representation of $\G(\ch)$ the space $L^{2}(Q, \d\mu)$ for
$Q=Q_{1}\times Q_{1}^{\perp}$, $\mu=\mu_{1}\otimes \mu_{1}^{\perp}$.
Accordingly we denote by $(q_{1}, q_{1}^{\perp})$ the elements of
$Q=Q_{1}\times Q_{1}^{\perp}$.

If $W\in B(\G(\ch))$ we set:
\[
B(\G(\ch))\ni\Pi_{1}W:=U\Big(
\G(\pi_{1})W\G(\pi_{1}^{*})\otimes \one\Big) U^{*}.
\]
The following lemma is shown  in \cite[Subsect. 7.1]{DG}.
\begin{lemma}
{\it i)} If $w\in B_{\fin}(\G(\ch))$ then
\beq
\Pi_{1}\Wick(w)= \Wick(\G(\pi_{1}^{*}\pi_{1})w\G(\pi_{1}^{*}\pi_{1})).
\label{sechigh.e2}
\eeq
{\it ii)} If $V$ is a multiplication operator by a function  in
$L^{2}(Q, \d\mu)$ then $\Pi_{1}V$ is the operator of multiplication by
the function
\beq
\Pi_{1}V(q_{1})=\int_{Q_{1}^{\perp}}V(q_{1},
q_{1}^{\perp})\d\mu_{1}^{\perp}.
\label{sechigh.e1}
\eeq
\label{sechigh.2}
\end{lemma}

In particular if $W=\Pi_{1}^{q}a^{*}(h_{i})\Pi_{1}^{p}a(g_{i})$, then
\beq
\Pi_{1}W=
\Pi_{1}^{q}a^{*}(\pi_{1}^{*}\pi_{1}h_{i})
\Pi_{1}^{p}a(\pi_{1}^{*}\pi_{1}g_{i}).
\label{P51}
\eeq
Let now $\{\pi_{n}\}_{n\in \nn}$  be a
sequence of orthogonal projections on
$\ch$ such that
\beq
\pi_{n}\leq \pi_{n+1},
\: [\pi_{n}, c]=0,\:\slim_{n\fld +\infty}\pi_{n}=\one,
\label{ep.11}
\eeq
and let $\Pi_{n}$ the associated maps defined by
(\ref{sechigh.e2}). Using the representation (\ref{sechigh.e1}) it is
shown in \cite[Prop. 4.9]{SHK} that
\beq
\begin{array}{l}
i)\: \Pi_{n}V\fld V\hbox{ in }L^{p}(Q, \d\mu),\hbox{ when }n\fld
\infty,\hbox{ if }V\in L^{p}(Q, \d\mu),\: 1\leq p<\infty\\[2mm]
ii)\: \|\e^{-t\Pi_{n}V}\|_{L^{1}(Q, \d\mu)}\leq \|\e^{-tV}\|_{L^{1}(Q,
d\mu)}.
\end{array}
\label{ep.9}
\eeq

\subsection{Notation}\label{sec4.4}

{\bf Index sets.}

An element $u\in \ch$ is of the form $(u_{l})\oplus u(k)\in
l^{2}(I)\oplus L^{2}(\rr, \d k)$. We  put together the variables
$l\in I$ and $k\in \rr$ into a single variable $K\in I\sqcup\rr$. We
denote by $\d K$ the measure on $I\sqcup\rr$ equal to the sum of the
counting measure on $I$ and the Lebesgue measure on $\rr$.
Then $\ch= L^{2}(I\sqcup \rr, \d K)$ and
\[
(u|v)_{\ch}= \sum_{l\in I}\overline{u}_{l}v_{l}+ \int_{\rr}
\overline{u}(k)v(k)\d k=\int_{I\sqcup \rr} \overline{u}(K)v(K)\d K.
\]
For $K\in I\sqcup\rr$ we set:
\[
\omega(K):=\left\{
\begin{array}{l}
(k^{2}+ m_{\infty}^{2})^{\12} \hbox{ if }K=k\in \rr,\\
(\lambda_{l}+ m_{\infty}^{2})^{\12}\hbox{ if }K=l\in I,
\end{array}
\right.
\]
so that the operator $\omega$ is the operator of multiplication by
$\omega(K)$ on $L^{2}(I\sqcup\rr, \d K)$. We set also:
\[
|K|:=\left\{
\begin{array}{l}
|k| \hbox{ if }K=k\in \rr,\\
l\hbox{ if }K=l\in I.
\end{array}
\right.
\]
\[
a^{\sharp}(K):=\left\{
\begin{array}{l}
 a^{\sharp}(k) \hbox{ if }K=k\in \rr,\\
a^{\sharp}(e_{l})\hbox{ if }K=l\in I,
\end{array}
\right.
\]
where $\{e_{l}\}_{l\in I}$ is the canonical basis of $l^{2}(I)$.

{\bf Lattices.}

For $\nu\geq 1$, we consider  the lattice $\nu^{-1}\zz$ and let
\[
\rr\ni k\mapsto [k]_{\nu}\in \nu^{-1}\zz
\]
be the integer part of $k$ defined by $-(2\nu)^{-1}<k-[k]_{\nu}\leq
(2\nu)^{-1}$.
We extend the function $[\cdot]_{\nu}$ to $I\sqcup\rr$ by setting
\[
[K]_{\nu}:=\left\{
\begin{array}{l}
[k]_{\nu} \hbox{ if }K=k\in \rr,\\
l\hbox{ if }K=l\in I.
\end{array}
\right.
\]
As above we put together the variables $l\in I$ and $\gamma\in
\nu^{-1}\zz$ into a single variable $\delta\in I\sqcup \nu^{-1}\zz$.
For
$\kappa\in [1, +\infty[$ an UV
cutoff parameter, we denote by $\Gamma_{\kappa,\nu}$ the finite lattice $\nu^{-1}\zz\cap
\{|\gamma|\leq \kappa\}$.

As in \cite[Sect. 7.1]{DG} we choose   increasing sequences $\kappa_{n}$,
$\nu_{n}$ tending to $+\infty$ in such a way that
\[
\Gamma_{\kappa_{n},
\nu_{n}}\subset \Gamma_{\kappa_{n+1}, \nu_{n+1}}.
\]
We denote by $\Gamma_{n}$ the finite lattice $\Gamma_{\kappa_{n},
\nu_{n}}$. The finite subset of $I\sqcup \nu^{-1}\zz$:
\[
T_{n}:=\{l\in I|\ \ l\leq \kappa_{n}\}\sqcup \Gamma_{n}
\]
can be rewritten as
\[
T_{n}= \{\delta\in I\sqcup \nu^{-1}\zz|\ \ |\delta|\leq \kappa_{n}\}.
\]

{\bf Finite dimensional subspaces.}

For $\gamma\in \nu^{-1}\zz$
we denote by $e_{\gamma}\in L^{2}(\rr, \d k)$ the
vector $e_{\gamma}(k)= \nu^{\12}\one_{]-(2\nu)^{-1},
(2\nu)^{-1}]}(k-\gamma)$.

Following our previous convention we set for $\delta\in
I\sqcup \nu^{-1}\zz$:
\[
e_{\delta}:=\left\{
\begin{array}{l}
0\oplus e_{\gamma}\hbox{ if }\delta=\gamma\in \nu^{-1}\zz,\\
e_{l}\oplus 0\hbox{ if }\delta=l\in I.
\end{array}
\right.
\]
Clearly $(e_{\delta})$ is an orthonormal family in $\ch$.

For $n\in \nn$
we denote by $\ch_{n}$ the finite dimensional subspace of
$\ch$ spanned  the $e_{\delta}$ for $\delta\in T_{n}$,
and denote by $\pi_{n}:\ch\to \ch_{n}$  the orthogonal
projection on the finite dimensional subspace $\ch_{n}$. Note that
$\ch_{n}$ is invariant under the conjugation $c$.

Finally we set
\[
a^{\sharp}(\delta):=\left\{
\begin{array}{l}
 a^{\sharp}(e_{\gamma}) \hbox{ if }\delta=\gamma\in \nu^{-1}\zz,\\
a^{\sharp}(e_{l})\hbox{ if }\delta=l\in I.
\end{array}
\right.
\]

\subsection{Proof of the higher order estimates}\label{sec4.6}

For $0\leq \tau\leq 1$ and $n\in \nn$ we set:
\[
N^{\tau}_{n}=\int \omega([K]_{\nu_{n}})^{\tau}a^{*}(K)a(K)\d K.
\]
Note that with the notation in Subsect. \ref{sec4.1}:
\[
N^{\tau}_{n}=\d\Gamma\left((\omega_{\rm d}\oplus
\omega_{\rm c}([k]_{\nu_{n}}))^{\tau}\right)=
\d\Gamma(\omega([K]_{\nu_{n}})^{\tau}).
\]
We set also
\[
H_{0, n}= N^{1}_{n}, \ \ H_{n}= H_{0, n}+ V_{n},
\]
where
\[
V_{n}=\Pi_{n}V.
\]
\begin{lemma}\label{smallem} there exists $C>0$ such that
\[
i)\quad (N^{\tau}_{n}+C)^{-1}\to (N^{\tau}+C)^{-1}, \ \ ii)\quad
(H_{n}+C)^{-1}\to (H+c)^{-1},
\]
in norm when $n\to \infty$.
\end{lemma}
\proof
To prove {\it i)} we note that  $\omega_{\rm c}([k]_{\nu_{n}})^{\tau}$ converges in
norm to $\omega_{\rm c}^{\tau}$ for $0\leq \tau\leq 1$, which implies
that $N^{-\12}(N^{\tau}_{n}-N^{\tau})N^{-\12}$ tends to $0$ in norm.
Since $\omega([K]_{\nu_{n}})\geq c>0$
uniformly in $n$, we know that $N^{\12}(N^{\tau}_{n}+1)^{-1}$ is
bounded uniformly in $n$. This implies that $(N^{\tau}_{n}+1)^{-1}$
converges in norm to $(N^{\tau}+1)^{-1}$ when $n\to \infty$.

To prove {\it ii)} we follow the proof of \cite[Prop. 4.8]{SHK}: we
have seen above that  $N^{-\12}(H_{0,
n}-H_{0})N^{-\12}$ tends to $0$ in norm.
Moreover $\omega_{{\rm d}}\oplus
\omega([k]_{\nu_{n}})\geq C>0$ uniformly w.r.t. $n$. This implies that
$\e^{-tH_{0, n}}$ is hypercontractive with hypercontractivity bounds
uniform in $n$. This implies that if $W\in L^{p}(Q,
\d\mu)$
and $\e^{-TW}\in L^{1}(Q, \d \mu)$ there exists C such
that $N\leq C(H_{0, n}+W+ C)$, uniformly in $n$.
Writing
\[
(H_{0, n}+ W+C)^{-1}-(H_{0}+ W+C)^{-1}= (H_{0,n}+
W+C)^{-1}(H_{0}-H_{0,n})(H_{0}+ W+C)^{-1}
\]
and using the above bound, we obtain that $(H_{0, n}+ W+C)^{-1}$
converges in norm to $(H+W+C)^{-1}$. Moreover it follows from
Theorem \ref{p.2} {\it ii)} that the constant $C$ above depend
only on $\|e^{-tW}\|_{L^{1}}$ for some $t>0$.

Since by (\ref{ep.9}) $\e^{-tV_{m}}$ is uniformly bounded in
$L^{1}(Q)$, we see that $(H_{0, n}+ V_{m}+C)^{-1}$ converges in norm to $(H_{0}+
V_{m}+C)^{-1}$ when $n\to \infty$, uniformly w.r.t. $m$. Again by
(\ref{ep.9}) $V_{m}\to V$ in $L^{p}$ for some $p>2$ and $\e^{-tV_{m}}$
is uniformly  bounded in $L^{1}$,  so by Prop. \ref{shk} we obtain that
$(H_{0}+ V_{m}+C)^{-1}$ converges to $(H_{0}+ V+C)^{-1}$ when $m\to
\infty$, wich
completes the proof of the lemma. \qed

\medskip

Let us denote simply by $\omega_{n}$ the operator $\omega_{\rm
d}\oplus \omega_{\rm c}(([k]_{\nu_{n}})$. Since $[\omega_{n},
\pi_{n}^{*}\pi_{n}]=0$, we have
\[
H_{0, n}=
U_{n}\left(\d\Gamma(\omega_{n}\big|_{\ch_{n}})\otimes\one+\one\otimes\d\Gamma(\omega_{n}\big|_{\ch_{n}^{\perp}})\right)U_{n}^{*},
\]
where $U_{n}: \Gamma(\ch_{n})\otimes\Gamma(\ch_{n}^{\perp})\to
\Gamma(\ch)$ is the exponential map.  This implies that
\[
U_{n}^{*}H_{n}U_{n}= \hat{H}_{n}\otimes\one+ \one\otimes
\d\Gamma(\omega_{n}\big|_{\ch_{n}^{\perp}}), \ \
U_{n}^{*}N^{\tau}_{n}U_{n}=\hat{N}^{\tau}_{n}\otimes\one
+\one\otimes\d\Gamma(\omega_{n}^{\tau}\big|_{\ch_{n}^{\perp}}),
\]
for $\hat{H}_{n}= \d\Gamma(\omega_{n}\big|_{\ch_{n}})+ V_{n}$,
$\hat{N}^{\tau}_{n}= \d\Gamma(\omega_{n}^{\tau}\big|_{\ch_{n}})$.

\begin{proposition}\label{blob}
Assume hypotheses {\it (BMi)} for $i=1,2,3$.
Set for $J=\{1, \dots, ,s\}\subset \nn$ and $K_{i}\in
I\sqcup\rr$:
\[
V_{n}^{J}:= {\rm ad}_{a(K_{1})}\cdots {\rm ad}_{a(K_{s})}V_{n}.
\]
Then there exists $b,c>0$ such that for all $\lambda_{1},
\lambda_{2}<-b$
\[
\|(H_{n}-\lambda_{2})^{-\12}V_{n} ^{J}(H_{n}-\lambda_{1})^{-\12}\|\leq
c\prod_{1}^{s}F(K_{i}),
\]
where $F:I\sqcup\rr\to \rr^{+}$ satisfies for each $\delta>0$:
\[
\int_{I\sqcup\rr}|F(x, K)|^{2}\omega(K)^{-\delta}\d K\leq C.
\]
\end{proposition}
\proof
We have using (\ref{P51}):
\beq
V_{n}=\int g(x) :\!
P(x,\varphi_{n}(x))\!:\d x,
\label{e4.4}
\eeq
where
\beq
:\!\varphi_{n}(x)^{p}\!:
=\sum_{r=0}^{p}\left(\begin{array}{c}p\\r\end{array}\right)
\prod_{1}^{r}a^{*}(\pi_{n}^{*}\pi_{n}W\omega^{-\12}\delta_{x})\prod_{r+1}^{p}a(\pi_{n}^{*}\pi_{n}
W\omega^{-\12}\delta_{x}).
\label{e4.5}
\eeq
We note that
\[
\pi_{n}^{*}\pi_{n}= \sum_{|\delta|\leq
\kappa_{n}}|e_{\delta}\rangle\langle e_{\delta}|,
\]
which yields
\beq
:\!
\varphi_{n}(x)^{p}\!:
=\sum_{r=0}^{p}\left(\begin{array}{c}p\\r\end{array}\right)\sum_{\delta_{1},
\ldots, \delta_{p}\in T_{n}}
\prod_{1}^{r}a^{*}(\delta_{i})\prod_{r+1}^{p}a(\delta_{i})\prod_{1}^{r}m_{n}(x,
\delta_{i})\prod_{r+1}^{p}\overline{m}_{n}(x, \delta_{i}),
\label{e4.6}
\eeq
where
\[
m_{n}(x, \delta)=(e_{\delta}|W\delta_{x})_{\ch}.
\]
Let  for $k\in \rr$:
\[
 C_{n}(k):=\Big[[k]_{\nu_{n}}-\12\nu_{n}^{-1},
[k]_{\nu_{n}}+\12\nu_{n}^{-1}\Big],
\]
be the cell of $\Gamma_{n}$ centered at $[k]_{\nu_{n}}$.
Using (\ref{e4.001}) we get:
\beq
m_{n}(x, \delta):=\left\{
\begin{array}{l}
\nu_{n}^{\12}\int_{C_{n}(\gamma)}(k^{2}+
m_{\infty}^{2})^{-\frac{1}{4}} \overline{\psi}(x, k)\d k \hbox{ if
}\delta=\gamma\in \Gamma_{n},
\\[2mm]
(\lambda_{l}+ m_{\infty}^{2})^{-\frac{1}{4}}\psi_{l}(x), \hbox{ if
}\delta=l\in I.
\end{array}
\right.
\label{e4.7}
\eeq

Then as in \cite{Ro}, \cite{DG}, we obtain that
\[
V_{n} ^{J}= \int_{\rr} g(x)\prod_{1}^{s}r_{n}(x, K_{i}):\!P^{(s)}(x,
\varphi_{n}(x))\!:\d x,
\]
where $P^{(s)}(x, \lambda)= (\frac{\d}{\d \lambda})^{s}P(x,\lambda)$
and
\[
r_{n}(x, K)=\left\{
\begin{array}{l}
\nu_{n}\int_{C_{n}(k)} \omega(k')^{-\12}\psi(x, k')\d k'
\hbox{ if }K=k\in \rr, \\[2mm]
\psi_{l}(x) \hbox{ if }K=l\in I.
\end{array}
\right.
\]
We note that 
\beq\label{e4.9b} 
V_{n} ^{J}= \Pi_{n}\int_{\rr}
g(x):\!R_{n}(x, K_{1}, \dots, K_{s}, \varphi(x))\!:\d x, 
\eeq 
for
\[
R_n(x, K_{1}, \dots, K_{s}, \lambda)= P^{(s)}(x, \lambda)\prod_{1}^{s}r_{n}(x,
K_{i}).
\]
Since assumptions {\it (BM1)},  {\it (BM2)} are satisfied,  we know that:
\[
|\psi(x, k)|\leq C M(x), \hbox{ uniformly for }x,k\in \rr,
\]
\[
|\psi_{l}(x)|\leq C\epsilon_{l}M(x), \hbox{ uniformly for }x\in \rr,
\:l\in I,
\]
where $\sum_{l\in I}\epsilon_{l}^{2}<\infty$.

Let us now prove corresponding bounds on the functions $r_{n}(x, K)$.
We consider first the case $K=l\in I$:
we have:
\beq
|r_{n}(x, l)|\leq C\epsilon_{l}M(x), \hbox{  uniformly in }x, l.
\label{e4.11}
\eeq
If $K=k\in \rr$ we get:
\[
\omega(k')^{-\12}|\psi(x, k')|\leq C\omega(k)^{-\12}M(x), \hbox{ uniformly for  }n\in
\nn, \ \ k'\in C_{n}(k),\ \ x\in \rr,
\]
which yields:
\beq
|r_{n}(x, k)|\leq C\omega(k)^{-\12}M(x), \hbox{ uniformly for }n\in
\nn,  \ \ k,  x\in \rr.
\label{e4.12}
\eeq
If we set:
\beq
F(K)=\left\{\begin{array}{l}
\omega(k)^{-\12} \hbox{ if }K=k\in \rr,\\[2mm]
\epsilon_{l}\hbox{ if }K=l\in I,
\end{array}
\right.
\label{e4.13b}
\eeq
and collect (\ref{e4.11}), (\ref{e4.12}) we get:
\beq
|r_{n}(x,K)|\leq CF(K)M(x), \hbox{ uniformly for }n\in \nn, K\in
I\sqcup\rr, \ \ x\in \rr
\label{e4.8}
\eeq
We note that by condition {\it (BM3)}, we have:
\[
ga_{p}M^{s}\in L^{2}, \ \ g(a_{p}M^{s})^{\frac{2n}{2n-p+s}}\in L^{1},
\ \ 0\leq s\leq p\leq 2n-1.
\]
If we apply the arguments in Prop.  \ref{lower-pert} to the polynomial
\[
Q_{n}(x, K_{1}, \dots, K_{s}, \lambda)=
P^{(s)}(x, \lambda)\prod_{1}^{s}F(K_{i})^{-1}r_{n}(x, K_{i}),
\]
using the bound (\ref{e4.8}), we obtain  that
\beq
\e^{-t(V\pm W_{n})}\hbox{ is uniformly bounded in }L^{1}(Q),
\label{e4.9}
\eeq
for
\[
W_{n}(K_{1}, \dots K_{s})=\int g(x):\!Q_{n}(x, K_{1},\dots, K_{s},
\varphi(x))\!:\d x.
\]
By (\ref{ep.9}) {\it ii)}, this implies that
\[
\e^{-t(V_{n}\pm \Pi_{n}W_{n})}\hbox{ is uniformly bounded in
}L^{1}(Q).
\]
Applying then Thm. \ref{XII} to $a=\omega_{n}$ and using (\ref{e4.9}) we get that there
exists $C>0$ such that
\[
\pm V_{n} ^{J}\leq \prod_{1}^{s}F(K_{i})(H_{n}+ C),\hbox{ uniformly in }n.
\]
To complete the proof of the proposition it remains to check that for
each $\delta>0$
\[
\int_{I\sqcup\rr}F(K)^{2}\omega(K)^{-\delta}\d K<\infty,
\]
which follows from (\ref{e4.13b}) since $\sum_{l\in I}\epsilon_{l}^{2}<\infty$. \qed

\medskip

{\bf Proof of Thm. \ref{4.1}.}

We follow the proof in \cite{Ro}. This proof consists in first proving
higher order estimates
for  the cutoff Hamiltonians $H_{n}$ and $N^{\tau}_{n}$, with
constants uniform in $n$. The corresponding estimates for the
Hamiltonians without cutoffs are then obtained by the principle of
cutoff independence (\cite[Prop. 4.1]{Ro}). The convergence results
needed to apply \cite[Prop. 4.1]{Ro} are proved in Lemma
\ref{smallem}. The estimates for the cutoff
Hamiltonians rely on three kinds of intermediate results:

the first
(\cite[Lemma4.2]{Ro}, \cite[Corollary 4.3]{Ro})
consists of identities expressing expectation values of (powers of) $N^{\tau}$ in terms of Wick monomials. These identities carry over
directly to our case, replacing $\rr$ by $I\sqcup\rr$, $a^{\sharp}(k)$
by $a^{\sharp}(K)$ and the mesure $\d k$ by $\d K$.

The second (\cite[Prop. 4.5]{Ro}) is the generalized pullthrough formula wich also carries
over  to our case. The last is the bound in \cite[Lemma
4.4]{Ro} which is replaced in our case by Prop. \ref{blob}. Carefully
looking at the proof
of the higher order estimates for the cutoff Hamiltonians in \cite[Thm.
4.7]{Ro} and  \cite[Corollary 4.8]{Ro} we see that  it relies on the fact that the
\[
\int_{I\sqcup \rr} F(K)^{2}\omega(K)^{-\delta}\d K<\infty,
\]
(in \cite{Ro} $F(K)$ equals simply
$\omega(k)^{-\12}$), which is checked in Prop. \ref{blob}. This
completes the proof of Thm. \ref{4.1}. \qed

\appendix

\section{Appendix A}\label{sec3}\init
In this section we will give sufficient conditions on the functions
$a, c$ in the definition of $\omega$ for conditions {\it (BM1)}, {\it
(BM2)} to hold.

\subsection{Sufficient conditions for  {\it (BM1)}}\label{sec3.0}
\begin{proposition}\label{suff}
Let $h= Da(x)D+ c(x)$ be as in Thm. \ref{basic}.
Then:

i)  condition {\it
(BM1)} is satisfied for $M(x)= \x^{\alpha}$ for $\alpha>\12$.

ii) if $h$ has a finite number of eigenvalues, condition {\it (BM1)} is
satisfied for $M(x)= 1$.
\end{proposition}
\proof
{\it ii)} is obvious. To prove {\it i)} we take
an orthonormal  basis $\{\psi_{l}\}_{l\in I}$ of the point
spectrum
subspace of $h$, and set $u_{l}= \langle
D\rangle^{s}\x^{-\alpha}\psi_{l}$ for some $s>\12$. By Sobolev's
theorem we have
\[
\|\x^{-\alpha}\psi_{l}\|^{2}_{\infty}\leq C\|u_{l}\|^{2}_{2}, \hbox{
uniformly in }l\in I.
\]
Next
\[
\begin{array}{rl}
&\sum_{l\in I}\|u_{l}\|^{2}_{2}={\rm Tr}\sum_{l\in I}|u_{l})(u_{l}|\\[2mm]
=&{\rm Tr}\langle D\rangle^{s}\x^{-\alpha}\sum_{l\in
I}|\psi_{l})(\psi_{l}|\x^{-\alpha}\langle D\rangle^{s}\\[2mm]
= &{\rm Tr}\langle D\rangle^{s}\x^{-\alpha}\one_{]-\infty,
m_{\infty}^{2}]}(h)\x^{-\alpha}\langle D\rangle^{s}.
\end{array}
\]
We use then that $\one_{]-\infty,
m_{\infty}^{2}]}(h)\langle D\rangle^{m}$ is bounded for all $m\in \nn$
by elliptic regularity, which implies that $\one_{]-\infty,
m_{\infty}^{2}]}(h)\x^{-\alpha}\langle D\rangle^{s}$ is
Hilbert-Schmidt if $\alpha>\12$. This completes the proof of the
proposition. \qed

\subsection{Generalized eigenfunctions}\label{sec3.1}
In this subsection we show that if $h= Da(x)D+ c(x)$
is a second order differential operator as in Thm. \ref{basic}
satisfying {\it (BM2)}, then the generalized eigenfunctions $\psi(x,
k)$ can be choosen to satisfy additionally the following reality
condition:
\[
\overline{\psi}(x, k)= \psi(x, -k), \ \ k \ \hbox{a.e.}
\]
\begin{lemma}\label{3.1}
Assume that the family  $\{\phi(\cdot,
k)\}_{k\in \rr}$ satisfies assumption  {\it (BM2)}. Then there
exists a family  $\{\psi(\cdot, k)\}_{k\in \rr}$ of generalized
eigenfunctions of $h$  satisfying  {\it (BM2)} and
additionally:
\beq\label{e.real1}
\overline{\psi}(x, k)= \psi(x, -k), \ \ k \ \hbox{a.e.}
\eeq
\end{lemma}
Let $\{\phi(x,k)\}_{k\in \rr}$ be a basis of generalized eigenfunctions for
$h$. To such a family one can associate a unitary map:
\[
W_{\phi}: \one_{[m_{\infty}^{2}, +\infty[}(h)L^{2}(\rr, \d x)\to  L^{2}(\rr, \d x),
\]
defined by \beq\label{e.real0} W_{\phi}u(x)=  (2\pi)^{-1}\int\int
\e^{\i k \cdot x } \overline{\phi}(y, k)u(y)\d y\d k, \eeq which
satisfies
\[
W_{\phi}h=  (D^{2}+ m^{2}_{\infty})W.
\]
Note that if $\psi$ satisfy (\ref{e.real1}) and $W_{\psi}$ is defined as in (\ref{e.real0}), then
$W_{\psi}$ is a {\em real }operator i.e.
\[
 \overline{W_{\psi}u}= W_{\psi}\overline{u}, \ \ u\in L^{2}(\rr).
\]
\proof Let us define the unitary operator $\Omega: L^{2}(\rr, \d x)\to
L^{2}(\rr^{+}, \d k)\otimes \cc^{2}$
obtained from $W_{\phi}$  by Fourier transform:
\[
\Omega u(k)=(2\pi)^{-1}\left((\phi(\cdot, k)|u), (\phi(\cdot, -k)|u)\right),
\]
satisfying $\Omega h= (k^{2}+m_{\infty}^{2})\otimes\one_{\cc^{2}}\Omega$.

Set \beq\label{e.real2}
\tilde{\phi}(x, k):= \overline{\phi}(x, -k)\ \ x, k\in \rr.
\eeq
 Clearly $\{\tilde{\phi}(\cdot,
k)\}_{k\in \rr}$ is a family of generalized eigenfunctions of $h$.
Therefore we can introduce the unitary map
\[
\tilde{\Omega} u(k)=(2\pi)^{-1}\left( (\tilde{\phi}(\cdot, k)|u),
(\tilde{\phi}(\cdot, -k)|u)\right).
\]
If $S=\Omega\tilde{\Omega}^{-1}$, then $S$ commutes with
$(k^{2}+m_{\infty}^{2})\otimes\one_{\cc^{2}}$ and is unitary, so:
\[
S=\int^{\oplus}_{\rr^{+}}S(k) \d k,
\]
for $S(k)\in U(\cc^{2})$. Using that $(\overline{\phi}(x, k), \overline{\phi}(x,
-k))= \Omega \delta_{x}$ for $x\in \rr$, the similar identity for
$\tilde{\phi}$ and (\ref{e.real2}), we obtain that
\beq
(\overline{\phi}(x, k), \overline{\phi}(x, -k))= S(k)T(\phi(x, k), \phi(x, -k)),
\ \ x\in \rr, \ k>0,
\label{e.real3}
\eeq
where $T(z_{1}, z_{2})= (z_{2}, z_{1})$. Iterating this formula we
obtain the identity:
\beq
T\overline{S}T= S^{-1}.
\label{e.real4}
\eeq
Let us find the generalized eigenfunctions $\psi(x, k)$ under the form
\[
(\psi(x, k), \psi(x,-k))= A(k)(\phi(x, k), \phi(x, -k)) \ \ x\in \rr,
\ k>0.
\]
Clearly $\{\psi(\cdot, k)\}_{k\in \rr}$ will be a basis of
generalized eigenfunctions of $h$ as soon as $A(k)\in U(\cc^{2})$.
Using (\ref{e.real3}) we see that it
will satisfy (\ref{e.real1}) if
\beq
\overline{A}(k)= TA(k)\overline{S}(k)T.
\label{e.real5}
\eeq
To solve (\ref{e.real5}), we deduce first from (\ref{e.real4}) that
\[
T\overline{S}^{n}T= S^{-n}, \ \ n\in \nn.
\]
Therefore $Tg(\overline{S})T= g(S^{-1})$ if $g$ is a polynomial. By the
standard approximation argument and the spectral theorem for unitary
operators this extends to all  measurable functions on the unit circle.

For $z=\e^{\i \theta}$, $-\pi<\theta\leq \pi$, we set $z^{\alpha}=\e^{i
\alpha \theta}$. Setting $A(k)= (\overline{S})^{-\12}(k)$, we get
\[
TA\overline{S}T= T\overline{S}^{\12}T= S^{-\12}.
\]
Since $\overline{z}^{-\12}= \overline{z^{-\12}}$, we obtain that
$S^{-\12}(k)=
\overline{A}(k)$. Therefore $A(k)$ satisfies (\ref{e.real5}). Moreover
since $z^{-\12}$ preserves the unit circle, $A(k)$ is unitary.
Therefore the family $\{\psi(\cdot, k)\}_{k\in \rr}$ is a basis of
generalized eingenfunctions of $h$. Moreover since the matrix $A(k)$ is unitary,
all entries have modulus less than $1$, which implies that
if $\{\phi(\cdot, k)\}_{k\in \rr}$   satisfies {\it (BM2)},
so  does $\{\psi(\cdot, k)\}_{k\in \rr}$. \qed

\subsection{Reduction to the case of the constant metric}\label{sec1.3}
We show in this subsection that in order to verify condition {\it
(BM2)} we can reduce ourselves to the case $a(x)\equiv 1$. We have
then
\[
h= D^{2}+ V +m_{\infty}^{2}, \hbox{ for }V(x)= c(x)-m_{\infty}^{2}\in S^{-\mu},
\]
which will allow to use standard results on generalized eigenfunctions for
Schr\"{o}dinger operators in one dimension.

Let $\psi:\rr\to \rr$  be a diffeomorphism with $\psi'>0$. We denote by
$\psi^{-1}$ the inverse of $\psi$. To $\psi$ we associate the unitary
map $T_{\psi}:
L^{2}(\rr)\to L^{2}(\rr)$
\[
T_{\psi}u(x):= \psi'(x)^{\12}u\circ \psi(x).
\]
\begin{lemma}\label{1.2}
Let $a,c$ satisfying (\ref{e1.1}) and set
 $g=a^{\12}$ and $\psi=\phi^{-1}$ for
\[
\phi(x)=\int_{0}^{x}\frac{1}{g(s)}\ ds.
\]
Then
\[
i) \ \ Da(x)D+ c(x)= T_{\psi}^{*}(D^{2}+
\tilde{c}(x))T_{\psi},
\]
where
\[
\tilde{c}(x)= c\circ \psi(x)+ (\frac{1}{4}(g')^{2}+ \12
gg")\circ \psi(x).
\]
{\it ii)} $\tilde{c}-m_{\infty}^{2}\in S^{-\mu}$.

{\it iii)} if $D^{2}+ \tilde{c}(x)$ satisfies  {\it (BM1)},
{\it (BM2)} for a weight function $\tilde{M}$, then $Da(x)D+ c(x)$
satisfies {\it (BM1)}, {\it (BM2)}  for $M(x)=
\left((\psi')^{-\12}\tilde{M}\right)\circ
\psi^{-1}(x)$. If $\tilde{M}(x)= \x^{\alpha}$, then $M(x)\simeq
\x^{\alpha}$.
\end{lemma}
\proof
Let $\psi:\rr\to \rr$ be a diffeomorphism as above. We have
\[
\p_{x}T_{\psi}u= T_{\psi}(\12\frac{\psi"}{\psi'}\circ \psi^{-1}u + \psi'\circ
\psi^{-1}\p_{x})u.
\]
Choosing $\psi$ as in the lemma we get $g(x)= \psi'\circ\psi^{-1}(x)$
and
\[
\p_{x}T= T(g(x)\p_{x}+\12 g'(x))=:TA.
\]
This yields
\[
\begin{array}{rl}
-\p_{x}a(x)\p_{x}=&(A^{*}-\12 g')(A-\12 g')\\
=&A^{*}A -\12(A^{*}g'+ g'A) + \frac{1}{4}(g')^{2}\\
=&A^{*}A + \frac{1}{4}(g')^{2}+ \12 gg".
\end{array}
\]
This easily implies the first statement of the lemma. Next from
(\ref{e1.1}) we get that $g-1\in S^{-\mu}$ hence $\phi(x)-x\in
S^{1-\mu}$ from which $\psi(x)-x\in S^{1-\mu}$ follows. This implies
that $(\frac{1}{4}(g')^{2}+ \12
gg")\circ \psi\in S^{-2-\mu}$ and $c\circ\psi-m_{\infty}^{2} \in
S^{-\mu}$. Statement {\it iii)} is obvious.
\qed
\section{Appendix B}\init\label{urk}
In this section we recall some results about generalized
eigenfunctions for one-dimensional Schr\"{o}dinger operators, taken from
\cite{Ya1}, \cite{Ya2}. For the reader's convenience, we will sketch
some of the proofs. These results are used to
obtain some  sufficient conditions for  {\it (BM2)}. We saw in
Subsect. \ref{sec1.3} that we can reduce ourselves to considering a
Schr\"{o}dinger operator:
\[
h= D^{2}+ V(x)+ m_{\infty}^{2}, \ \ V\in S^{-\mu} \hbox{ for }\mu>0.
\]
It turns out that condition {\it (BM2)} is really a condition on the
behavior of generalized eigenfunctions $\psi(x, k)$ for $k$ near $0$.
In this respect the potentials fall naturally into two classes,
depending on whether $\mu>2$ or $\mu\leq 2$.

This distinction is also relevant to condition {\it (BM1)}. In fact by
the Kato-Agmon-Simon theorem (see \cite[Thm. XIII.58]{RS}) if $V\in
S^{-\mu}$ for $\mu>0$ $h$ has no strictly positive eigenvalues. As is
well known  $h$ has a finite number of negative eigenvalues if
$\mu>2$.  Therefore condition {\it (BM1)} is always satisfied for
$M(x)\equiv 1$ if $\mu>2$.

Results of Subsects. \ref{sec3.2}, \ref{sec3.2bis}, \ref{sec3.3} are
standard results. We used the reference \cite{Ya1}. Results of Subsects. \ref{sec3.4},
\ref{sec3.5}, \ref{sec3.6} are easy adaptations from those in
\cite{Ya2}.

\def\Arg{{\rm Arg}}
For $-\pi\leq a<b\leq \pi$, we denote by $\Arg]a, b[$ the open sector
$\{z\in \cc|\ \ a<{\rm arg}z<b\}$. The corresponding closed
sector (with $0$ excluded) will be denoted by $\Arg[a,b]$.  For
$\alpha\in \rr$ the function $z^{\alpha}$ is defined by $(r\e^{\i
\theta})^{\alpha}= r^{\alpha}e^{\i \alpha\theta}$, for
$-\pi<\theta\leq \pi$.
\subsection{Jost solutions for quickly decreasing potentials}\label{sec3.2}
For two solutions $f, g$ of the equation
\[
-u"+ Vu= \zeta^{2}u,
\]
the Wronskian $W(f, g)= f'(x)g(x)-f(x)g'(x)$ is independent on $x$.
\def\Im{{\rm Im}}
We start  by recalling a well-known fact about existence of Jost solutions.
\begin{proposition}\label{3.2}
Assume $V\in S^{-\mu}(\rr)$ for $\mu>2$.
Then for any $\zeta\in \Arg[0, \pi]$ there exist unique solutions
$\theta_{\pm}(x, \zeta)$ of
\[
-u"+ Vu = \zeta^{2}u,
\]
with asymptotics
\[
 \theta_{\pm}(x, \zeta)= \e^{\pm \i \zeta \cdot x}(1+ o(1)), \ \
\theta_{\pm}'(x, \zeta)= \pm\i \zeta\e^{\pm \i \zeta \cdot x}(1+
o(1))
\]
when $x\to \pm \infty$. They satisfy the estimates
\[
|\theta_{\pm}(x, \zeta)-\e^{\pm\i \zeta x}|\leq \e^{\mp{\rm
Im}\zeta \cdot x}C\langle x\rangle^{-\mu +1},
\]
uniformy for $\pm x\geq 0$ and $\zeta\in \Arg[0, \pi]$.

Moreover one has:
\[
\overline{\theta_{\pm}}(x, \zeta)= \theta_{\pm}(x, -\overline{\zeta}).
\]
\end{proposition}
\proof Uniqueness of $\theta_{\pm}$ is obvious since the Wronskian
of two solutions vanishes at  $\pm\infty$. We look for
$\theta_{\pm}(x, \zeta)$ as  solutions of the Volterra equations:
\beq\label{e3.01} \theta_{\pm}(x, \zeta)= \e^{\pm \i  \zeta \cdot
x}+ K_{\pm}\theta_{\pm}(x, \zeta), \eeq where:
\[
K_{+}(\zeta)u(x)= \int_{x}^{+\infty}\zeta^{-1}\sin
( \zeta(y-x))V(y) u(y)\d y,
\]
\[
K_{-}(\zeta)u(x)= \int_{-\infty}^{x}\zeta^{-1}\sin
( \zeta(x-y))V(y) u(y)\d y.
\]
Using the bound
\[
|\zeta^{-1}\sin (\zeta(y-x))\e^{-{\rm Im} \zeta \cdot y}|\leq C
y\e^{-{\rm Im} \zeta \cdot x}, \ \ 0\leq x\leq y,
\]
we obtain that
\[
|(K_{+}(\zeta))^{n}u(x)|\leq \e^{-{\rm Im}\zeta \cdot
x}(n!)^{-1}(C\int_{x}^{+\infty}y |V(y)|\d y)^{n}, \hbox{ for
}x\geq 0,
\]
which gives the estimate
\[
|\theta_{+}(x, \zeta)-\e^{\i \zeta \cdot x}|\leq \e^{-\Im \zeta
\cdot x}(\e^{C\int_{x}^{+\infty} |y||V(y)| \d y}-1),
\]
proving the desired bound for $\theta_{+}(x, \zeta)$. The case of
$\theta_{-}(x, \zeta)$ is treated similarly.  The last identity
follows from uniqueness.
\qed

\medskip

We recall additional identities between Jost solutions
$\theta_{\pm}(\cdot, \zeta)$ if $\zeta=k>0$.
We first set:
\[
w(k):= W(\theta_{+}(\cdot, k), \theta_{-}(\cdot, k)).
\]
Next  by computing the
Wronskian below at $\pm\infty$, we get that:
\[
W(\theta_{\pm}(\cdot, k), \theta_{\pm}(\cdot, -k))= \pm 2\i k.
\]
Clearly
\beq
\begin{array}{l}
\theta_{-}(x, k)= m_{++}(k)\theta_{+}(x, k)+ m_{+-}(k)\theta_{+}(x,
-k), \\
\theta_{+}(x, k)= m_{--}(k)\theta_{-}(x, k)+ m_{-+}(k)\theta_{-}(x,
-k).
\end{array}
\label{e3.1}
\eeq
We set
\beq
m(k):= (2\i k)^{-1}w(k).
\label{e3.0}
\eeq
We can express the coefficients in (\ref{e3.1}) using Wronskians  and
get
\beq
\begin{array}{l}
m_{+-}(k)= m_{-+}(k)= (2\i k)^{-1}W(\theta_{+}(\cdot,
k)\theta_{-}(\cdot, k))= m(k),\\
m_{++}(k)=-(2\i k)^{-1}W( \theta_{+}(\cdot, -k), \theta_{-}(\cdot, k)),
\\
m_{--}(k)= -(2\i k)^{-1}W( \theta_{+}(\cdot, k), \theta_{-}(\cdot,
-k)).
\end{array}
\label{e3.2}
\eeq
Using the identity $\theta_{\pm}(x, -k)= \overline{\theta}_{\pm}(x, k)$
and iterating the identities (\ref{e3.1}), we obtain
\beq
\begin{array}{l}
m(-k)= \overline{m}(-k), \ \ \overline{m}_{++}(k)= -m_{--}(k), \\
|m(k)|^{2}= 1+ |m_{++}(k)|^{2}= 1+ |m_{--}(k)|^{2}.
\end{array}
\label{e3.3}
\eeq
\subsection{Resolvent and spectral family}\label{sec3.2bis}
\begin{proposition}\label{3.3}
The family $\{\psi(\cdot, k)\}$ defined by
\beq\label{e3.7}
\psi(x, k):=\left\{\begin{array}{l}
m(k)^{-1}\theta_{+}(x, k)\ \ k>0, \\
m(-k)^{-1}\theta_{-}(x, -k) \ \ k<0
\end{array}\right.
\eeq
is a family of generalized eigenfunctions of $h$.
\end{proposition}
\proof
Since $\theta_{\pm}(\cdot, \zeta)\in L^{2}(\rr^{\pm})$ for $\Im
\zeta>0$, we obtain by
the standard argument that the resolvent $(h-z)^{-1}$ has kernel
\[
R(x, y, z)= \left\{
\begin{array}{l}
-w(\zeta)^{-1} \theta_{+}(x, \zeta) \theta_{-}(y, \zeta), \ \ y\leq x,
\\
-w(\zeta)^{-1} \theta_{-}(x, \zeta) \theta_{+}(y, \zeta), \ \ x\leq y.
\end{array}
\right.
\]
for $\zeta^{2}= z$, $\Im \zeta>0$ and
\[
w(\zeta)= W(\theta_{+}(\cdot, \zeta), \theta_{-}(\cdot, \zeta)).
\]
The zeroes of $w$ lie on $\i \rr^{+}$ and correspond to negative
eigenvalues of $h$.

If $E(\lambda)=\one_{]-\infty, \lambda]}(h)$, then from
\[
(2\i \pi)\frac{\d E}{\d \lambda}(\lambda)= R(\lambda+ \i
0)-R(\lambda-\i 0),
\]
We obtain that for $\lambda>0$ $\frac{\d E}{\d \lambda}(\lambda)$ has
a kernel satisfying:
\[
4\pi k \frac{\d E}{\d \lambda}(x, y, \lambda)= m(k)^{-1}\theta_{+}(x,
k)\theta_{-}(y, -k) + m(-k)^{-1}\theta_{+}(x, -k)\theta_{-}(y, k), \ \
\hbox{for }y\leq x,
\]
where $k^{2}= \lambda$.
Note that $\frac{\d E}{\d \lambda}(\lambda)$ is both real and
selfadjoint hence $\frac{\d E}{\d \lambda}(x, y, \lambda)=\frac{\d
E}{\d \lambda}(y,x, \lambda)$.

Using the identities (\ref{e3.1}) and (\ref{e3.3}), we obtain that
\beq
4\pi k \frac{\d E}{\d \lambda}(x, y, \lambda)= |m(k)|^{-2}\left(
\theta_{+}(x, k)\theta_{+}(y, -k)+ \theta_{-}(x, k)\theta_{-}(y,
-k)\right)
\label{e3.4}
\eeq
for $k^{2}= \lambda$.
Setting
\beq
\psi_{\pm}(x, k):= m(k)^{-1}\theta_{\pm}(x, k) \ \ k>0,
\label{e3.5}
\eeq
we obtain
\beq
4\pi k \frac{\d E}{\d \lambda}(x, y, \lambda)= \psi_{+}(x,
k)\overline{\psi}_{+}(y, k)+ \psi_{-}(x,
k)\overline{\psi}_{-}(y, k),
\label{e3.6}
\eeq
for $k^{2}= \lambda$, which shows that $\{\psi(\cdot, k)\}_{k\in \rr}$
defined in (\ref{e3.7}) is a family of generalized eigenfunctions of
$h$. \qed

\subsection{Condition {\it (BM2)} for quickly decreasing
potentials}\label{sec3.3}
Let us now consider in more details the Volterra integral equations
(\ref{e3.01}) for $\zeta=k>0$. Let $F_{\pm}$ be the Banach space of
$C^{1}$ functions on $\rr^{\pm}$  bounded with bounded derivatives
equipped with the obvious norm.

The operators $(\one -K_{\pm}(k))^{-1}$ are bounded on $F_{\pm}$
and $]0, +\infty[\ni k\mapsto (\one -K_{\pm}(k))^{-1}\in
B(F_{\pm})$ is norm continuous. It follows that $k\to
\theta_{\pm}(\cdot, k)\in F_{\pm}$ is continuous on $]0,
+\infty[$, and hence $w(k)$ is continuous on $]0, +\infty[$.

Moreover when $k\to 0$, $(\one
-K_{\pm}(k))^{-1}$ converges in $B(F_{\pm})$ to $(\one
-K_{\pm}(0))^{-1}$, where
\[
\begin{array}{l}
K_{+}(0)u(x)= \int_{0}^{+\infty}(y-x)V(y)u(y)\d y, \\[2mm]
K_{-}(0)u(x)=\int _{-\infty}^{x}(x-y)V(y)u(y)\d y.
\end{array}
\]
Therefore
\[
\lim_{k\to 0}w(k)=:w(0)\hbox{ exists}
\]
and $w(0)=0$ iff there exists a solution $u$
of
\[
-u"+ Vu=0,
\]
with asymptotics:
\[
u(x)\to u_{\pm}, \ \ u'(x)\to 0\hbox{ for }x\to \pm
\infty, \ \ u_{\pm} \neq 0.
\]
Such a solution is called a {\em zero energy resonance} for $h$.
Recall that condition {\it (BM2')} is introduced in Remark \ref{remi}.
\begin{proposition}\label{3.3bis}
Assume that $V\in S^{-\mu}$ for $\mu>2$. Then:

1) if  $h$ has no zero
energy resonance, then  $h$ satisfies
{\it (BM2)} for $M(x)\equiv 1$.

2) if $h$ has a zero energy resonance and $|w(k)|\geq
C|k|^{3/2-\epsilon}$
in $|k|\leq 1$ for some $\epsilon>0$ then   $h$
satisfies {\it (BM2')} for $M(x)\equiv 1$.
\end{proposition}
\begin{remark} Assume $V\in S^{-\mu}$ for $\mu>3$. Then if $h$ has a
resonance, $|w(k)|\geq C|k|$ (see \cite[Prop.7.13]{Ya1}).
\end{remark} \proof For $k>0$ we deduce from (\ref{e3.1}) that:
\[
\psi(x, k)=\left\{
\begin{array}{l}
\frac{1}{m(k)} \theta_{+}(x, k), \ \ x\geq 0, \\
\frac{m_{--}(k)}{m(k)} \theta_{-}(x, k)+ \theta_{-}(x, -k) \ \ x\leq 0
\end{array}
\right.
\]
and
\[
\psi(x, -k)=\left\{
\begin{array}{l}
\frac{1}{m(k)} \theta_{-}(x, k), \ \ x\leq 0, \\
\frac{m_{++}(k)}{m(k)} \theta_{+}(x, k)+ \theta_{+}(x, -k) \ \ x\geq 0
\end{array}
\right.
\]
By Prop. \ref{3.1} the functions $\theta_{\pm}(x, k)$ are
uniformly  bounded in $k> 0$ and $x\in \rr^{\pm}$. Moreover by
(\ref{e3.3})
\[
|\frac{m_{++}(k)}{m(k)}|\leq 1, \ \ |\frac{m_{--}(k)}{m(k)}|\leq 1.
\]
Therefore it suffices to bound  $m(k)^{-1}$.
 Using the integral equations (\ref{e3.01}), we
obtain that:
\[
\theta_{\pm}(0, k)= 1+ O(k^{-1}), \ \ \theta_{\pm}'(0, k)= \pm \i k+
O(1)
\]
when $k\to +\infty$. Therefore $w(k)= 2\i k(1+ O(k^{-1}))$ and
$m(k)^{-1}$ is bounded near $+\infty$.  If $w(0)=\lim_{k\to 0}w(k)\neq
0$, then $\lim_{k\to 0}m(k)^{-1}=0$ and hence $m(k)^{-1}$ is uniformly
bounded on $]0, +\infty[$. If $|w(k)|\geq C|k|^{-3/2+\epsilon}$ for
$|k|\leq 1$, we get instead that $|m(k)|^{-1}\leq C|k|^{-\12+ \epsilon}$ for $|k|\leq
1$. This completes the proof of the theorem.
\qed

\subsection{Quasiclassical solutions for slowly decreasing
potentials}\label{sec3.4}
Let $V\in S^{-\mu}$ for $0<\mu\leq 2$.  For $\Im\zeta\geq 0$, we set
\[
F(x, \zeta):=(V(x)-\zeta^{2})^{\12},
\]
where $z^{\12}$ is defined as in the beginning of Sect. \ref{urk}.
We see that $F(x, \zeta)$ is holomorphic in $\zeta$ in the two sectors
$\Arg]0, \pi/2[$ and $\Arg]\pi/2, \pi[$ and $C^{\infty}$ in $x$ if $\zeta$
belongs to the above sectors. It is continuous in $\zeta$ in the
closed sectors $\Arg[0, \pi/2]$ and $\Arg[\pi/2, \pi]$ but may not be
continuous across $\Arg\zeta=\pi/2$ depending on the value of $V(x)$.

Note that $\overline{(z^{\12})}= \overline{z}^{\12}$ if $\Arg z\neq
\pi$, which implies that
\beq
\overline{F}(x, \zeta)= F(x, -\overline{\zeta}), \ \ \zeta\in \Arg]0,
\pi/2[\cup \Arg]\pi/2, \pi[, \ \ \overline{F}(x, k)= F(x, -k) \ \ k>0.
\label{e3.10}
\eeq
We set also
\beq
S(a, x,\zeta):=\int_{a}^{x}F(y, \zeta)\d y, \ \ a\in \rr.
\label{e3.11}
\eeq
\begin{proposition}\label{3.4}
For $\zeta\in \Arg[0, \pi]$, $\zeta\neq 0$
there exist unique solutions $\eta_{\pm}(x, \zeta)$ of
\[
-u"+ Vu= \zeta^{2}u, \ \ \ \ \hbox{(E)}
\]
with asymptotics
\[
\eta_{+}(x, \zeta)= F(x, \zeta)^{-\12}\e^{-S(0, x, \zeta)}(1+ o(1)), \
\ \eta_{+}'(x, \zeta)= -F(x, \zeta)^{\12}\e^{-S(0, x, \zeta)}(1+ o(1)),
\ \ x\to +\infty,
\]
\[
\eta_{-}(x, \zeta)= F(x, \zeta)^{-\12}\e^{S(0, x, \zeta)}(1+ o(1)), \
\ \eta_{-}'(x, \zeta)= F(x, \zeta)^{\12}\e^{S(0, x, \zeta)}(1+ o(1)),
\ \ x\to -\infty.
\]
We have
\[
\overline{\eta}_{\pm}(x, \zeta)= \eta_{\pm}(x, -\overline{\zeta}).
\]
For $\epsilon>0$ let $R(\epsilon)$ be such that $|V(y)|\leq
\epsilon/2$ for $|y|\geq R(\epsilon)$. Then the following estimates
are valid:
\beq\label{e3.15}
|\eta_{\pm}(x, \zeta)|\leq C(\epsilon)|\zeta|^{-\12}, \ \
|\eta_{\pm}'(x, \zeta)|\leq C(\epsilon)|\zeta|^{\12}, \hbox{uniformly
in }\pm x\geq \pm R(\epsilon), \ \ |k|\geq \epsilon.
\eeq
\end{proposition}
\proof
We follow \cite{Ya2} and treat only the case of $\eta_{+}$. For
$|\zeta|\geq \epsilon$ and $x\geq R(\epsilon)$, we look for
$(\eta_{+}(x, \zeta), \eta'_{+}(x, \zeta))$ of the form
\beq\label{ark}
\begin{array}{rl}
\eta_{+}(x, \zeta)=&  F(x, \zeta)^{-\12}\e^{-S(0, x, \zeta)}\left(u_{1}(x,
\zeta)+ u_{2}(x, \zeta)\right), \\[2mm]
\eta'_{+}(x, \zeta)= & F(x, \zeta)^{-\12}\e^{-S(0, x,
\zeta)}\left((F(x,\zeta)-V'(x)(4F^{2})^{-1})u_{1}(x, \zeta)\right.\\[2mm]
&\left.-(F(x,
\zeta)+V'(x)(4F^{2})^{-1})u_{2}(x, \zeta)\right).
\end{array}
\eeq
We find that $(u_{1}(\cdot, \zeta), u_{2}(x, \zeta))$ has to satisfy
the following Volterra equation:
\beq\label{e3.12}
\begin{array}{l}
u_{1}(x, \zeta)= -\int_{x}^{+\infty} \e^{-2S(x, y,\zeta)}M(y,
\zeta)(u_{1}(y, \zeta)+ u_{2}(y, \zeta))\d y, \\[2mm]
u_{2}(x, \zeta)= 1+ \int_{x}^{+\infty}M(y,
\zeta)(u_{1}(y, \zeta)+ u_{2}(y, \zeta))\d y,
\end{array}
\eeq
for
\[
M(x, \zeta)= (32)^{-1}(4V"(x)F(x, \zeta)^{-3} -5(V')^{2}(x)F(x,
\zeta)^{-5}).
\]
Uniformly for $|\zeta|\geq \epsilon$ and $y\geq x\geq R(\epsilon)$, we have:
\[
|M(x, \zeta)|\leq C(\epsilon)\langle x\rangle^{-2-\mu}, \ \
|\e^{-S(x, y, \zeta)}|\leq 1.
\]
The equation (\ref{e3.12}) can be solved by iteration and we obtain as
in the proof of Prop. \ref{3.2} that:
\[
\begin{array}{l}
|u_{1}(x, \zeta)|\leq (\e^{C(\epsilon)\langle x\rangle^{-1-\mu}}-1),
\\[2mm]
|u_{2}(x, \zeta)|\leq\e^{C(\epsilon)\langle x\rangle^{-1-\mu}},
\end{array}
\]
uniformly for $|\zeta|\geq \epsilon$ and $y\geq x\geq R(\epsilon)$.
Since
\[
C_{1}(\epsilon)|\zeta|^{\12}\leq |F(x, \zeta)|\leq
C_{2}(\epsilon)|\zeta|^{\12},
\]
we obtain the desired bounds on $\eta_{+}(\cdot, \zeta)$,
$\eta'_{+}(\cdot, \zeta)$.

To prove  uniqueness of $\eta_{+}(\cdot, \zeta)$, we verify that  the
Wronskian of two solutions computed at $x=+\infty$ vanishes.
The fact that $\overline{\eta}_{\pm}(\cdot, \zeta)= \eta_{\pm}(\cdot,
-\overline{\zeta})$ follows from (\ref{e3.10}). \qed

\medskip

As in Subsect. \ref{sec3.2}, we compute some Wronskians.

\begin{lemma}\label{3.5}
For $|k|\geq\epsilon$ we have:
\[
\begin{array}{l}
W(\eta_{+}(\cdot, k), \eta_{+}(\cdot, -k))= 2\i {\rm sgn}(k)\e^{-2{\rm
Re}\int_{0}^{R(\epsilon)}(V(y)-k^{2})^{\12}\d y},\\[2mm]
W(\eta_{-}(\cdot, k), \eta_{-}(\cdot, -k))= -2\i {\rm sgn}(k)\e^{-2{\rm
Re}\int^{0}_{-R(\epsilon)}(V(y)-k^{2})^{\12}\d y}.
\end{array}
\]
\end{lemma}
\proof From Prop. \ref{3.4} we obtain that:
\[
\eta_{+}(x, k)\sim (-\i k)^{-\12}\e^{-S(0, x, k)}, \ \ \eta_{+}(x,
k)\sim -(-\i k)^{\12} \e^{-S(0, x,k)}, \ \ x\to +\infty.
\]
Using that
\[
(\i k)^{\12}(-\i k)^{-\12}= \i, \ \ (-\i k)^{\12}(\ i k)^{-\12}=
-\i\hbox{ for }k>0,
\]
we obtain
\beq\label{e3.13}
\eta'_{+}(x, k)\eta_{+}(x, -k)- \eta_{+}(x, k)\eta'_{+}(x, -k)\sim 2\i
\e^{-2{\rm Re}S(0, x, k)}, \ \ x\to +\infty.
\eeq
If $k\geq\epsilon$ and $y\geq R(\epsilon)$ we have $V(y)-k^{2}<0$ so
${\rm Re}(V(y)-k^{2})=0$. Letting $x\to +\infty$ in (\ref{e3.12}) we
obtain the first identity for $k>0$ and replacing then $k$ by $-k$ for
all $k\neq 0$. The proof of the second identity is similar, using
instead
\[
\eta_{-}(x, k)\sim (-\i k)^{-\12}\e^{S(0, x, k)}, \ \ \eta_{-}(x,
k)\sim (-\i k)^{\12} \e^{S(0, x,k)}, \ \ x\to -\infty. \ \ \Box
\]

\medskip

\begin{proposition}\label{3.6}
Set for $|k|\geq \epsilon$:
\beq\label{e3.16}
\begin{array}{l}
\theta_{+}(x, k):= \eta_{+}(x, k)|k|^{\12}\e^{{\rm
Re}\int_{0}^{R(\epsilon)}(V(y)-k^{2})\d y}, \\[2mm]
\theta_{-}(x, k):=
\eta_{-}(x, k)|k|^{\12}\e^{{\rm
Re}\int^{0}_{-R(\epsilon)}(V(y)-k^{2})\d y}.
\end{array}
\eeq
Then we have:
\[
\begin{array}{l}
\overline{\theta_{\pm}}(x, k)= \theta_{\pm}(x, -k) \ \
W(\theta_{\pm}(\cdot, k), \theta_{\pm}(\cdot, -k))=\pm 2\i k,
\end{array}
\]
and
\[
|\theta_{\pm}(x, k)|\leq C_{\epsilon}, \ \ |\theta_{\pm}'(x, k)|\leq
C_{\epsilon}|k|, \hbox{ uniformly for }|k|\geq
\epsilon, \ \ \pm x\geq 0.
\]
\end{proposition}
\proof
The first statement follows from Lemma \ref{3.5}. To prove the second
statement we use Prop. \ref{3.4}. In fact  by (\ref{e3.15}), the bounds in the second statement
are valid uniformly for $|k|\geq \epsilon$ and $\pm x\geq \pm
R(\epsilon)$. Let us first fix $C\gg 1$ such that for $\epsilon\geq C\gg 1$, we have
$R(\epsilon)=0$. Hence the bounds in the second statement are valid
uniformly for $|k|\geq C$ and $\pm x\geq 0$.

It remains to check the
bounds uniformly for $\epsilon\leq |k|\leq C$ and $\pm x\in [0,
R(\epsilon)]$. We have
\[
|\theta_{\pm}(\pm R(\epsilon), k)|+ |\theta_{\pm}'(\pm R(\epsilon),
k)|\leq C(\epsilon).
\]
Writing the differential equation satisfied by $\theta_{\pm}$ as a
first order system, we see that this bound extends to $\pm x\in [0,
R(\epsilon)]$ uniformly for $\epsilon\leq |k|\leq C$. \qed

\subsection{Resolvent and spectral family}\label{sec3.5}
\begin{proposition}\label{3.7}
We set as in Subsect. \ref{sec3.2}:
\[
w(k):= W(\theta_{+}(\cdot, k),
\theta_{-}(\cdot, k)), \ \ m(k):= (2\i k)^{-1}w(k).
\]
The family $\{\psi(\cdot, k)\}$ defined by
\beq\label{e3.7bis}
\psi(x, k):=\left\{\begin{array}{l}
m(k)^{-1}\theta_{+}(x, k)\ \ k>0, \\
m(-k)^{-1}\theta_{-}(x, -k) \ \ k<0
\end{array}\right.
\eeq
is a family of generalized eigenfunctions of $h$ in $|k|\geq
\epsilon$.

\end{proposition}
\proof
As in Subsect. \ref{sec3.2bis}, we can since $\eta_{\pm}(\cdot,
\zeta)\in L^{2}(\rr^{\pm})$ for ${\rm Im}\zeta>0$ write the
 kernel of $(h-z)^{-1}$ as:
\[
R(x, y, z)= \left\{
\begin{array}{l}
-r(\zeta)^{-1} \eta_{+}(x, \zeta) \eta_{-}(y, \zeta), \ \ y\leq x,
\\
-r(\zeta)^{-1} \eta_{-}(x, \zeta) \eta_{+}(y, \zeta), \ \ x\leq y.
\end{array}
\right.
\]
for $\zeta^{2}= z$, $\Im \zeta>0$ and
\[
r(\zeta)= W(\eta_{+}(\cdot, \zeta), \eta_{-}(\cdot, \zeta)).
\]
The zeroes of $w$ lie on $\i \rr^{+}$ and correspond to negative
eigenvalues of $h$.  We write the kernel of the spectral family
$\frac{\d E}{\d \lambda}(x, y, \lambda)$ using the functions
$\theta_{\pm}(x, \pm k)$. Using (\ref{e3.16}) we obtain:
\[
4\pi k \frac{\d E}{\d \lambda}(x, y, \lambda)= m(k)^{-1}\theta_{+}(x,
k)\theta_{-}(y, -k) + m(-k)^{-1}\theta_{+}(x, -k)\theta_{-}(y, k), \ \
\hbox{for }y\leq x,
\]
where $k^{2}= \lambda$ and

By Prop. \ref{3.6} the algebraic identities used in the proof of Prop.
\ref{3.3} are  satisfied by $\theta_{\pm}(\cdot, k)$. Repeating the
above proof we obtain the proposition. \qed

\subsection{Bounds on generalized  eigenfunctions away from $k=0$.}\label{sec3.6}
The following result shows that generalized eigenfunctions are always
uniformly bounded in $|k|\geq \epsilon$ for $\epsilon>0$.
\begin{proposition}\label{3.8}
Assume $V\in S^{-\mu}$ for $\mu>0$. Then for $\{\psi(x, k)\}_{k\in
\rr}$ defined in  (\ref{e3.7bis}) one has for all $\epsilon>0$:
\[
\|\psi(\cdot, k)\|_{\infty}\leq C_{\epsilon} \hbox{ uniformly
for }|k|\geq \epsilon.
\]
\end{proposition}

\proof  Arguing as in the proof of Thm. \ref{3.3bis} it suffices by
Prop. \ref{3.6} to verify that
\[
|m(k)|^{-1}= \frac{2|k|}{w(k)},
\]
 is uniformly bounded for
$|k|\geq \epsilon$.

We first claim that $w(k)$ is a continuous
function of $k$ in $|k|\geq \epsilon$. In fact writing the Volterra
integral equation (\ref{e3.12}) as a fixed point equation in an
appropriate Banach space of continuous functions, we see that
for a fixed $x\geq R(\epsilon)$, $u_{1}(x, k)$ and $u_{2}(x, k)$ are
continuous functions of $k$ in $|k|\geq \epsilon$. The same holds for
$\eta_{+}(x, k)$, $\eta_{+}'(x, k)$. Using the differential equation
satisfied by $\eta_{+}(\cdot, k)$, we see that $k\mapsto (\eta_{+}(0,
k), \eta'_{+}(0, k))$ is continuous in $k$. Using the same argument
for $\eta_{-}(\cdot, k)$, we obtain the continuity of $w(k)$ in
$|k|\geq \epsilon$. We note that $w(k)$ does not vanish in $|k|\geq
\epsilon$ since $w(k)=0$ would imply that $k^{2}$ is an eigenvalue of
$h$ which is impossible if $V\in S^{-\mu}$.

Therefore $|m(k)|^{-1}$ is locally bounded in $|k|\geq \epsilon$. It
remains to bound $|m(k)|^{-1}$ near infinity. We use the notation in
the proof of Prop. \ref{3.4}. Let us pick $C\gg 1$ such that
$R(C)=0$.
Then for $k\geq C$ we have:
\[
F(x,k)= -\i k(1 + 0(\langle x\rangle^{-\mu}|k|^{-2})), \ \ M(x,
k)=O(\langle x\rangle^{-2-\mu}|k|^{-3}).
\]
Using the fact that $u_{1}$, $u_{2}$ are uniformly bounded in $x\geq
0$ and $k\geq C$, we obtain from (\ref{e3.12}) that
\[
u_{1}(0, k)=O(|k|^{-3}), \ \ u_{2}(0, k)= 1+ O(|k|^{-3}),
\]
which yields
\[
\eta_{+}(0, k)= (-\i k)^{-\12}(1+ O(|k|^{-2})), \ \ \eta'_{+}(0, k)=
-(-\i k)^{\12}(1+ O(|k|^{-2})).
\]
The same argument gives
\[
\eta_{-}(0, k)= (-\i k)^{-\12}(1+ O(|k|^{-2})), \ \ \eta'_{-}(0, k)=
(-\i k)^{\12}(1+ O(|k|^{-2})),
\]
and hence $W(\eta_{+}(\cdot, k), \eta_{-}(\cdot, k))= -2+
0(|k|^{-2})$. Using that $\theta_{\pm}(x, k)= \eta_{\pm}(x,
k)|k|^{\12}$ for $k\geq C$, we obtain that $|w|(k)\sim 2|k|$ when
$k\to \infty$, which shows that $|m|^{-1}(k)$ is uniformly bounded
near infinity. \qed

\subsection{Condition {\it (BM2)} for slowly decreasing potentials}\label{sec3.6b}
In this subsection we give some classes of slowly decreasing
potentials for wich condition {\it (BM2)} holds.

As in Subsect.
\ref{sec3.3} the possible existence of {\em zero energy resonances}
has to be taken into account.
For quickly decreasing potentials, the definition of zero
energy resonances is connected with  the
integral equation (\ref{e3.01}) for $\zeta=0$.
For slowly decreasing potentials, we have to consider instead the integral
equations (\ref{e3.12}).
This leads to the following definition:

Assume that $v\in S^{-\mu}$ for $0<\mu<2$ is such that $|V(x)|\geq
c\x^{-\mu}$ for $|x|$ large enough.
We will say that $h$ has a {\em zero energy resonance} if there exists
a solution of
\[
-u"+ Vu=0,
\]
with asymptotics:
\[
\begin{array}{rl}
u(x)=& u_{\pm}V(x)^{-1/4}\e^{\mp  \int_{0}^{x}(V(s))^{\12}\d s}(1+
o(1)), \ \ x\to \pm\infty,\\[2mm]
 u'(x)=& \mp u_{\pm}V(x)^{1/4}\e^{\pm
\int_{R}^{x}(V(s))^{\12}\d s}(1+ o(1)), \ \ x\to \pm\infty,
\end{array}
\]
for constants $u_{\pm}\neq 0$.
\medskip

{\bf Potentials negative near infinity.}

\medskip

We consider first  the case of potentials which are {\em negative } near
infinity.
We assume that $V\in S^{-\mu}$ for $0<\mu<2$ and:
\beq
V(x)\leq -c \langle x\rangle^{-\mu} \hbox{ in }|x|\geq R, \hbox{ for
some }c, R>0.
\label{negat}
\eeq
\begin{proposition}\label{suffsuff}
Assume that $V\in S^{-\mu}$ for $0<\mu<2$ satisfies (\ref{negat}) and
has no zero energy resonance.
Then condition {\it (BM2)} holds for $M(x)=\x^{\mu/4}$.
\end{proposition}
\proof
By Prop. \ref{3.8} it suffices to consider the region $|k|\leq 1$.
We fix $R$ as (\ref{negat}) and  define the functions
$\eta_{\pm}(x, k)$ using  the phase $S(\pm R, x, \zeta)$. We will
consider only the $+$ case.
We first claim that
\beq
\theta_{+}(x, k)\in O(\x^{\mu/4}), \ \ \theta_{+}'(x, k)\in O(1),
\hbox{ uniformly in } x\geq -R, \ \ |k|\leq 1.
\label{eti}
\eeq
Clearly it suffices to prove the statement in $x\geq R$, since we can
extend the bound  to $[-R, R]$ using the differential equation
satisfied by $\theta_{+}$.  Let us prove (\ref{eti}).
We will simply write  "$f(x, k)\in O(\x^{\epsilon})$" for "$f(x, k)\in
O(\x^{\epsilon})$ uniformly in $x\geq R$, $|k|\leq 1$".

The function $F(x, k)$ is smooth in $|x|\geq R$ and one
has $|F(x, k)|\geq c\x^{-\mu/2}$. This
implies that $M(x, k)\in O(\x^{-2+ \mu/2})$,
from which we get
\[
u_{1}(x, k)\in O(\x^{-1+\mu/2}), \ \ u_{2}(x, k)\in O(1),
\]
and
\[
\eta_{+}(x, k)\in O(\x^{\mu/4}), \ \ \eta'_{+}(x, k)\in O(1).
\]
This proves (\ref{eti}). Next as in Subsect. \ref{sec3.3}, we can
set $U=(u_{1}, u_{2}-1)$ and consider the equations (\ref{e3.12}) as a fixed
point equation:
\[
(\one + T(k))U= F,
\]
in the Banach space
\[
{\cal B}=\{U=(v_{1}, v_{2})\:| v_{i}\hbox{ continuous},\ \ \sup_{[R,
+\infty[}|\x^{1-\mu/2}v_{i}(x)|<\infty \}.
\]
For $R$ large enough, $\|T(k)\|<\12$ uniformly in $|k|\leq 1$ and
$k\mapsto T(k)$ is norm continuous. It follows that $k\mapsto U(k)\in
{\cal B}$ is continuous up to $k=0$. Therefore  $(u_{1}(\cdot, k),
u_{2}(\cdot, k)-1)$  has a limit $(u_{1}(\cdot, 0), u_{2}(\cdot, 0)-1)$ in ${\cal B}$ when $k\to 0$.

This implies also
that $(\eta_{+}(\cdot, k), \eta_{+}'(\cdot, k))$
converges locally uniformly when $k\to 0$ to the pair
$(\eta_{+}(\cdot, 0), \eta'_{+}(\cdot, 0))$ obtained from  $(u_{1}(\cdot, 0),
u_{2}(\cdot, 0))$ by formula (\ref{ark}) for $k=0$.

We see that $\eta_{+}(x, 0)$ is a solution of
\[
-u''+ V(x)u=0,
\]
with asymptotics:
\[
\begin{array}{l}
\eta_{+}(x, 0)= V(x)^{-1/4}\e^{\i \int_{R}^{x}(-V(s))^{\12}\d s}(1+
o(1)), \\[2mm]
 \eta'_{+}(x, 0)= -V(x)^{1/4}\e^{\i
\int_{R}^{x}(-V(s))^{\12}\d s}(1+ o(1)),
\end{array}\ \ x\to +\infty.
\]
By the convergence result above (and its analog for $\eta_{-}(\cdot,
k)$), we also see that
\[
\lim_{k\to 0}m(k)=:m(0)=cW(\eta_{+}(\cdot, 0), \eta_{-}(\cdot, 0)),
\]
 for some $c\neq 0$. Clearly $m(0)=0$ iff $h$ admits a zero energy
resonance. Using (\ref{e3.7bis}),
(\ref{eti}) and Prop. \ref{3.8}, we obtain then that
\[
|\psi(x, k)|\leq C\x^{\mu/4}, \hbox{ uniformly for }x\in \rr, k\in
\rr,
\]
which completes the proof of the proposition. \qed

\medskip

{\bf Potentials positive near infinity.}

\medskip

Let us now consider the case of potentials wich are {\em positive near
infinity}.
The following lemma is shown in
\cite[Thm. 4]{Ya2}.
\begin{lemma}\label{3.9}
 Assume that $V\in S^{-\mu}$ for $0<\mu<2$ is {\em positive
near infinity}, more precisely:
\beq\label{e3.17}
V(x)\sim q_{0}|x|^{-\mu}, x\to \infty, \ \ q_{0}>0.
\eeq
Then there exists unique solutions $\eta_{\pm}(x, 0)$ of
\[
-u"+ Vu=0,
\]
with asymptotics:
\[
\eta_{+}(x, 0)\sim V(x)^{-\frac{1}{4}}\e^{-\int_{a}^{x}(V(y))^{\12}\d
y}, \ \ \eta'_{+}(x, 0)\sim -V(x)^{\frac{1}{4}}\e^{-\int_{a}^{x}(V(y))^{\12}\d
y}, \ \ x\to +\infty,
\]
\[
\eta_{-}(x, 0)\sim V(x)^{-\frac{1}{4}}\e^{-\int_{a}^{x}(V(y))^{\12}\d
y}, \ \ \eta'_{-}(x, 0)\sim V(x)^{\frac{1}{4}}\e^{-\int_{-a}^{x}(V(y))^{\12}\d
y}, \ \ x\to -\infty,
\]
where $a\gg 1$ is such that $V(x)>0$ in $|x|\geq a$.
\end{lemma}

 \begin{lemma}\label{3.10}
Assume in addition to (\ref{e3.17}) that there exists $\theta, R>0$
such that $V$ extends
holomorphically to $D(R, \theta)=\{z\in \cc| |z|>R, \ \ |{\rm Arg}
z|<\theta\}$ and satisfies
\[
|V(z)|\leq C(1+ |z|)^{-\mu}, \ \ z\in D(R, \theta).
\]
Then for any $\pm x\geq R$, $(\eta_{\pm}(x, s),
\eta'_{\pm}(x, s))$ converges to $(\eta_{\pm}(x, 0),
\eta'_{\pm}(x, 0))$ when $s\to 0$.
\end{lemma}
\proof:
We check  that the assumptions of \cite[Thm. 7]{Ya2} are satisfied.
We consider the two parts $D^{\pm}(R, \theta)=D(R, \theta)\cap
\{\pm{\rm Re}z>0\}$ of $D(R, \theta)$ and set $z'={\rm log}(\pm z)$ for $z\in D^{\pm}(R,
\theta)$. Applying
 Hadamard three lines theorem to $F(z')= V(\e^{z'})(\pm
\e^{\mu z'})-q_{0}$, we obtain
\[
V(z)\sim q_{0}(\pm z)^{-\mu} \hbox{ when }|z|\to +\infty,
\]
uniformly in $D(R, \theta_{0})\cap\{\pm {\rm Re}z>0\}$ for all
$0<\theta_{0}<\theta$. Similarly it follows from Cauchy's inequalities that
\[
|\p_{z}^{k}V(z)|\leq C(1+ |z|)^{-k-\mu}, \ \ z\in D(R,
\theta_{0}).
\]
Applying then \cite[Thm. 7]{Ya2}, we obtain the lemma. \qed

\begin{lemma}\label{ilt}
The functions $\eta_{\pm}(x, k)$ are uniformly bounded for $|x|\leq
R$, $|k|\leq 1$.
\end{lemma}
\proof We consider only the case of $\eta_{+}(x, k)$. Let $\phi_{0}(x,
k)$, $\phi_{1}(x, k)$ the two regular solutions of {\it (E)} with boundary
conditions:
\[
\phi_{0}(0,k)=1, \: \phi'_{0}(0, k)=0, \ \ \phi_{1}(0, k)=0, \:
\phi'_{1}(0, k)=1.
\]
Clearly $\phi_{i}(x, k)$,  $\phi'_{i}(x,k)$ are uniformly bounded
and continuous in $\{(x, k)|\: |x|\leq R, \ \ |k|\leq C\}$. We have
\[
\eta_{+}(\cdot, k)= a_{1}(k)\phi_{0}(\cdot, k)+
a_{0}(k)\phi_{1}(\cdot, k),\hbox{ for }a_{i}(k)= W(\eta_{+}(\cdot, k),
\phi_{i}(\cdot, k)).
\]
By Lemma \ref{3.10}, $a_{i}(k)$ converges to $a_{i}(0)$ when $k\to 0$.
\qed

\medskip

\begin{proposition}\label{suffsuffsuff}
Assume that $V\in S^{-\mu}$ for $0<\mu<2$ satisfies the hypotheses of Lemma
\ref{3.10} and has no zero energy resonance. Then for each $R>0$ condition {\it
(BM2)} is satisfied for
\[
M(x)=\left\{
\begin{array}{l}
1\hbox{ for }|x|\leq R, \\[2mm]
+\infty\hbox{ for }|x|>R,
\end{array}
\right.
\]
\end{proposition}
We refer the reader to Remark \ref{infin} for the meaning of {\it
(BM2)} if $M$ takes its values in $[0, +\infty]$.

\medskip

\proof The uniform boundedness of $\eta_{\pm}(x, k)$ and hence of
$\theta_{\pm}(x, k)$ for $|x|\leq R$
and $|k|\leq 1$ is shown in Lemma \ref{ilt}. We have to show that
$\lim_{k\to 0}m(k)=:m(0)\neq 0$. The limit exists and equals
$cW(\eta_{+}(\cdot, 0), \eta_{-}(\cdot, 0))$ for some $c\neq 0$
by Lemma \ref{3.10}. By Lemma \ref{3.9} $m(0)=0$ iff $h$ has a zero
energy resonance. \qed

\section{Appendix C}\label{appc}
\subsection{Proof of Prop. \ref{1.1}}\label{appc.1}
 We forget the superscript ${\rm w}$ to simplify notation. 
We recall the following caracterization of $\Op(S^{p,m})$ known
as the {\em Beals criterion}:

$M\in \Op(S^{p,m})$ iff $M: {\cal S}(\rr^{d})\to {\cal S}(\rr^{d})$
and
\beq\label{pdoe.2}
\langle D\rangle^{-p+ |\alpha|}\langle
x\rangle^{-m+ |\beta|}\ad_{x}^{\alpha}\ad_{D}^{\beta}M\hbox{ is
bounded on }L^{2}(\rr^{d}), \hbox{ for all }\alpha, \beta\in \nn^{d}.
\eeq
The topology given by  the norms of the multicommutators
with $\Op(m)$ in (\ref{pdoe.2}) is the same as the original topology
on $S^{p,m}$.

We will need also similar objects for symbols and operators depending
on a real parameter $s\geq 0$. We say that $m(s, x,\xi)$ belongs to
$S^{p,m,k}$ if
\[
|\p_{x}^{\alpha}\p_{k}^{\beta}m(s, x, k)|\leq C_{\alpha,
\beta}(\langle k\rangle^{2}+ \langle s\rangle )^{k}\langle
k\rangle^{-p+|\alpha|}
\langle x\rangle^{-m+|\beta|}), \ \ \alpha, \beta\in
\nn^{d},
\]
uniformly for $s\geq 0$. By the result recalled above, we see that
$M(s)\in \Op(S^{p,m, k})$
iff $M(s):{\cal S}(\rr^{d})\to {\cal S}(\rr^{d})$ and
\beq
(\langle D\rangle^{2}+ \langle
s\rangle)^{-k}\langle D\rangle^{-p+ |\alpha|}\langle
x\rangle^{-m+ |\beta|}\ad_{x}^{\alpha}\ad_{D}^{\beta}M(s)\hbox{ is
bounded on }L^{2}(\rr^{d}),
\label{pdoe.3}
\eeq
uniformly for $s\geq 0$.

Let us now prove Prop.  \ref{1.1}.
By elliptic regularity, we know that $h$ is selfadjoint  and bounded
below on
$H^{2}(\rr^{d})$
 and $(h+s)^{-1}$ preserves $\cS(\rr^{d})$. Computing multicommutators
$\ad^{\alpha}_{x}\ad^{\beta}_{D}(h+s)^{-1}$ on $\cS(\rr^{d})$, we
first see
 by induction on $\alpha, \beta$ that $(\langle D\rangle^{2}+
\langle s\rangle)\x^{\alpha}\tD^{\beta}(h+s)^{-1}
\x^{-\alpha}\tD^{-\beta}\in O(1)$, uniformly in $s\geq 0$.

The same computations show then  that $(\tD^{2}+ \langle
s\rangle)\tD^{|\alpha|}\x^{|\beta|}\ad^{\alpha}_{x}\ad^{\beta}_{D}(h+s)^{-1}$
is uniformly bounded on $L^{2}(\rr^{d})$, which by the Beals criterion
show that
\beq\label{sit}
(h+s)^{-1}\in \Op(S^{0,0, -2}).
\eeq
Using the formula
\beq\label{pdoe.6}
\lambda^{-\alpha}= c_{\alpha}\int_{0}^{+\infty}s^{-\alpha}(\lambda+s)^{-1}\d s,
\hbox{ for }\lambda\geq 0, \ \ \alpha\in ]0, 1[
\eeq
we obtain that $h^{-\alpha}\in \Op(S^{-2\alpha, 0})$ for
$\alpha\in ]0, 1[$. Using also that $h^{n}\in \Op(S^{2n, 0})$ for
integer $n$, we obtain {\it ii)}.

\subsection{Proof of Prop. \ref{exemple1}}

Let us first prove {\it i)}.  We use the notation in Subsect.
\ref{appc.1}. Set $T(s)= \Op((b+s)^{-1})$. By pdo calculus
and (\ref{sit}), we get that
\[
(h+s)T(s)-\one \in \Op(S^{0, -1-\mu, -\12})
\]
hence
\[
(h+s)^{-1}- T(s)\in \Op(S^{0, -1-\mu, -3/2}).
\]
Using (\ref{pdoe.6}) for $\epsilon=\12$ this implies that
\[
h^{-\12}-\Op(b^{-\12})\in \Op(S^{-2, -1-\mu}).
\]
Next we write using again pdo calculus:
\[
h^{\12}= hh^{-\12}= h\Op(b^{-\12})+ \Op(S^{0, -1-\mu})=
\Op(b^{\12})+ \Op(S^{0, -1-\mu}),
\]
which proves {\it i)}. Let us now prove {\it ii)}.
 By Prop. \ref{1.1},
we know that
\[
\omega= \Op(c), \hbox{ for }c-b^{\12}\in S^{0, -1-\mu}, 
\]
where $b$ is defined in Prop. \ref{1.1}. By pseudodifferential
calculus, we obtain that:
\[
[\omega, \i [\omega, \i \x]]= \Op(\{c, \{c, \x\}\})+ \Op(S^{0, -2}).
\]
Since $c-\langle k\rangle\in S^{1, -\mu}$, we get:
\[
\{c, \{c, \x\}\}=\{\langle k\rangle, \{\langle k\rangle, \x\}\} + S^{0, -1-\mu}=
\x^{-1}(\frac{\xi^{2}}{\langle k\rangle^{2}}-\frac{(\xi|x)^{2}}{\langle k\rangle^{2}\x^{2}})+ S^{0, -1-\mu}.
\]
We pick $0<\epsilon\ll 1$ and write:
\[
(\frac{\xi^{2}}{\langle k\rangle^{2}}-\frac{(\xi|x)^{2}}{\langle
k\rangle^{2}\x^{2}})= d^{2}(x, \xi) -\x^{-2\epsilon},
\hbox{ for }d(x, \xi)=
(\frac{\xi^{2}}{\langle k\rangle^{2}}-\frac{(\xi|x)^{2}}{\langle k\rangle^{2}\x^{2}}+\x^{-2\epsilon})^{\12}.
\]
Using that $d^{2}\in S^{0, 0}$ and $d^{2}\geq \x^{-2\epsilon}$, we see
easily that $d\in S^{0, 0}_{\epsilon}$, hence $\x^{-\12}d\in S^{0,
-\12}_{\epsilon}$.

Using again (\ref{calc-bis}), we get:
\[
\Op(\x^{-1}d^{2})= \Op(\x^{-\12}d)^{2}+ \Op(S_{\epsilon}^{0, -3+4\epsilon}).
\]
Choosing $\epsilon>0$ small enough and setting $\gamma= \x^{-\12}d$,  we obtain the proposition. \qed

\subsection{A technical lemma}
\begin{lemma}\label{trouduc}
Let $h= Da(x)D + c(x)$ for $a,c$ as in (\ref{e1.1}),
$h_{\infty}= D^{2}+
m_{\infty}^{2}$.
Set $\omega= h^{\12}$, $\omega_{\infty}= h_{\infty}^{\12}$.
Let $\chi\in \coinf(\rr)$ and $F\in \cinf(\rr)$ with $F\equiv 0$ near
$0$ and $F\equiv 1$ near $\infty$. Then for $C$ large enough
\[
\chi(\frac{\omega}{\kappa})F(\frac{\omega_{\infty}}{C\kappa})\omega_{\infty}\in
O(1).
\]
\end{lemma}
\proof  We know from Prop.  \ref{1.1} {\it ii)} that $\omega$ and
$\omega_{\infty}$ and hence $[\omega, \omega_{\infty}]$ belong to
$\Op^{\rm w}(S^{1, 0})$. Using formula (\ref{HS}), we deduce from this
fact that for $\chi\in\coinf(\rr)$:
\beq
[\chi(\frac{\omega}{\kappa}), \omega_{\infty}]\in O(1).
\label{etoto.1}
\eeq
We take $\chi_{1}\in \coinf(\rr)$ with $\chi_{1}\chi=\chi$ and set
\[
\tilde{\omega}_{\infty}= \chi_{1}(\frac{\omega}{\kappa})\omega_{\infty}\chi_{1}(\frac{\omega}{\kappa}).
\]
We first see  that
\beq
\chi(\frac{\omega}{\kappa})(\omega_{\infty}-\tilde{\omega}_{\infty})=
[\chi(\frac{\omega}{\kappa}),
\omega_{\infty}](1-\chi_{1})(\frac{\omega}{\kappa})\in O(1),
\label{etoto.3}
\eeq
by (\ref{etoto.1}). We claim also that for $F\in \coinf(\rr)$:
\beq
\chi(\frac{\omega}{\kappa})(F(\frac{\omega_{\infty}}{\kappa})-F(\frac{\tilde{\omega}_{\infty}}{\kappa}))\in
O(\kappa^{-1}).
\label{etoto.2}
\eeq
In fact we write using (\ref{HS}):
\[
\begin{array}{rl}
&\chi(\frac{\omega}{\kappa})(F(\frac{\omega_{\infty}}{\kappa})-F(\frac{\tilde{\omega}_{\infty}}{\kappa}))\\[3mm]
=& \frac{\i}{2\pi}\int_{\cc}\partial_{\,\overline z}\tilde{F}(z)
\chi(\frac{\omega}{\kappa})(z-\frac{\omega_{\infty}}{\kappa})^{-1}\kappa^{-1}(\omega_{\infty}-\tilde{\omega}_{\infty})
(z-\frac{\tilde{\omega}_{\infty}}{\kappa})^{-1}\d z\wedge \d\,\overline
z\\[3mm]
=&\frac{\i}{2\pi}\int_{\cc}\partial_{\,\overline z}\tilde{F}(z)
\chi(\frac{\omega}{\kappa})(z-\frac{\omega_{\infty}}{\kappa})^{-1}\kappa^{-1}\chi(\frac{\omega}{\kappa})(\omega_{\infty}-\tilde{\omega}_{\infty})
(z-\frac{\tilde{\omega}_{\infty}}{\kappa})^{-1}\d z\wedge \d\,\overline
z\\[3mm]
&+\frac{\i}{2\pi}\int_{\cc}\partial_{\,\overline z}\tilde{F}(z)
(z-\frac{\omega_{\infty}}{\kappa})^{-1}\kappa^{-1}[\chi(\frac{\omega}{\kappa}),
\omega_{\infty}](z-\frac{\omega_{\infty}}{\kappa})^{-1}\kappa^{-1}(\omega_{\infty}-\tilde{\omega}_{\infty})
(z-\frac{\tilde{\omega}_{\infty}}{\kappa})^{-1}\d z\wedge \d\,\overline
z.
\end{array}
\]
This is easily seen to be $O(\kappa^{-1})$ using the fact that
 $(z-a)^{-1}$, $a(z-a)^{-1}$ are $O(|{\rm Im}z|^{-1})$ for $z\in \supp
\tilde{F}$.

We note then that
\[
\tilde{\omega}_{\infty}\leq c_{1}\kappa,
\]
for some $c_{1}>0$ since $\omega_{\infty}\leq c_{0}\omega$. Hence if
$G(s)= F(C^{-1}s)$ for $F$ as in the lemma and $C$ is large enough, we
have $G(\frac{\tilde{\omega}_{\infty}}{\kappa})=0$.
Applying then (\ref{etoto.2}) to $F= 1-G$, we obtain that
\beq\label{etoto.5}
\chi(\frac{\omega}{\kappa})G(\frac{\omega_{\infty}}{\kappa})\in
O(\kappa^{-1}).
\eeq
We write:
\[
\begin{array}{rl}
\chi(\frac{\omega}{\kappa})G(\frac{\omega_{\infty}}{\kappa})\omega_{\infty}
=&\omega_{\infty}
\chi(\frac{\omega}{\kappa})G(\frac{\omega_{\infty}}{\kappa})+
[\chi(\frac{\omega}{\kappa}),
\omega_{\infty}]G(\frac{\omega_{\infty}}{\kappa})\\[2mm]
=&\omega_{\infty}\omega^{-1}\omega
\chi(\frac{\omega}{\kappa})G(\frac{\omega_{\infty}}{\kappa})+
[\chi(\frac{\omega}{\kappa}),
\omega_{\infty}]G(\frac{\omega_{\infty}}{\kappa}).
\end{array}
\]
The first term in the last line is $O(1)$ using (\ref{etoto.5}), the
second also using (\ref{etoto.1}). This completes the proof of the
Lemma. \qed

\end{document}